\documentclass[prb,twocolumn,notitlepage,superscriptaddress,longbibliography]{revtex4}
\usepackage{amsmath,amsfonts,amssymb,amsthm,epsfig,array}
\usepackage{dsfont}
\usepackage{amssymb}
\usepackage{slashed}
\usepackage{graphics}
\usepackage[caption=false]{subfig}
\usepackage{float}
\usepackage{verbatim}%\begin{comment} ... \end{comment}
\usepackage{color}
\usepackage[dvipsnames]{xcolor}
\usepackage[hidelinks,colorlinks,linkcolor=blue,
citecolor=blue,urlcolor=blue]{hyperref}
\usepackage[titletoc,title]{appendix}
\usepackage{multirow}
\usepackage{bibentry}

\setcitestyle{number,sort,open={(},close={)}}

\newcommand{\dd}{\mathrm{d}}

\newcommand{\bra}[1]{\langle #1|}
\newcommand{\ket}[1]{|#1 \rangle}

\newcommand{\eqnref}[1]{Eq.\,\eqref{#1}}
\newcommand{\figref}[1]{Fig.\,\ref{#1}}

\let\oldAA\AA
\renewcommand{\AA}{\text{\normalfont\oldAA}}
\newcommand{\sgn}[1]{\text{sgn}(#1)}

\newcommand{\TR}{\mathcal{T}}

\usepackage[mathscr]{euscript}
\renewcommand{\vec}[1]{{\textbf{\textit{#1}}}}
\newcommand{\addCXL}[1]{{\color{blue} #1}}
\newcommand{\addKJ}[1]{{\color{blue} #1}}

\begin{document}

	\title{$\mathbb Z_2$-Nontrivial Moir\'e Minibands and Interaction-Driven Quantum Anomalous Hall Insulators in Topological Insulator Based Moir\'e Heterostructures}
	\author{Kaijie Yang}
	\affiliation{Department of Physics, the Pennsylvania State University, University Park, PA 16802, USA}
	\author{Zian Xu}
	\affiliation{School of Materials Science and Engineering, Beihang University, Beijing, 100191, China}
	\author{Yanjie Feng}
	\affiliation{School of Materials Science and Engineering, Beihang University, Beijing, 100191, China}
	\author{Frank Schindler}
	\affiliation{Princeton Center for Theoretical Science, Princeton University, Princeton, NJ 08544, USA}
    \affiliation{Blackett Laboratory, Imperial College London, London SW7 2AZ, United Kingdom}
	\author{Yuanfeng Xu}
%	\affiliation{Max Planck Institute of Microstructure Physics, Halle, Germany}
    \affiliation{Center for Correlated Matter and School of Physics, Zhejiang University, Hangzhou, 310058, China}
	\affiliation{Department of Physics, Princeton University, Princeton, NJ 08544, USA}
	\author{Zhen Bi}
	\affiliation{Department of Physics, the Pennsylvania State University, University Park, PA 16802, USA}
	\author{B. Andrei Bernevig}
	\affiliation{Department of Physics, Princeton University, Princeton, NJ 08544, USA}
	\affiliation{Donostia International Physics Center, P. Manuel de Lardizabal 4, 20018 Donostia-San Sebastian, Spain}
	\affiliation{IKERBASQUE, Basque Foundation for Science, Bilbao, Spain}
	\author{Peizhe Tang}
	\affiliation{School of Materials Science and Engineering, Beihang University, Beijing, 100191, China}
	\affiliation{Max Planck Institute for the Structure and Dynamics of Matter and Center for Free Electron Laser Science, Hamburg 22761, Germany}
	\author{Chao-Xing Liu}
	\email{cxl56@psu.edu}
	\affiliation{Department of Physics, the Pennsylvania State University, University Park, PA 16802, USA}
	\affiliation{Department of Physics, Princeton University, Princeton, NJ 08544, USA}
	\begin{abstract}
We studied electronic band structure and topological property of a topological insulator thin film under a moir\'e superlattice potential to search for two-dimensional (2D) $\mathbb Z_2$ non-trivial isolated mini-bands. % can generally appear when the moir\'e potential approximately forms a hexagonal lattice with six-fold rotation symmetry. % \addKJ{or three -fold rotation symmetry under small two-fold symmetry breaking}.
%\addPZTang{Delta (r) in Equ.1 keeps C3 rotational symmetry. Why here, the morie potential must be six-fold. In DFT, we confirm that AB and BA stackings are different, it should be C3. In Figure3, Z2 phase appears when phi is closed to C6 rotational region, which means that C6 symmetry could be broken, but the breaking term should not be too large. I think we need to revise this part, it is a little misleading.}
To model this system, we assume the Fermi energy inside the bulk band gap and thus consider an effective model Hamiltonian with only two surface states that are located at the top and bottom surfaces and strongly hybridized with each other. The moir\'e potential is generated by another layer of 2D insulating materials on top of topological insulator films. 
%The topological insulator thin films have two surface states forming moir\'e minibands and hybridization between them with the Fermi level lying in the gap.
In this model, the lowest conduction (highest valence) mini-bands can be $\mathbb Z_2$ non-trivial when the minima (maxima) of the moir\'e potential approximately forms a hexagonal lattice with six-fold rotation symmetry.
%the min \addKJ{approximate hexagonal moir\'e potential} has two minima (maxima). 
%\addPZTang{such expression is misleading. What do you mean? the potential?}
For the nontrivial conduction mini-band cases, the two lowest Kramers' pairs of conduction mini-bands both have nontrivial $\mathbb Z_2$ invariant in presence of inversion, while applying external gate voltages to break inversion leads to only the lowest Kramers' pair of mini-bands to be topologically non-trivial. 
%when the inversion symmetry is broken \addKJ{by  asymmetric potentials comparable to moir\'e mini-band width from gate voltages}. 
The Coulomb interaction can drive the lowest conduction Kramers' mini-bands into the quantum anomalous Hall state when they are half-filled, which is further stabilized by breaking inversion symmetry. We propose the monolayer Sb$_{2}$ on top of Sb$_2$Te$_3$ thin films to realize our model based on results from the first principles calculations. 
	\end{abstract}
	\maketitle

%\addCXL{CX: please unify the notation. }
%\addCXL{CX: for CB1, CB2, VB1, VB2 here, we mean a Kramers' pair of mini-bands due to TR symmetry. I draft the statement very carefully in the introduction part, but not that carefully in the later part. One needs to revise the later part to make it clear. }
{\it Introduction -}
Recent research interests have focused on the moir\'e superlattice in 2D Van der Waals heterostructures, including graphene\cite{bistritzer2011moire,cao2018correlated,cao2018unconventional,sharpe2019emergent,yankowitz2019tuning,serlin2020intrinsic,lu2019superconductors,kennes2021moire} and transition metal dichalcogenide (TMD) multilayers\cite{zhang2017interlayer,mak2022semiconductor,wu2018hubbard,regan2020mott,tang2020simulation,alexeev2019resonantly,jin2019observation,seyler2019signatures,tran2019evidence}
%[\addCXL{cite the earliest paper and Review papers}]
, due to the strong correlation effect in the presence of flat bands. The flat bands formed by low-energy gapless Dirac fermions in magic angle twisted bilayer graphene typically have a bandwidth $\sim 5$ meV, much smaller than the band gap $25\sim 35$ meV that separates flat bands from higher energy bands and the Coulomb interaction of order $30$ meV\cite{cao2018unconventional,cao2018correlated}. 
%\addCXL{CX: can you read the literature about TMD Moire and give the typical values, e.g. bandwitdh, energy gap and Coulomb interaction in TMD Moire?} 
In contrast, the flat bands in TMD moir\'e heterostructures are formed by electrons with parabolic dispersion and have a typical bandwidth $\sim 10$ meV, separated by a comparable gap from other energy bands, and a huge on-site Coulomb interaction $\sim 100$ meV\cite{mak2022semiconductor,devakul2021magic,wu2018hubbard}. Besides the above materials, moir\'e superlattice has also been found in another family of van der Waals heterostructure consisting of topological insulators (TIs) \cite{chang2015band,salvato2022nanometric,schouteden2016moire,yin2022moire,song2010topological,wang2012scanning,liu2014tuning,xu2015van,vargas2017tunable,liu2022moir}. 
%\addCXL{CX: cite all the existing experiments that show Moire lattice in TI films}. 
These TI-based moir\'e heterostructures show different features. 
TIs have the anomalous gapless surface bands that connect the bulk conduction and valence bands due to non-trivial bulk topology.
The spin splitting of surface bands has a typical energy scale of hundreds meV due to the strong spin-orbit coupling (SOC). 
%In contrast, the graphene has a negligible SOC while TMD has spin-orbit splitting for hundreds meV\cite{zhu2011giant,reyes2016spin}.
% \addCXL{CX: please check the TMD's SOC}. 
Previous studies\cite{cano2021moire,dunbrack2022magic,wang2021moire} show that a single surface state remains gapless upon the moir\'e superlattice potential, leading to satellite Dirac cones and van Hove singularities, instead of isolated flat bands. Furthermore, the moir\'e superlattice in magnetic TI materials, e.g. MnBi$_2$Te$_4$, is predicted to host Chern insulator phase\cite{lian2020flat}.
%However, it remains a question how to obtain isolated topologically non-trivial moir\'e minibands in non-magnetic TI systems. 

% The difference between the twisted bilayer graphene (TBG) and our topological insulator (TI) based moir\'e heterostructure comes from the anomalous surface Dirac cones of topological insulators. 
% Unlike TBG couples the Dirac cones on both layers to open a gap, the moir\'e super-lattice in our model cannot provide an anomalous surface Dirac cones and open gaps between moir\'e minibands.
% A additional inter-surface coupling between Dirac cones on top and bottom surfaces is needed for isolated moir\'e minibands as shown in \figref{fig:system}.
% The gap is comparable to bandwidth and Coulomb interaction, while gaps are larger than both in TBG.
% The anomalous surface Dirac cones has strong spin-orbit coupling and located at $\Gamma$, lacking the spin and valley degrees of freedom in TBG.
% The interaction driven quantum anomalous hall effects (QAH) in our system will go beyond the valley polarized ground state in TBG.

\begin{figure}
	\centering
	\includegraphics[width=\columnwidth]{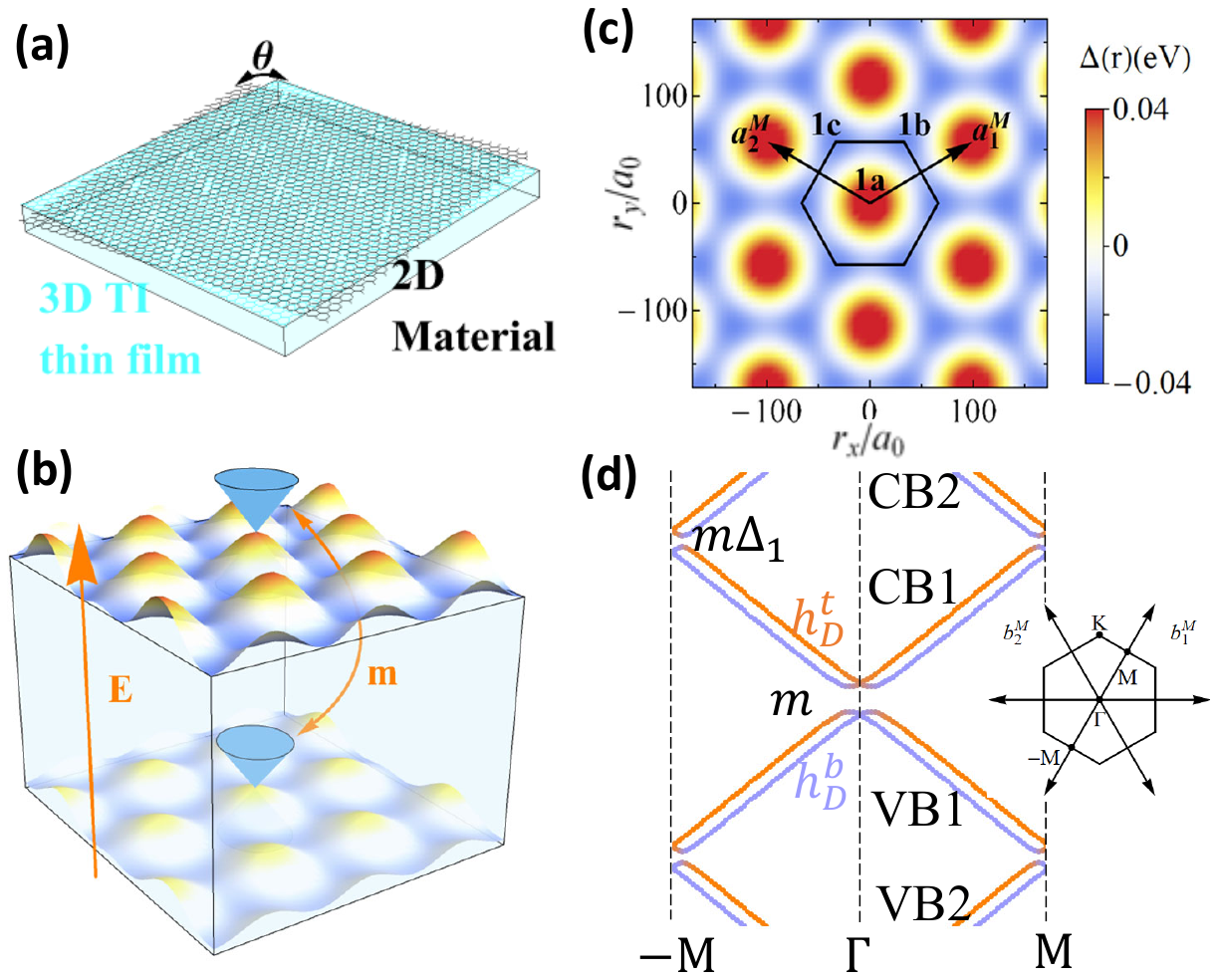}
	\caption{
		(a) A schematic figure for the twisted 2D materials (black) on top of a topological insulator thin film (cyan).
		%KJ: I change figure label accordingly which i does not color them.
		(b) Schematic illustration of the moir\'e potentials from twisted 2D materials on the top and bottom surface of a TI thin film. The blue Dirac cones represent the top and bottom surface states coupled by $m$. An out-of-plane external electrical field $E$ creates the potential $V_0$. 
		%\addCXL{inset is too small, difficult to see any words. One should make this inset larger.  }
		(c) The moir\'e potential $\Delta(\vec r)$ with $\phi = 0$. $a_1^M, a_2^M$ are primitive vectors for a moir\'e unit cell. $1a,1b,1c$ are Wyckoff positions under the point group $C_{3v}$.
		(d) Schematic view of the spectrum. The orange (blue) lines are top (bottom) surface Dirac cones at $\Gamma,b_1^\text M$. %They are shifted by a small external electrical field for distinguishability. 
		Inset is the moir\'e BZ with the first shell moir\'e reciprocal lattice vectors.
		%{\color{red}{PZ: The plot(a) is a little misleading. The Morie potential is applied to one side or both side? The proposed setup is an interface structure or sandwich structure? I think your model could describe both structures. If you consider both, there should be a relative angle between top and bottom Moire potential.Please set enough inter-space between a and b. $\frac{r}{a_y}$ is too closed to figure(a).} }
	}
	\label{fig:system}
\end{figure}

\begin{figure}
	\centering
	\includegraphics[width=\columnwidth]{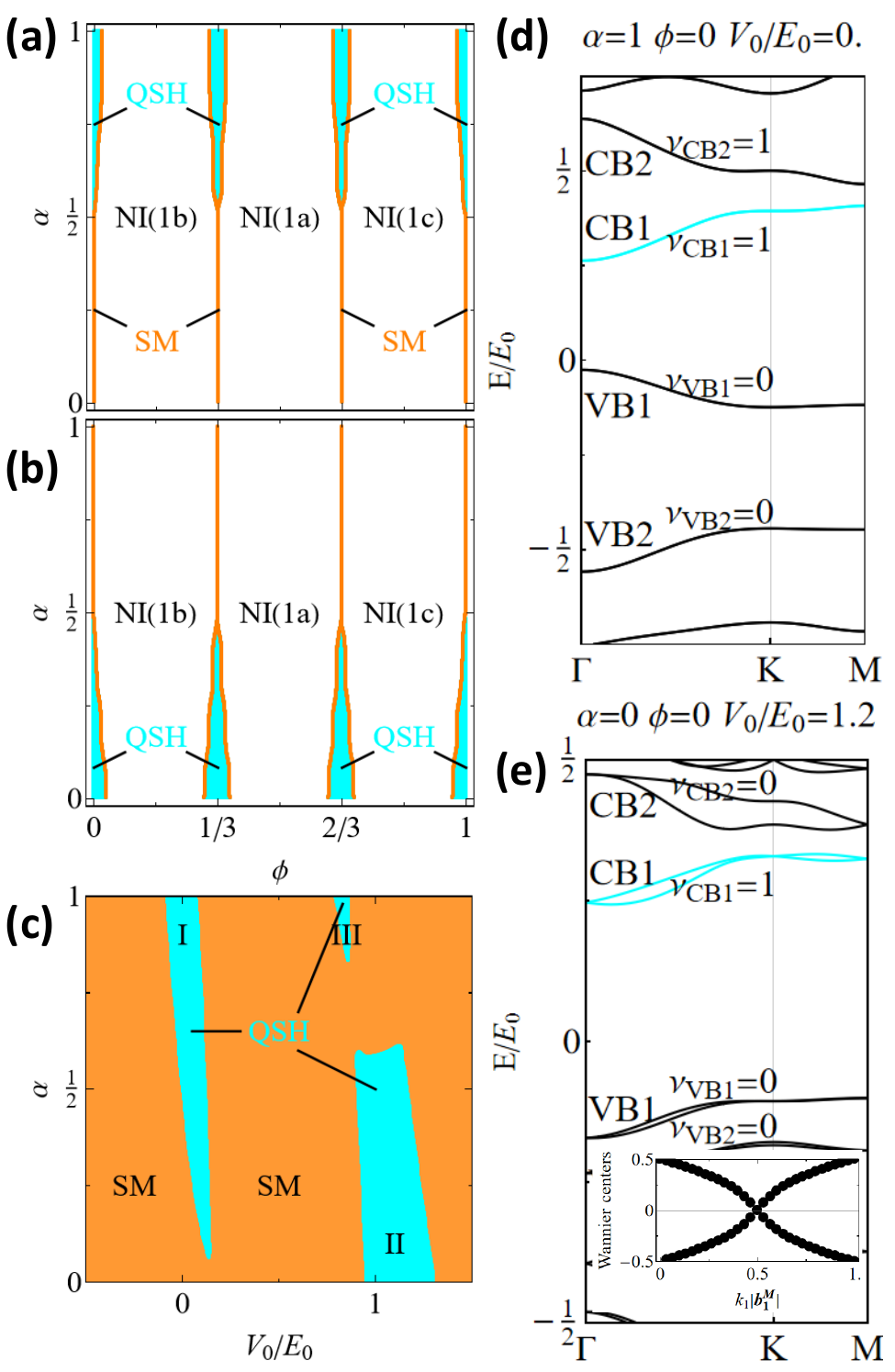}
	\caption{
	    (a)(b) The topological phase diagrams of the lowest conduction bands CB1 for different moir\'e potentials with $V_0 /E_0 = 0$ for (a) and $V_0/E_0=1.2$ for (b).
		(c) The phase diagram for different uniform asymmetrical potentials with $\phi = 0$. Regions \MakeUppercase{\romannumeral 1}, \MakeUppercase{\romannumeral 2} and \MakeUppercase{\romannumeral 3} are three parameter regimes with $\nu_{CB1}=1$ for CB1. 
		(d)(e) Example spectra with nontrivial CB1 in the regions \MakeUppercase{\romannumeral 1} and \MakeUppercase{\romannumeral 2}, respectively. %The moir\'e potential is shown in \figref{fig:system}(c). 
        The spectrum in (d) has both TR and inversion, and is thus doubly degenerate.
		The inset of (e) is the Wannier center flow for CB1.
	}
	\label{fig:spectrum}
\end{figure}

%\addCXL{CX: see the statement below is accurate or not. }
In this work, we studied a model of the TI thin film  (e.g. (Bi,Sb)$_2$Te$_3$ film) with the moir\'e superlattice potential (See \figref{fig:system}). Different from a bulk TI, a strong hybridization between two surface states is expected for the TI thin film. 
%\addCXL{With a moir\'e superlattice constant around $\sim XXX$ nm and a typical hybridization strength $\sim XXX$ meV, we find the moir\'e bandwidth $\sim XXX$ meV, mini-band gap $\sim XXX$ meV, and Coulomb interaction $\sim XXX$ meV, all being comparable and smaller than the typical spin-orbit coupling splitting $\sim XXX$ meV of surface bands at the moir\'e scale, which serves as the energy unit of our problem. } 
The hybridization between two surface states can create isolated minibands that possess non-trivial $\mathbb Z_2$ topological invariant, denoted by $\nu$ below, in the low-energy moir\'e spectrum in a wide parameter space, particularly when the moir\'e potential approximately has six-fold rotation symmetry. % \addKJ{or three -fold rotation symmetry under small two-fold symmetry breaking}. 
In the presence of inversion symmetry, an emergent chiral symmetry in the low energy sector of surface states gives rise to $\nu_\text{CB1} + \nu_\text{VB1} =1$ for the lowest Kramers' pair of conduction mini-bands, denoted as CB1, and the highest Kramers' pair of valence minibands, denoted as VB1, in \figref{fig:system}(d). We find $\nu_\text{CB1}=1, \nu_\text{VB1} =0$ ($\nu_\text{CB1}=0, \nu_\text{VB1} =1$) when the minima (maxima) of the moir\'e potential approximately form a hexagonal lattice.
In the case of non-trivial CB1 ($\nu_\text{CB1}=1, \nu_\text{VB1} =0$), the lowest two Kramers' pairs of conduction mini-bands (CB1 and CB2 in \figref{fig:system}(d)) together can be adiabatically connected to the Kane-Mele model\cite{kane2005quantum} when increasing quadratic terms, and thus CB2 is also topologically non-trivial, $\nu_{CB2} =1$. 
An asymmetric potential between two surface states can be generated by external gate voltages to break inversion but preserve six-fold rotation and generally induce the gap closing between different conduction mini-bands, leading to nodal phases. 
In the parameter regions where the conduction mini-bands are gapped from other mini-bands (parameter regions I, II, III in \figref{fig:spectrum}), the CB1 is always topologically non-trivial, $\nu_{CB1}=1$. 
We further study the influence of the Coulomb interaction via Hartree-Fock mean field theory when the CB1 carries $\nu_{CB1} =1$ and is half filled, and find that the quantum anomalous Hall (QAH) state competes with a trivial insulator state in region I of \figref{fig:spectrum}(c) and it can be robustly energetically favored by the asymmetric potential in region II. 
Finally, we propose a possible experimental realization of the TI-based moir\'e heterostructure consisting of a monolayer Sb$_{2}$ layer on top of Sb$_2$Te$_3$ thin films based on results from the first principles calculations. 

{\it Model Hamiltonian -}
We show a schematic of a hetero-structure consisting of TI thin films and another 2D material (e.g. 2D Sb thin films) in \figref{fig:system}(a) and (b), and the moir\'e potential induced by the 2D material can affect both the top and bottom surface states with different strength.
We assume the Fermi energy is within the bulk gap of TI thin film, and thus model this system with the Hamiltonian
%on the top and bottom surfaces with the inter-surface hybridization, labelled by $m$ \cite{yu2010quantized}.
%The moir\'e potential does not have electronic states near the surface states.
%The low energy effective Hamiltonian for two surface states with moir\'e potentials in real space is 
% \begin{equation}\label{eq:Hamiltonian}
% \begin{split}
%     & H_0(\vec r)=H^\text{TI}+H^\text M(\vec r),\\
%     & H^\text{TI}= 
%     \begin{pmatrix}
%     h^t_D (\vec r) & m s_0  \\
%     m s_0 & h^b_D (\vec r)
%     \end{pmatrix},\\
%     & H^\text M(\vec r)=
%     \begin{pmatrix}
%     \Delta(\vec r)s_0 +V_0 s_0  & 0  \\
%     0 & \alpha \Delta(\vec r)s_0 -V_0 s_0
%     \end{pmatrix}.  
% \end{split}
% \end{equation}
\begin{equation}\label{eq:Hamiltonian}
\begin{split}
    & H_0(\vec r)=H^\text{TI}+H^\text M(\vec r),\\
    & H^\text{TI}= v\tau_z (-i\partial_y s_x + i \partial_x s_y) + m \tau_x s_0 ,\\
    & H^\text M(\vec r)= \frac{1 + \alpha}{2} \Delta(\vec r)\tau_0 s_0 + \frac{1-\alpha}{2} \Delta(\vec r) \tau_z s_0 + V_0 \tau_z s_0.
\end{split}\raisetag{2.5\baselineskip}
\end{equation}
$H^\text{TI}$ denotes two surface states of a TI thin film with the inter-surface hybridization $m = m_0 + m_2 (-\partial_x^2 - \partial_y^2)$, and $h^{t/b}_D (\vec r) = \pm v(-i\partial_y s_x + i \partial_x s_y)$ is the top/bottom surface Dirac Hamiltonian\cite{yu2010quantized}. 
$s_{0,x,y,z}(\tau_{0,x,y,z})$ are the identity and Pauli matrices for spin (surfaces) and $v$ is the Fermi velocity.
$H^\text M$ denotes the potential term, in which the $V_0$ term is the uniform asymmetric potential between two surfaces by gate voltages,
%\addCXL{KJ: is it okay to use $V_0$ rather than $V_0 / 2$ in $H^\text M?$}
the $\Delta(\vec r)$ term is the moir\'e potential, and the $\alpha$ parameter ($0\le\alpha\le 1$) represents the asymmetry between top and bottom surfaces. 
$\Delta(\vec r)$ is real, spin-independent\cite{cano2021moire}, % and hermicity
and assumed to possess the $C_{3v}$ symmetry coinciding with the atomic crystal symmetry of TI thin films.
With the basis of the Hamiltonian, the corresponding symmetry operators are $C_{3z} = \exp(-i \pi  \tau_0 s_z  / 3)$ for three-fold rotation, $\mathcal M_y = \tau_0 s_y $ for y-directional mirror, and $\mathcal T = i\tau_0  s_y  \mathcal K$ with $\mathcal K$ as complex conjugate for time-reversal.
%A typical potential form of $\Delta(\vec r)$ is exemplified in \figref{fig:system}(b).
The moir\'e superlattice potential can be expanded as
\begin{equation}\label{eq:Morie potential}
    \Delta(\vec r) = \sum_{\vec G} \Delta_{\vec G} e^{i \vec G \cdot \vec r},
\end{equation}
where ${\vec G} = n_1 {\vec b}_1^\text M + n_2 {\vec b}_2^\text M$ is the moir\'e reciprocal lattice vectors with ${\vec b}_{1}^\text M = \frac{ 4 \pi}{ \sqrt 3 |{\vec a}_{1}^\text M| }( 1/2, \sqrt 3 /2), {\vec b}_{2}^\text M = \frac{4 \pi}{ \sqrt 3 |{\vec a}_{1}^\text M|} (-1/2, \sqrt 3 /2)$ and $n_{1,2}$ as integers.
${\vec a}_{1,2}^\text M$ are the primitive vectors for moir\'e superlattice (see \figref{fig:system}(c)).
%\addCXL{CX: specify the values of G}. 
The uniform part $\Delta_{\vec G=0}$ can be absorbed into the chemical potential $\mu$ and the asymmetric potential $V_0$.
To the lowest order, we only keep the first shell reciprocal lattice vectors $\pm {\vec b}_1^\text M, \pm {\vec b}_2^\text M, \pm({\vec b}_1^\text M - {\vec b}_2^\text M)$
%\addCXL{CX: also specify the values}
, as shown in \figref{fig:system}(d).
The values of $\Delta_{\vec G}$ for different ${\vec G}$s are connected by three-fold rotation $C_{3z}$ and $\TR$, so there is only one independent complex parameter, chosen to be $\Delta_{\vec b_1^\text M} = \Delta_1 e^{i 2\pi \phi}$, where
%$\vec b_1^\text M$ is the primitive vector of the moir\'e reciprocal lattice.
$\Delta_1$ is real and $\phi$ is the phase that tunes the relative strengths of potentials at three Wyckoff positions $1a,1b,1c$ in one moir\'e unit cell.
\figref{fig:system}(c) shows the moir\'e potential at $\phi = 0$ with an additional six-fold rotation symmetry $C_{6z} = \exp(-i\pi \tau_0s_z /6)$, and the corresponding potential minima form the multiplicity-2 Wyckoff positions of the hexagonal lattice. 
%a hexagonal lattice with two equivalent Wyckoff positions $1b$ and $1c$, \addKJ{which are }
The parameters used in our calculations below are $\vert \vec a_1^\text M \vert = 28 $nm, $E_0 = v \vert \vec b_1^\text M\vert = 38.5$ meV,\cite{liu2010model} $m_0 = 0.4 E_0,~\Delta_1 = 0.24 E_0$.
%\addPZTang{Why do you choose these parameter? Please cite the reference. citation deals with $v$. $\Delta_1$ is taken from DFT and $\vert \vec a_1^\text M \vert$ is chosen to make $E_0$ and $\Delta_1$ comparable.}
%\addKJ{$\Delta_1$ and $m$ are taken as comparable to $E_0$ for gaps shown in \figref{fig:system}(d).}
%\addCXL{CX: I think we should discuss the range of parameters here. }
The $m_2$ term and other quadratic terms are negligible for the low energy mini-bands in realistic materials as the relevant energy scale is around $1$ meV with a typical moir\'e momentum $10^{-2} \AA^{-1}$, much smaller than other terms in $H^{\text{TI}}$. But we still keep this term in low energy Hamiltonian as it plays an important role for connecting this model to the Kane-Mele model discussed below.
%{\color{red}PZ: The original phase without Moire potential is a QSH insulator or a trivial insulator? I think it would be better to add one sentence here. If the original phase is trivial, Moire potential change the topological property, it is an interesting point to emphasize.}
% an under estimation as there are some numbers ~ 2 before for momenta and m_2 ~ 20 eV \AA^-2.

{\it $\mathbb Z_2$ nontrivial moir\'e minibands -}
We first illustrate the crucial role of inter-surface hybridization in inducing isolated moir\'e minibands in TI thin films 
through the schematic view of the spectrum in \figref{fig:system}(d). 
For a single Dirac surface state, it is known\cite{cano2021moire, mora2019flatbands, wang2021moire} that moir\'e potential can fold the Dirac dispersion and the band touchings at the TR-invariant momenta, e.g. $\Gamma$ and $M$,
%\addPZTang{M point? Yes, as shown in Fig.1d the gap at M needs hybridization to open.}
in the moir\'e BZ remain gapless due to the Kramers' theorem of TR symmetry. 
This leads to satellite Dirac cones, but prevents the formation of gaps and hence of isolated moir\'e minibands. For TI thin films, the inter-surface hybridization $m$ can directly result in a gap at $\Gamma$ while its combined effect with the moir\'e potential $\Delta(\vec r)$ can lead to a gap (proportional to $m\Delta_1$) at $M$ (\figref{fig:system}(d)). 
The gap openings at both $\Gamma$ and $M$ lead to the isolated moir\'e minibands, as demonstrated in \figref{fig:spectrum}(d) and (e) for the moir\'e spectrum of the model Hamiltonian (\ref{eq:Hamiltonian}) with different sets of parameters.
%States at $M$ comes from Dirac cones at $\Gamma$ and $b_1^\text M$ coupled by $\Delta_1$.
%The gaps are opened between two surface states through $m\Delta_1$.
%An isolated moir\'e minibands are thus formed and the inter-surface coupling $m$ plays a crucial role.
% To determine roughly the bandwidth of the isolated moir\'e minibands, the constants in the Hamiltonian used in the following sections are $E_0 = \vert v \vec b_{1}^\text M \vert =38.5$ meV, $\Delta_1 / E_0 = 0.24$ and $m / E_0 =0.4$. 
% $\vert v \vec b_{1}^\text M/2 \vert$ is approximately the bandwidth of Dirac cones in the moir\'e Brillouin zone (BZ).
% It has the same order as the moir\'e potential and the inter-surface coupling, making an isolated flat bands promising.

%\addCXL{ CX: we probably should have some labeling for the lowest-energy conduction mini-bands and the highest-energy valence minibands. }
We are interested in the possibility of realizing ${\mathbb Z}_2$-nontrivial moir\'e mini-bands, particularly the low-energy Kramers' pairs of conduction (valence) mini-bands, labelled by CB1, CB2 (VB1, VB2) in \figref{fig:spectrum}(d) and (e). 
%Thus, we focus on the two lowest-energy conduction minibands, labelled by CB1 (blue color) in \figref{fig:spectrum}(d)(e), which form a Kramers' pair due to TR symmetry, and calculate the ${\mathbb Z}_2$-invariant $\nu$. 
%\addCXL{CX: I think we should talk about the phase diagram first and then go to the details of how the nontrivial bands are formed. }
%\addKJ{From the chiral symmetry  $\mathcal C=\tau_z s_z$ of $H^\text{TI}$, which satisfies $\{H^\text{TI},\mathcal C\} = 0$ and $[H^\text M, \mathcal C] = 0$,  the topology of conduction bands for $\Delta(\vec r), V_0$ is the same as that of valence bands for $-\Delta(\vec r), -V_0$.
%Answering Frank' question: It would be good to justify why we focus on CB1 for the phase diagrams and Hartree-Fock calculations — why not (also) on VB1, which is even closer to E=0 in Fig 2d,e?
%\addKJ{We find CB1 and CB2 topologically nontrivial and focus on conduction bands in the main text. The $\mathbb Z_2$ topology of valence bands for $\Delta(\vec r), V_0$ is the same as conduction bands for $-\Delta(\vec r), -V_0$ (See Supplemental Material (SM) Sec.XXX.)}
For the parameters in \figref{fig:spectrum}(d), CB1 and CB2 are topologically non-trivial while VB1 and VB2 are trivial ($\nu_\text{CB1}=\nu_\text{CB2}=1$, $\nu_\text{VB1}=\nu_\text{VB2}=0$). For the parameters in \figref{fig:spectrum}(e), only CB1 is non-trivial while other mini-bands are trivial ($\nu_\text{CB1}=1, \nu_\text{CB2}=\nu_\text{VB1}=\nu_\text{VB2}=0$). 
Fig. \ref{fig:spectrum}(a) and (b) show the ${\mathbb Z}_2$-invariant $\nu_\text{CB1}$ for CB1 %
%\addCXL{CX: make changes of the notation below}
as a function of $\alpha$ and $\phi$ for a fixed $V_0/E_0=0$ and $1.2$, respectively. 
The blue regions correspond to $\nu_\text{CB1}=1$ while the white regions to  $\nu_\text{CB1}=0$, and these two regions are separated by metallic lines (orange color). For both $V_0$ values, the $\nu_\text{CB1}=1$ blue regions appear around $\phi = 0,1/3,2/3$. 
%\addPZTang{It is around these values, but not the exact values. So the C6 is not necessary. Once |phi is closed to these value, we still could have Z2=1.}
At these $\phi$ values, there is an additional $C_{6z}$ rotation symmetry, leading to a hexagonal lattice with the $C_{6v}$ group.
%\addCXL{is this the correct group? Does it include inversion? This is the right group and does not include inversion}. 
\figref{fig:spectrum}(c) shows $\nu_\text{CB1}$ at $\phi = 0$ as a function of $\alpha$ and $V_0$, and we find three different parameter regions \MakeUppercase{\romannumeral 1}, \MakeUppercase{\romannumeral 2}, \MakeUppercase{\romannumeral 3} with $\nu_\text{CB1}=1$. 
%\addCXL{We should label these three regions.} 
These topologically non-trivial regions are separated by semi-metal phases that have band touchings between CB1 and CB2. 
$\nu_\text{CB1}$ for other $\phi$ is discussed in \addKJ{SM} Sec.\MakeUppercase{\romannumeral 1}.B and normal insulator phases are discussed in SM Sec.\MakeUppercase{\romannumeral 1}.D.
%{\color{red} Once we fit DFT results with the effective model, how about the parameter range?}

% In another limit with moir\'e potential on one surface by $\alpha=0$, the lowest conduction bands can also be isolated and have nontrivial $\mathbb Z_2$ topology as shown in \figref{fig:spectrum}(e).
% A similar topological phase transition also happens for different $\phi$ as shown in the phase diagram \figref{fig:spectrum}(g).

% From the topological phase diagrams, moir\'e potentials with $\phi = 0,1/3,2/3$ with an additional $C_{6z}$ symmetry has nontrivial $\mathbb Z_2$ for the lowest conduction bands.
% For different $\alpha$, topological phases can be achieved by tuning the external electrical field as shown in \figref{fig:spectrum}(h).
% Thus, QSH phases can be realized for nearly $C_{6z}$ symmetric moir\'e potentials on one or both surfaces under external electrical fields.

% We study topology of the lowest conduction bands in \figref{fig:system}(c) in this section. The topological phase diagrams of the heterostructure for different moir\'e potentials and external electrical fields are summarized in \figref{fig:spectrum}(a)-\ref{fig:spectrum}(c). 

The region \MakeUppercase{\romannumeral 1} can be adiabatically connected to the parameter set $\alpha = 1,~V_0/E_0= 0$ with the band dispersion shown in \figref{fig:spectrum}(d), where the inversion symmetry $\mathcal I = \tau_x s_0$ and the horizontal mirror symmetry $\mathcal M_z=-i\tau_x s_z$ are present ($D_{6h}$ group).
%\addKJ{The inversion symmetry is helpful in checking and understanding the $\mathbb Z_2$ topology for the region \MakeUppercase{\romannumeral 1}.}
From the Fu-Kane parity criterion\cite{fu2007topological}, the $\mathbb Z_2$-invariant $\nu$ can be determined by $(-1)^\nu=\prod_i\lambda_{\Gamma_i}$ and $\lambda_{\Gamma_i}$ is the parity of eigen-states at the TR invariant momenta $\Gamma_{i=1,...,4}$. In 2D moir\'e BZ, they are corresponding to one $\Gamma$ point and three $M$ points, %\addCXL{CX: unify the notation for parity using $\lambda$ below.} 
their values can be derived analytically in the weak \addCXL{$\Delta_1$} limit (See SM Sec. \MakeUppercase{\romannumeral 1}.A).
%\addPZTang{choose a suitable expression? The notation is the same as that used in Eq. 1.}
%, as depicted in \figref{fig:spectrum}(a). 
% To understand the existence of the non-trivial ${\mathbb Z}_2$ number for these minibands, we studied the evolution of the miniband dispersion and possible band crossing in \figref{fig:spectrum}(a)-(e). For  (\figref{fig:spectrum}(a)), , so we can determine the $\mathbb Z_2$ topology from Fu-Kane parity criterion\cite{fu2007topological}. 
% We start with \figref{fig:spectrum}(a) for .
% moir\'e potential of $\phi=0$ has a 2-fold rotation symmetry $ C_{2z}$ and $\alpha=1, V_0/E_0 =0$ for symmetric moir\'e potential on two surfaces has a horizontal mirror symmetry $ m_z$.
% The inversion symmetry $ I=m_z  C_{2z}$ is present for the system.
% Fu-Kane parity criterion can be used to determine the $\mathbb Z_2$ topology of the system \cite{fu2007topological}. 
% The parities of the states at TR-invariant momenta at $\Gamma$ and $M$ can be determined from $m$ and $\Delta_1$, as depicted in \figref{fig:spectrum}(a), 
% which can be derived analytically by perturbations in the weak $m$ and $\Delta_1$ limit (See Supplemental Material Sec. \MakeUppercase{\romannumeral 1}.A) and shown in .
% Here, we focus on the lowest conduction bands (cyan) and the highest valence bands.
%\addCXL{CX: $m_z$ has not been defined. }
The four eigen-states of $H^\text{TI}$ are denoted as $\ket {\psi^\text{TI}_{I,m_z}(\vec k) }$ with the gauge choice to satisfy $\mathcal I \ket {\psi^\text{TI}_{I,m_z}(\vec k)}  = I \ket {\psi^\text{TI}_{I,m_z}(- \vec k) }$ and $\mathcal M_z \ket {\psi^\text{TI}_{I,m_z}(\vec k)}  = m_z \ket {\psi^\text{TI}_{I,m_z}(\vec k) }$, where $I=\pm$ and $m_z = \pm i$.
The eigen-energies for $\ket {\psi^\text{TI}_{I,m_z}(\vec k) }$ is $E^\text{TI}_{I,m_z}(\vec k) = \sgn m I \sqrt{m^2+v^2 k^2}$ and two opposite mirror-eigen-value states $\ket {\psi^\text{TI}_{I,m_z=\pm i}(\vec k) }$ are degenerate.
At $\Gamma$, $\ket{\psi^\text{TI}_{+,m_z}(\Gamma)}$ and $\ket{\psi^\text{TI}_{-,m_z}(\Gamma)}$ are just the bonding and anti-bonding states formed by the top and bottom surface states, respectively. As the eigen-energies depend on the sign of $\sgn m I$, the eigen-state of CB1 is $\ket{\psi_{\text{CB1},m_z}(\Gamma)}=\ket{\psi^\text{TI}_{+\sgn m,m_z}(\Gamma)}$ with the energy $E_{\text{CB1},m_z}(\Gamma) = \vert m\vert$ and parity $\lambda^\text{CB1}_{\Gamma} = \sgn m$, while the eigen-state of VB1 is $\ket{\psi_{\text{VB1},m_z}(\Gamma)}=\ket{\psi^\text{TI}_{-\sgn m,m_z}(\Gamma)}$ with the energy $E_{\text{VB1},m_z}(\Gamma) = -\vert m\vert$ and parity $\lambda^\text{VB1}_{\Gamma} = - \sgn m$, so we get $\lambda^\text{VB1}_{\Gamma} = -\lambda^\text{CB1}_{\Gamma}$.
At $\vec M$, the potential term $\Delta_1$ that can be treated as perturbation couples the states $\ket{\psi^\text {TI}_{I,m_z}(\vec M)}$ and $\ket{\psi^\text {TI}_{I,m_z}(-\vec M)}$. Based on the degenerate perturbation theory, the eigen-state of CB1 is $\ket{\psi_{\text{CB1},m_z}(\vec M)}=(\ket{\psi^\text {TI}_{+\sgn m,m_z}(\vec M)} - \sgn{\Delta_1} \ket{\psi^\text {TI}_{+\sgn m,m_z}(-\vec M)})/\sqrt 2$ with the energy $E_{\text{CB1},m_z}(\vec M) = \sqrt{m^2 + v^2 k_\text M^2} - \vert \Delta_1 m\vert / \sqrt{m^2 + v^2 k_\text M^2}$ and  parity $\lambda^\text{CB1}_{\vec M}=+\sgn m (-\sgn {\Delta_1})$, where $k_\text M = \vert \vec M \vert$.
The eigen-state of VB1 is $\ket{\psi_{\text{VB1},m_z}(\vec M)}=(\ket{\psi^\text {TI}_{-\sgn m,m_z}(\vec M)} + \sgn{\Delta_1} \ket{\psi^\text {TI}_{-\sgn m,m_z}(-\vec M)})/\sqrt 2$ with the energy $E_{\text{VB1},m_z}(\vec M) =- \sqrt{m^2 + v^2 k_\text M^2} + \vert \Delta_1 m\vert / \sqrt{m^2 + v^2 k_\text M^2}$ and  parity $\lambda^\text{VB1}_{\vec M}=-\sgn m (+\sgn {\Delta_1})$ (See SM Sec. \MakeUppercase{\romannumeral 1}.A for more details).
Thus, we have $\lambda^\text{CB1}_{\vec M} = \lambda^\text{VB1}_{\vec M}$.
CB1 and VB1 have the same parity at $\vec M$ and opposite parities at $\Gamma$, resulting in $\nu_\text{CB1} + \nu_\text{VB1} = 1$ mod 2, implying that one of them is $\mathbb Z_2$-nontrivial while the other is trivial.
%\addCXL{CX: This statement is wrong, the minibands at M cannot be from Dirac cones at $\Gamma$, also in your talk. Also think about how to present it clearly. The following discussion is difficult to follow.}.
%\addCXL{CX: not clear. You mentioned four pairs of bands here CB1, CB2, VB1 and VB2. Specify which two bands.
%Another question is why you want to mention CB2 and VB2. Can one also determine the Z2 number for them from the argument here? }
%{\color{red}PZ: So the topological properties of Moire structure are determined by both intrinsic system without Moire potential and the sign of Moire potential? Yes.}
As discussed in SM Sec. \MakeUppercase{\romannumeral 1}.A, the relation of $\mathbb{Z}_2$ invariant between the CB1 and VB1 mini-bands can be understood as the consequence of the emergent chiral symmetry operator $\mathcal C=\tau_z s_z$ of $H^\text{TI}$,
which satisfies $\{\mathcal C,H^\text{TI}\}=0$, $[\mathcal C, H^\text{M}]=0$ and $\{\mathcal C, \mathcal I\}=0$.

At $\phi=0$ and $\alpha=1$ in \figref{fig:spectrum}(d), we notice that the CB2 mini-bands are also topologically non-trivial ($\nu_\text{CB2}=1$), so $\nu_\text{CB1}+\nu_\text{CB2}=0$ mod 2. According to the irreducible representations of CB1 and CB2 at high-symmetry momenta (See SM Sec.\MakeUppercase{\romannumeral 1}.C), these two mini-bands can together form an elementary band representation (EBR) $\bar E^{2b}_1 \uparrow G$ induced in the space group $P6mm$\cite{bradlyn2017topological}, which corresponds to the atomic limit with two s-wave atomic orbitals at the symmetry-related Wyckoff positions $1b$ and $1c$ in \figref{fig:system}(c). %\addCXL{CX: please check what is the language used in Bernevig's paper for this EBR; SM Page 9, Main text below Figure 1}. 
Indeed, as demonstrated in SM Sec.\MakeUppercase{\romannumeral 1}.C, when the $m_2$ term is tuned to dominate over other terms in $H_0$, we can adiabatically connect the CB1 and CB2 together in \figref{fig:spectrum}(d) to the effective Kane-Mele model\cite{kane2005quantum}. 
%\addCXL{CX: I think we should explicitly demonstrate our system can be connected to the Kane-Mele model. That's possible. }
This provides an alternative explanation of non-trivial ${\mathbb Z}_2$ numbers for both CB1 and CB2 in \figref{fig:spectrum}(b). 
%{\color{red}PZ: I did not fully understand what you are talking about here. Kane-Mele model is also an effective kp model, it should not be restricted to hexagonal lattice, is that right?}

For the nontrivial region II in \figref{fig:spectrum}(c), we consider the parameter set $\phi=0,\alpha=0, V_0/E_0=1.2$ with the energy dispersion shown in \figref{fig:spectrum}(e). The Fu-Kane criterion cannot be applied as inversion is broken, so we directly calculate the Wannier center flow\cite{yu2011equivalent} for the CB1 in the inset of \figref{fig:spectrum}(e), which corresponds to $\nu_\text{CB1}=1$. Different from the case of \figref{fig:spectrum}(d), CB2 is now topologically trivial $\nu_\text{CB2}=0$. 
%{\color{red}PZ: Without inversion symmetry, as you shown \figref{fig:spectrum}(e), double degeneracy for each E(k) is splitted, so CB1 stands for two states? The expression here is a little misleading.}
We also examine the band evolution with respect to $m_2$ in the model, which is quite different from the case with inversion symmetry, as discussed in SM Sec.\MakeUppercase{\romannumeral 1}.C. When the $m_2$ term dominates in $H_0$, CB1 and CB2 can be mapped to the Kane-Mele model with a Rashba SOC term from the inversion symmetry breaking, which leads to the gap closing between CB1 and CB2 around $K$ in moir\'e BZ with the overall ${\mathbb Z}_2$ number $\nu_\text{CB1}+\nu_\text{CB2}=0$ mod 2 since CB1 and CB2 together form an EBR. 
%\addCXL{CX: check this statement by projecting the Hamiltonian to the Kane-Mele model. } 
When reducing $m_2$, a Dirac type of gap closing between CB2 and higher-energy conduction mini-bands occurs at certain critical value of $m_2$ and changes $\nu_\text{CB1}+\nu_\text{CB2}$ to 1, which is persisted to $m_2=0$ ($\nu_\text{CB1}=1$ and $\nu_\text{CB2}=0$). The other ${\mathbb Z}_2$ non-trivial mini-band is found to appear in a much higher energy when $m_2$ is small (See Fig. S6 in SM Sec.\MakeUppercase{\romannumeral 1}.C). 
%\addCXL{KJ: Can the Dirac type gap closing happen for four bands?}
This is in sharp contrast to the inversion-symmetric case in which CB1 and CB2 together have $\nu_\text{CB1}+\nu_\text{CB2}=0$ when varying $m_2$. 

\begin{figure}
	\centering
	\includegraphics[width=\columnwidth]{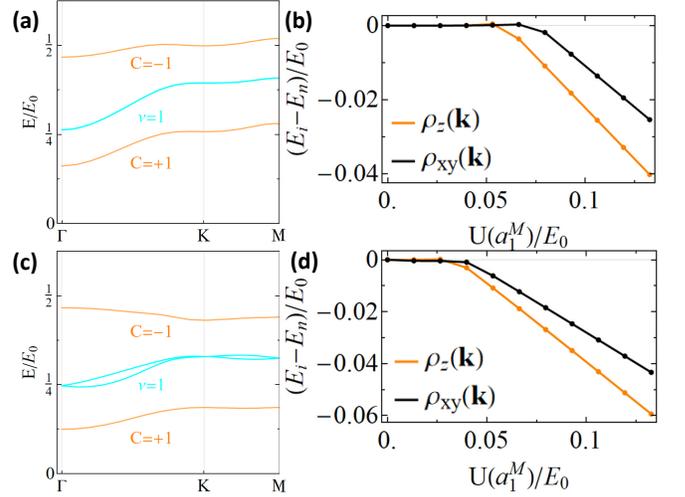}
	\caption{
		(a) The spectra for the Hartree-Fock mean-field Hamiltonian with the order parameter $\rho_z(\vec k)$ at half filling of CB1 for the case with $\phi=0,\alpha=1,V_0/E_0=0$. $C$ is the Chern number of each band.
		(b) The difference in energy per particle between the self-consistent Hartree-Fock states $E_i$ and the non-interacting state $E_n$ as a function of Coulomb interaction strengths for the order parameters $\rho_z(\vec k)$ (orange) and $\rho_{xy}(\vec k)$ (black).
        (c) The spectra for the Hartree-Fock mean-field Hamiltonian with the order parameter $\rho_z(\vec k)$ at half filling of CB1 for the case with $\phi=0,\alpha=0,V_0/E_0=1.2$.
        (d) The energy difference $E_i-E_n$ for the order parameters $\rho_z(\vec k)$ (orange) and $\rho_{xy}(\vec k)$ (black).
	}
	\label{fig:hartreefock}
\end{figure}

{\it Interaction-driven QAH state $-$}
%\addCXL{I think you need to mention $E_0$ somewhere in the early part. }
The Coulomb interaction of electrons in the moir\'e superlattice can be estimated as $U_0 = e^2 / 4 \pi \varepsilon_0 \varepsilon_r \vert \vec a_1^M \vert \approx 5.11$meV $\sim 0.13 E_0$, in which $e$ is the electron charge, $\varepsilon_0$ is vacuum permittivity, and dielectric constant $\varepsilon_r$ is about 10. \cite{bernevig2021twisted} 
%{\color{red}PZ: please cite an reference here.}
The value of $U_0$ is comparable to both the moir\'e mini-band width $\sim 0.1 E_0 \approx 3.85$ meV and mini-band gaps $\sim 0.1 E_0$.
We next study the effects of the Coulomb interaction with the Hartree-Fock mean-field theory\cite{zhang2020correlated,lian2021twisted,liu2021nematic,xie2020nature,bultinck2020ground,bernevig2021twisted}.
%Thus, unlike the twisted bilayer graphene and TMD moir\'e superlattice, which should be described by the Hubbard model\cite{koshino2018maximally,wu2018hubbard} as the Coulomb interaction is much larger than mini-band width, the moderate strength of Coulomb interaction in TI moir\'e heterostructures suggests that the Hartree-Fock mean field theory \cite{zhang2020correlated,lian2021twisted,liu2021nematic,xie2020nature,bultinck2020ground,bernevig2021twisted} should provide a good description. 
%\addCXL{CX: please check my description below is accurate or not. } 
We first project the moir\'e Hamiltonian and the Coulomb interaction into the low-energy subspace spanned by either CB1 (a two-band model) or both CB1 and CB2 (a four-band model). By treating the density matrix $\rho_{n_1 n_2} (\vec k) = \langle c^\dagger_{n_1}(\vec k) c_{n_2} (\vec k)\rangle$ as the order parameter with $c^\dagger_{n}(\vec k)$ for the creation operator of the $n$th eigenstate in the two-band or four-band subspace, we can decompose the Coulomb interaction Hamiltonian into two-fermion terms so that the order parameter $\rho (\vec k)$ can be solved self-consistently (See SM Sec.\MakeUppercase{\romannumeral 2}).

In the two-band model, we generally consider two types of order parameters, (1) $\rho_z (\vec k) \propto f_z(\vec k) \sigma_z$ and (2) $\rho_{xy}(\vec k) \propto f_{x}(\vec k) \sigma_{x} + f_{y}(\vec k) \sigma_{y}$,
%\addPZTang{what is the relationship between sigma z and s z?}
where the $\sigma$ matrix is for the Kramers' pair %spin states 
of CB1 and $f_{x,y,z}(\vec k)$ represents the momentum-dependent part of the order parameter. 
The order parameter $\rho_0 \propto \sigma_0$ is directly related to the band occupation and we always consider half-filling for the Kramers' pair bands of CB1. 
At $\phi=0, \alpha=1,V_0/E_0 = 0$, the spin basis of CB1 also corresponds to mirror eigen-values $\pm i$ of horizontal mirror symmetry $\mathcal M_z$ of $D_{6h}$ group, and these two mirror-eigenstates carry nonzero mirror Chern number $\pm 1$ from the nontrivial $\mathbb Z_2$ topology. 
Thus, $\rho_z(\vec k)$ and $\rho_{xy}(\vec k)$ correspond to the mirror-polarized and mirror-coherent ground states. 
The self-consistent calculations suggest that both $\rho_z(\vec k)$ and $\rho_{xy}(\vec k)$ can be non-zero solutions when the Coulomb interaction exceeds certain critical values $U_c\sim 0.05 E_0\approx 1.92$ meV, as shown in \figref{fig:hartreefock}(b), where the ground state energies of self-consistent $\rho_z(\vec k)$ and $\rho_{xy}(\vec k)$ are shown as a function of interaction strength $U(a_1^\text M)$, which is treated as a tuning parameter and equal to $U_0$ for the realistic moir\'e superlattice. Our estimate of Coulomb interaction $0.13E_0$ in TI moir\'e systems is larger than this critical value. From \figref{fig:hartreefock}(b), we also see that the mirror-polarized state $\rho_z(\vec k)$ has a lower ground state energy than the mirror-coherent state $\rho_{xy}(\vec k)$. 
The energy spectrum of the CB1 before (blue lines) and after (orange lines) taking into account the $\rho_z(\vec k)$ order parameter is shown in \figref{fig:hartreefock}(a), in which the metallic state of CB1 (blue lines) is fully gapped out by $\rho_z(\vec k)$ at half-filling. 
Due to non-zero mirror Chern number of non-interacting CB1 state, the mirror-polarized state $\rho_z(\vec k)$ carries Chern number $\pm 1$ and thus gives rise to the QAH state. 
As shown in SM Sec.\MakeUppercase{\romannumeral 2}.C, the mirror coherent state $\rho_{xy}(\vec k)$ has nodes in its spectrum due to the $C_{2z}{\mathcal T}$ symmetry. % for the low energy model.
%\addCXL{CX: please check if this statement is correct or not. }. 
This explains why the mirror-polarized state has a lower ground state energy than the mirror-coherent state. Thus, the mirror-polarized QAH state can be driven by Coulomb interaction in this system.  

We also studied the case of $\phi=0, \alpha=0, V_0/E_0 = 1.2$ within the two-band model, in which the mirror $\mathcal M_z$ is broken at the single particle level and six-fold rotation remains, in SM Sec.\MakeUppercase{\romannumeral2}.C and find the $\rho_z(\vec k)$ is still energetically favored, as shown in \figref{fig:hartreefock}(d). 
The spectra with the order parameter $\rho_z(\vec k)$ is shown in \figref{fig:hartreefock}(c) and the ground state is a Chern insulator.

\begin{figure}
	\centering
	\includegraphics[width=\columnwidth]{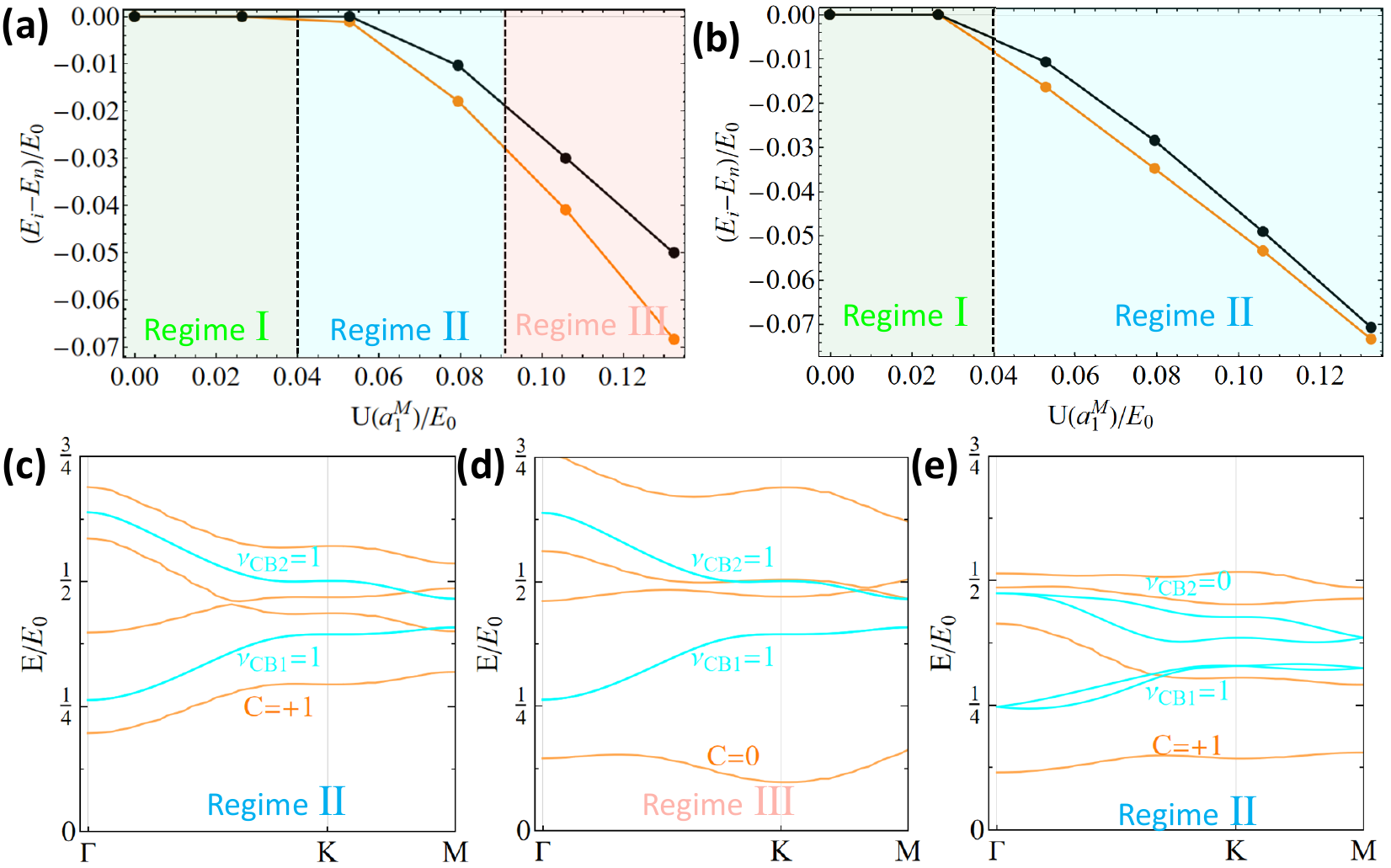}
	\caption{(a)The energy difference per particle $E_i-E_n$ at $1/4$ filling of the four-band model with both CB1 and CB2 for the case $\phi=0,\alpha=1,V_0/E_0=0$. Here $E_i$ and $E_n$ is the interacting ground state energy and non-interacting metallic state energy, respectively. The orange (black) line is for the $C_2\TR$ symmetry breaking  (preserving) density matrix. The interacting ground states in the regime \MakeUppercase{\romannumeral 1}, \MakeUppercase{\romannumeral 2}, and \MakeUppercase{\romannumeral 3} correspond to a metallic phase, an insulating phase with $C=\pm 1$, and an insulating phase with $C=0$, respectively. 
        (b) $E_i-E_n$ for the case with $\phi=0,\alpha=0,V_0/E_0=1.2$.
		(c)(d) The spectra of the Hartree-Fock mean-field Hamiltonian for the Coulomb interaction strength in regime \MakeUppercase{\romannumeral 2} and \MakeUppercase{\romannumeral 3} of (a). $C$ is the Chern number of each band. The spectra for the Hartree-Fock mean-field Hamiltonian for the case with $\phi=0,\alpha=1,V_0/E_0=0$.
        (e) The spectra of the mean-field Hamiltonian for the Coulomb interaction strength in Regime \MakeUppercase{\romannumeral 2} of (b).
	}
	\label{fig:hartreefockfourband}
\end{figure}

As the mini-band gap is comparable to Coulomb interaction, one may ask if the inter-mini-band mixing due to Coulomb interaction can change the topological nature of the ground state. Thus, we study the Coulomb interaction effect in a four-band model including both CB1 and CB2, as discussed in SM Sec.\MakeUppercase{\romannumeral2}.D. 
%\addCXL{CX: we need to add the discussion on Coulomb interaction in four-band model below. }
For the inversion-symmetric case $\phi=0, \alpha=1,V_0/E_0 = 0$, the ground state of the four-band model is still the mirror polarized $C = \pm 1$ state in regime \MakeUppercase{\romannumeral 2} (blue) of \figref{fig:hartreefockfourband}(a), when $U(a_1^\text M) = 0.08 E_0$ is smaller than the mini-band gap $\sim 0.1 E_0$, with the spectra shown in  \figref{fig:hartreefockfourband}(c).
When $U(a_1^\text M) = 0.13 E_0$ is larger than the mini-band gap (regime \MakeUppercase{\romannumeral 3} (red) of \figref{fig:hartreefockfourband}(a)), the strong Coulomb interaction can induce mixing between CB1 and CB2 within one mirror parity sector and drive a topological phase transition to the $C = 0$ state shown in \figref{fig:hartreefockfourband}(d) (More details in SM Sec.II.D). 
However, the situation for the inversion-asymmetric case $\phi=0, \alpha=0,V_0/E_0 = 1.2$ is different as $\nu_\text{CB1}=1$ and $\nu_\text{CB2}=0$. For the realistic estimated value $U(a_1^\text M) \approx 0.13 E_0$ that is larger than mini-band gap, the interacting ground state of the four-band model carries $C=\pm 1$ and thus remains the same as that of the two-band model, as shown by the regime \MakeUppercase{\romannumeral 2} (blue) in \figref{fig:hartreefockfourband}(b). The energy spectra in this case is shown in \figref{fig:hartreefockfourband}(e).
By comparing the phase diagrams for the inversion symmetric and asymmetric cases, we conclude that the asymmetric potential $V_0$ stabilizes the interaction-driven QAH state in TI moir\'e heterostructures.

\begin{figure*}
	\centering
	\includegraphics[width=0.9\textwidth]{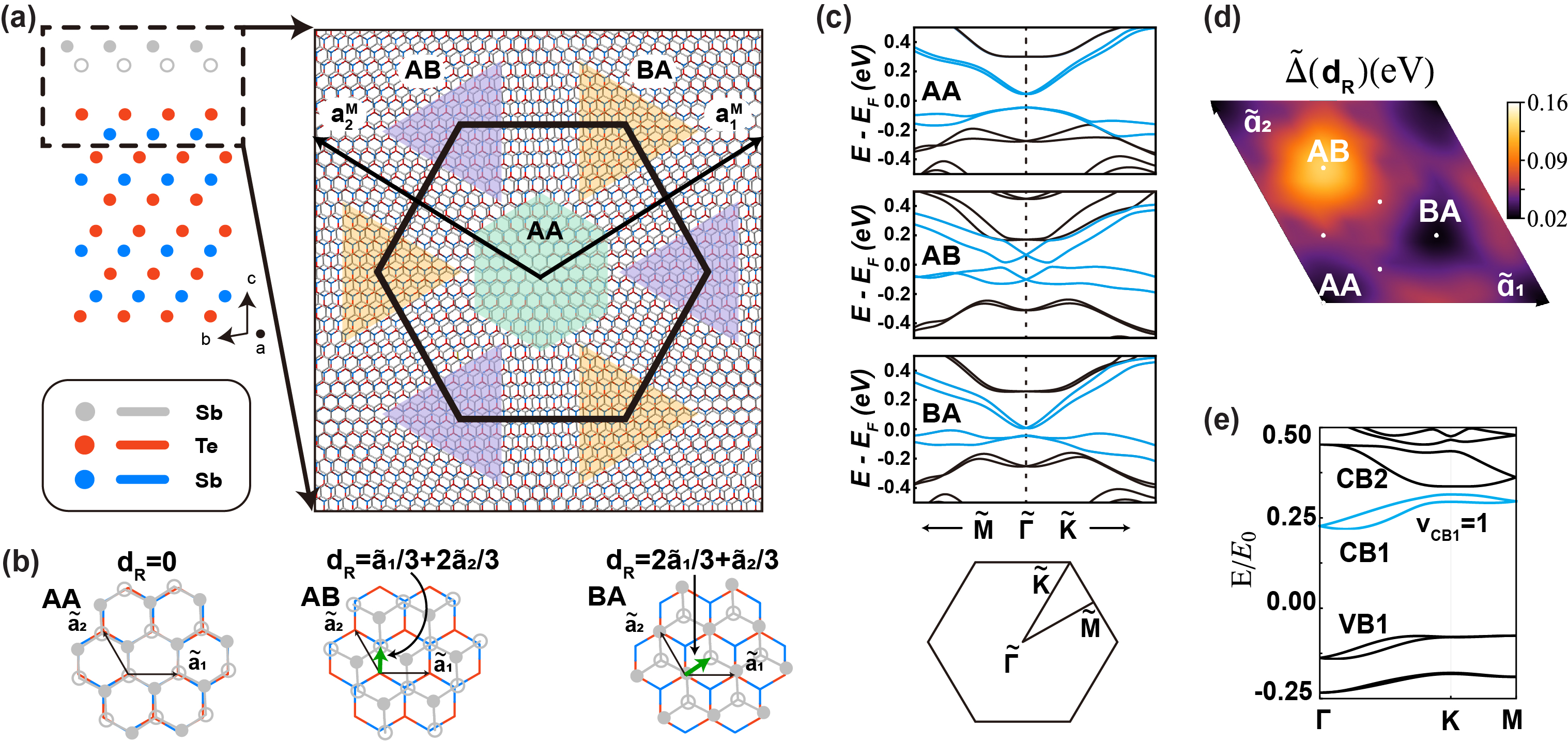}
	\caption{ %\addCXL{CX: can we add brackets for a, b, c, ... in each subplot? }
		(a) Side view of Sb$_{2}$/2QL Sb$_2$Te$_3$ heterostructure with AA stacking (left panel) and the moir\'e pattern for twisted Sb$_{2}$ on top of Sb$_2$Te$_3$ thin film (right panel). To show the moir\'e pattern clearly, we only plot atoms in the region marked by black dashed lines in the left panel. The triangle regions with green, purple, and yellow background label structures with AA, AB, and BA stacking respectively. The primitive vectors for moir\'e supercell $a_1^{M}$ and $a_2^{M}$ are marked by black arrows. (b) Top views of configurations with AA, AB, and BA stacking. The atomic primitive lattice vectors of the 2QL Sb$_2$Te$_3$ thin film are labeled as $\tilde{\vec a}_1 \text{ and } \tilde{\vec a}_2 $. The green arrow labels the shift $\vec d_\vec R$ between the Sb$_2$Te$_3$ layer and Sb$_{2}$ monolayer in each stacked configuration.
		(c) Band structures around the $\Gamma$ point for heterostructures with AA, AB, and BA stacking from DFT calculations. The Brillouin zone is plotted for the slab model used in DFT calculations with atomic primitive lattices. The Fermi levels are set as zero.
		(d) The superlattice potential $\tilde \Delta(\vec d_R)$ as a function of $\vec d_R$ shown in the moir\'e superlattice. $\tilde{\vec a}_{1,2}$ are marked by the black arrows.
		(e) Energy spectrum for twisted monolayer Sb$_{2}$ and 2QL Sb$_2$Te$_3$ with the superlattice potential shown in (d). 
		%\addKJ{we may need to specify $\vec d$ in (b) and change $a_{1,2}$ in (e) to $\tilde{a}_{1,2}$}
	}
	\label{fig:dft}
\end{figure*}
%\addZianXu{Zian: I will check this part and add some necessary information later, I still need some time}
%Twisted monolayer Sb$_{2}$ on top of 2QL Sb$_2$Te$_3$ thin film with an enlarged moir\'e super-cell within the black hexagon.
%Different regions in the super-cell can be approximated by different stackings of commensurate monolayer Sb$_{2}$ on top of Sb$_2$Te$_3$ thin film represented by a constant shift $\vec d_\vec R$. \addKJ{$\vec R$ is the atomic lattice vectors of the the Sb$_2$Te$_3$ layer.

{\it Sb$_2$/Sb$_2$Te$_3$ moir\'e heterostructure.$-$}
We propose a possible experimental realization of TI based moir\'e heterostructure with twisted Sb$_{2}$ monolayer on top of Sb$_2$Te$_3$ thin film. The moir\'e lattice structure is shown in \figref{fig:dft}(a). 
Sb$_2$Te$_3$ is a prototype of three dimensional TI with layered structures. 
Within one quintuple layer (QL, see the red and green dots in \figref{fig:dft}(a)), there is strong chemical binding formed by the sequential Te-Sb-Te-Sb-Te atomic layers and the van der Waals coupling is between adjacent QLs \cite{zhang2009topological}. 
Precise control of layer thickness of the Sb$_2$Te$_3$ thin film has been achieved via molecular beam epitaxy (MBE) method experimentally\cite{JiangYP2012,ZhangTong2013}. 
On the top of Sb$_2$Te$_3$ thin film, Sb$_{2}$ monolayer could be deposited\cite{zhu2019evidence,BianGuang2012,ChangCZ2015}, forming Sb$_2$/Sb$_2$Te$_3$ heterostructure. 
By using density functional theory (DFT) calculations, we confirm that Sb$_{2}$ monolayer with buckled honeycomb structure marked as the gray in \figref{fig:dft}(a) is a semiconductor with a band gap larger than that of Sb$_2$Te$_3$ thin films. 
Furthermore, we put Sb$_{2}$ monolayer on the top of 2QL Sb$_2$Te$_3$ thin films with different stackings, including the AA, AB, and BA stackings (see \figref{fig:dft}(a)). 
The corresponding electronic band structures are shown in \addCXL{\figref{fig:dft}(c)}. The work function of monolayer Sb$_2$ and Sb$_2$Te$_3$ thin film matches with each other, forming the type \MakeUppercase{\romannumeral 1} semiconductor hetero-junction. 
Around the Fermi level, the conduction and valence bands are both mainly contributed by two strongly hybridized surface states of the 2QL Sb$_2$Te$_3$ thin film. 
The role of Sb$_{2}$ monolayer is to provide a potential along the out-of-plane direction, leading to a Rashba type of spin-split bands. 
Thus, the twisted Sb$_2$/Sb$_2$Te$_3$ moir\'e heterostructure satisfies the requirements mentioned above for the $\mathbb Z_2$ nontrivial moir\'e minibands.  

To connect the theoretical moir\'e model Hamiltonian in \eqnref{eq:Hamiltonian} to electronic band structure from DFT calculations, we first introduce a uniform shifting vector $\vec d_\vec R$ %\addCXL{CX: make sure the consistency between the notation here and that in the figure. } 
between monolayer Sb$_{2}$ and 2QL Sb$_2$Te$_3$ thin film, and AA, AB, and BA stackings correspond to $\vec d_R= 0, \tilde{\vec a}_1 /3 + 2 \tilde{\vec a}_2 /3,\text{ and }  2\tilde{\vec a}_1/3 + \tilde{\vec a}_2 /3 $, respectively (\figref{fig:dft}(b)). 
$\tilde{\vec a}_{1,2}$ are atomic primitive lattice vectors for Sb$_2$Te$_3$ lattice shown in \figref{fig:dft}(b).
The spectrum from DFT calculations with different stacking is fitted by the dispersion of two-surface-state atomic Hamiltonian 
\begin{equation} \label{eq:HDFT}
\begin{split}
    H^\text{DFT}&(\vec k, \vec d_\vec R) = H^\text{TI}(\vec k) \\
    & + \frac{1+\alpha}{2} \tilde{\Delta}(\vec d_\vec R)\tau_0 s_0 + \frac{1-\alpha}{2} \tilde{\Delta}(\vec d_\vec R)\tau_z s_0,
\end{split}
\end{equation}
where $s_0 (\tau_{0,z})$ are the Pauli matrices for the spin (surfaces). 
$\tilde{\Delta}(\vec d_\vec R)$ is a uniform atomic potential induced by the Sb$_2$ monolayer for a fixed $\vec d_\vec R$ and different $\vec d_\vec R$ values correspond to different stacking configurations, shown in \figref{fig:dft}(b). For the $\vec d_\vec R$ values corresponding to the AA, AB, BA and several other stackings in SM Sec.\MakeUppercase{\romannumeral 3}, we fit the energy dispersion of the model Hamiltonian $H^\text{DFT}(\vec k, \vec d_\vec R)$ to that from the DFT calculations to extract $\tilde{\Delta}(\vec d_\vec R)$, which can be further interpolated as a continuous function of $\vec d_\vec R$ shown in \figref{fig:dft}(d). $\tilde{\Delta}(\vec d_\vec R)$ has the periodicity of the atomic unit-cell defined by $\tilde{\vec a}_{1,2}$. All other parameters in $H^\text{DFT}(\vec k, \vec d_\vec R)$ are treated as constants and can also obtained by fitting to the DFT bands.
%Based on $\tilde{\Delta}(\vec d_\vec R)$ for AA, AB, BA and several other stackings in SM Sec.\MakeUppercase{\romannumeral 3}, $\tilde{\Delta}(\vec d_\vec R)$ is interpolated as a function of $\vec d_\vec R$ in the whole atomic BZ of the Sb$_2$Te$_3$ shown in \figref{fig:dft}(e).
% We replace the superlattice potential $\Delta(\vec r)$ in the model Hamiltonian (\eqnref{eq:Hamiltonian}) by a potential $\Tilde \Delta(\vec d_\vec R)$ with a constant $\vec d_\vec R$ and fit the energy spectrum of this model to the DFT band structure for different stackings (see \figref{fig:dft}(d)). 
% Then, we interpolate the potential $\Tilde \Delta(\vec d_\vec R)$ as a function of $\vec d_\vec R$, as shown in \figref{fig:dft}(e).
After obtaining the parameters for $H^\text{DFT}(\vec k, \vec d_\vec R)$, the next step is to connect them to those of the moir\'e Hamiltonian $H_0$ in Eq. (\ref{eq:Hamiltonian}).   
For the moir\'e TI with the twist angle $\theta$, the local shift between two layers at the atomic lattice vector $\vec R$ of the Sb$_2$Te$_3$ layer is $\vec d_\vec R = \mathcal R(\theta) \vec R - \vec R$ 
%shown in \figref{fig:dft}(c)
, where $\mathcal R(\theta)$ is the rotation operator, so we can obtain the potential
\begin{equation}
    \Delta(\vec R) \approx \Tilde \Delta(\vec d_\vec R)
\end{equation}
at the location $\vec R$. The last step is to treat $\Delta(\vec r)$ as a function of continuous $\vec r$ by interpolating the function $\Delta(\vec R)$ (See SM.\MakeUppercase{\romannumeral 4}), and $\Delta(\vec r)$ serves as the mori\'e superlattice potential for the model Hamiltonian $H_0(\vec r)$. Besides, all the other parameters in $H_0$ are chosen to be the same as those in $H^\text{DFT}$. 
%The superlattice potential $\Delta(\vec r)$ can be approximated by $\Tilde \Delta(\vec d_\vec R)$ through $\Delta(\vec r)$ for any $\vec r$ in moir\'e unit cells can be interpolated from $\Delta(\vec R)$ (See SM.\MakeUppercase{\romannumeral 4}).
In \figref{fig:dft}(d), the potential maximum of  $\tilde{\Delta}(\vec d_\vec R)$ appears at the AB stacking while two local minima exist at the BA and AA stackings and are close in energy. The parameters for the moir\'e potential at $\theta=0.5^\circ$ is given by $\Delta_1 / E_0 = 0.22, \alpha=0.16$, and $\phi = 0.68 \pi$, close to $\phi = 2 \pi/3$ for the $C_6$-rotation symmetric potential in \figref{fig:spectrum}(g). \figref{fig:dft}(e) shows the energy dispersion of moir\'e mini-bands for $V_0 / E_0 =1.2$, in which the lowest conduction bands (cyan) indeed are isolated mini-bands with nontrivial $\nu_{CB1}$=1.

{\it Conclusion and Discussion $-$}
In summary, we demonstrate that the superlattice potential in a TI thin film can give rise to $\mathbb Z_2$ non-trivial isolated moir\'e mini-bands and Coulomb interaction can drive the system into the QAH state when the Kramer's pair of non-trivial mini-bands are half filled. Besides the twisted Sb$_{2}$ monolayer on top of Sb$_2$Te$_3$ thin film, our model can be generally applied to other TI heterostructures with the in-plane superlattice potential, which can come from either the moir\'e pattern of another 2D insulating material or by gating a periodic patterned dielectric substrate\cite{forsythe2018band,yankowitz2018dynamic,li2021anisotropic,shi2019gate,xu2021creation}.
%\addKJ{KJ: whether a material for moire potential is insulating near the Fermi energy of TI depends on their band alignment. The materials for moire potential in cited paper are not used for TI, meaning they may not be insulating for TI. In this sense, do we still cite these papers? Insulating materials are used as patterned dielectric. So i merge the citation.}
%\addCXL{CX: we may also need to comment how our results depend on the superlattice unit-cell size. We may also comment on how our results are different from the recent other works on twisted TIs. }
The 2D TI thin films can be in a quantum spin Hall state or trivial insulator state, depending on the relative sign between $m_0$ and $m_2$ in the model Hamiltonian (see \eqnref{eq:Hamiltonian}) \cite{liu2010oscillatory}. Our calculations suggest that the Moir\'e potential can lead to $\mathbb Z_2$ non-trivial mini-bands no matter the sign of $m_2$, once this term is negligible compared to the linear term in the moir\'e scale. Such a result implies the possibility of realizing isolated $\mathbb Z_2$ non-trivial mini-bands in other 2D topologically trivial systems with strong Rashba SOC. In our calculation, a large moir\'e superlattice constant ($\vert \vec a_1^\text M \vert \sim 28$nm) leads to small energy scales, around a few meV, for mini-band widths, mini-band gaps and Coulomb interactions, which may be disturbed by disorders. In See SM Sec.\MakeUppercase{\romannumeral 2}.E, we reduce $\vert \vec a_1^\text M \vert$ to $\sim 14$nm, which yields larger energy scales (around $10$meV) of mini-bands and Coulomb interaction, and our Hartree-Fock calculations suggests the estimated Coulomb interaction is still strong enough to drive the system into the QAH state. For a smaller moir\'e lattice constant $\vert \vec a_1^\text M \vert$, it is desirable to reduce the bandwidth of moir\'e mini-bands while keeping the Coulomb energy, and this can be achieved by twisting two identical TIs or with in-plane magnetic fields, as proposed recently \cite{chaudhary2022twisted,dunbrack2022magic}. 

{\it Acknowledgement} --
We would like to acknowledge Liang Fu, Jainendra Jain, Ribhu Kaul, Binghai Yan, Yunzhe Liu, Lunhui Hu and Jiabin Yu for the helpful discussion.  KJY and CXL acknowledge the support through the Penn State MRSEC–Center for Nanoscale Science via NSF award DMR-2011839.
CXL and BAB also acknowledges the support from the Princeton NSF-MERSEC (Grant No. MERSEC DMR 2011750). 
FS was supported by a fellowship at the Princeton Center for Theoretical Science.
BAB was furthermore supported by Simons Investigator Grant No. 404513, ONR Grant No. N00014-20-1-2303, the Schmidt Fund for Innovative Research, the BSF Israel US Foundation Grant No. 2018226, the Gordon and Betty Moore Foundation through Grant No. GBMF8685 towards the Princeton theory program and Grant No. GBMF11070 towards the EPiQS Initiative, and the Princeton Global Network Fund. BAB acknowledges additional support through the European Research Council (ERC) under the European Union's Horizon 2020 research and innovation program (Grant Agreement No. 101020833).
PZT was supported by the National Natural Science Foundation of China (Grants No. 12234011) and the Open Research Fund Program of the State Key Laboratory of Low-Dimensional Quantum Physics.
%\addCXL{ADD your support}
\bibliography{ref}

\clearpage
%========================================
% Supplemental materials
%========================================

\pagebreak
\widetext
\begin{center}
	{\large{\bf Supplemental Materials}}
\end{center}

\setcounter{equation}{0}
\setcounter{figure}{0}
\setcounter{secnumdepth}{2}
\renewcommand{\theequation}{S\arabic{equation}}
\renewcommand{\thefigure}{S\arabic{figure}}
\renewcommand{\thetable}{S\arabic{table}}
\renewcommand{\tocname}{Supplementary Materials}
\renewcommand{\appendixname}{Supplement}

\section{Topology of the lowest conduction bands}

\subsection{Perturbation Theory and emergent chiral symmetry}
%\addCXL{CX: can one distinguish the operator by its values by adding a hat for the operators and its eigen-value without a hat? You have changed the inversion operator, but we need to do that for ALL symmetry operators. Otherwise, it is confusing that you use the same notation to label both operators and their eigen-values.  }
In this section, the topology of inversion symmetric moir\'e system in \figref{fig:spectrum}(a) of the main text with $\phi=0, \alpha = 1,V_0/E_0=0$ is studied under the perturbation of moir\'e potential strength $\Delta_1$. 

By the Fu-Kane parity criterion\cite{fu2007topological}, the $\mathbb Z_2$ invariant $\nu$ can be determined by the parities $\lambda_i$ at time reversal ($\TR$) invariant momenta one $\Gamma$ and three M of moir\'e Brillouin Zone (MBZ) for one of the degenerate states by $(-1)^\nu = \prod_i \lambda_i$
%\addCXL{CX: I think it is better to use math formula to show this relation, which is more accurate. }
At $\phi=0, \alpha=1, V_0/E_0=0$, the crystal symmetry of this system is described by the point group $D_{6h}$ with six-fold rotation $C_{6z}$ about the z-axis, the inversion $\mathcal I$, the y-directional mirror $\mathcal M_y$ and the z-directional mirror $\mathcal M_z$. 
We label the original basis of our model Hamiltonian Eq. 1 in main text by $\ket{{\vec k}, J_z, \alpha}$, where ${\vec k}$ is the momentum, $J_z=\pm 1/2$ labels two spin states of surface states and $\alpha=\text t,\text b$ labels the top and bottom surfaces. 
The z-directional mirror $\mathcal M_z$ transforms the top surface to the bottom surface and thus it relates the basis wave-functions on two surfaces by 
\begin{equation}
	\mathcal M_z \ket{\vec k, J_z, \text t} = e^{-i \pi J_z}  \ket{\vec k, J_z, \text b} \quad \mathcal M_z \ket{\vec k, J_z, \text b} = e^{-i \pi J_z}  \ket{\vec k, J_z, \text t}.
\end{equation}
We may transform the basis wave-functions to the bonding and anti-bonding states of two surface states as
\begin{equation}
	\ket{\vec k, J_z, I} = \frac{1}{\sqrt 2} (\ket{\vec k, J_z, \text t} + I \ket{\vec k, J_z, \text b})
\end{equation}
with $I=\pm$ labels the transformation property under the inversion parity 
\begin{equation}
	 \mathcal I \ket{\vec k, J_z,\pm} = \pm \ket{-\vec k, J_z,\pm}
\end{equation}
and the eigen-values of the      $\mathcal M_z$ operator
\begin{equation}
	\mathcal M_z \ket{\vec k, J_z, \pm} = \pm e^{-i \pi J_z}   \ket{\vec k, J_z,\pm}.
\end{equation}
% The Hamiltonian \eqnref{eq:Hamiltonian} of the main text can be transformed into a block diagonal form
% Applying $ m_z$ to the basis gets
% \begin{equation}
% 	m_z \ket{k J_z \text t} = e^{-i \pi J_z}  \ket{k J_z \text b} \quad m_z \ket{k J_z \text b} = e^{-i \pi J_z}  \ket{k J_z \text t} .
% \end{equation}
% $J_z=\pm 1/2$ is the spin angular momenta and $t/b$ is the top/bottom surface. The bonding and anti bonding states of the top and bottom surface states
% \begin{equation}
% 	\ket{k J_z I} = \frac{1}{\sqrt 2} (\ket{k J_z \text t} + I \ket{k J_z \text b}) 
% \end{equation}
% with $I=\pm$ are diagonal in $m_z$ as
% \begin{equation}
% 	m_z \ket{k J_z I} = e^{-i \pi J_z} I  \ket{k J_z I}.
% \end{equation}
On these bonding and anti-bonding basis 
\begin{equation}\label{eq:basis}
    \ket{\vec k, J_z, I} = \ket{\vec k, +\frac{1}{2}, +}, \ket{\vec k, -\frac{1}{2}, -}, \ket{\vec k, -\frac{1}{2}, +}, \ket{\vec k, +\frac{1}{2}, -}, 
\end{equation}
the Hamiltonian $H_0$ in the main text Eq.1 can be written in a block diagonal form, 
\begin{equation}
	H_0(\vec r) = H^\text{TI} + H^{\text{M}}(\vec r) = 
	\begin{pmatrix}
		H^\text{TI}_{m_z=-i} & 0 \\ 
		0 & H^\text{TI}_{m_z=+i}\\ 
	\end{pmatrix} + \Delta(\vec r) I_{4\times 4} .
\label{eq:hammoire}
\end{equation}
with 
\begin{equation} \label{eq:HTI_mz}
	H^\text{TI}_{m_z=\pm i} = 
	\begin{pmatrix}
		m  & \mp i v \hat k_{\pm} \\ 
		\pm i v \hat k_{\mp} & -m \\ 
	\end{pmatrix},
\end{equation}
where
$\hat k_\pm = - i (\partial_x \pm i \partial_y) $, $m = m_0 + m_2 (-\partial_x^2 - \partial_y^2)$, $I_{4\times 4}$ is a $4\times 4$ identity matrix, and $m_z=\pm i$ labels the eigen-values of 
the mirror operator $\mathcal {M}_z$. $H^{\text{M}}$ is the moir\'e potential with $\phi = 0, \alpha=1, V_0/E_0 = 0$.
%The above Hamiltonian is block diagonal with upper and lower $2\times 2$ blocks labelled by mirror eigen-values $-i$ and $+i$, respectively. 
% The block of $\ket{k J_z=+1/2 \ I=+} \ket{k J_z=-1/2 \ I=-}$ with $ m_z$ eigenvalue $-i$ does not couple with the other block $\ket{k J_z=-1/2 \ I=+}\ket{k J_z=+1/2 \ I=-}$ with $ m_z$ eigenvalue $+i$. 
% The Hamiltonian becomes block diagonal.  
% The two block are related by $\TR$. 
% \addCXL{I do not understand this sentence. What is parity determination? }
% The upper block in \eqnref{eq:hammoire} is picked for the parity determination and $m_z$ Chern number are calculated for the Hamiltonian
% \begin{equation}
% 	H_{m_z}(\vec r) = 
% 	\begin{pmatrix}
% 		m + \Delta(\vec r) & i v \hat k_- \\ 
% 		-i v \hat k_+& -m + \Delta(\vec r)\\ 
% 	\end{pmatrix}.
% \end{equation}

We next determine the parities of lower-energy mini-bands at $\TR$ invariant momenta, including one $\Gamma$ and three $M$ in the moir\'e BZ, of the Hamiltonian $H_0$ in the limit $|\Delta_1| \ll  \vert v \vec b_1^\text M \vert,  |m|$ via perturbation theory. 
As the $H_0$ is block diagonal in the $m_z=\pm i$ subspace, we may perform the perturbation calculation for the $m_z=-i$ block while the mini-band parity of the $m_z=+i$ block can be related by TR symmetry.
For the $m_z=-i$ block, the unperturbed Hamiltonian is $H^\text{TI}_{m_z=-i}$ while $H^{\text{M}}$ is treated as the perturbation. 
We choose the eigen-wavefunctions of $H^\text{TI}_{m_z=-i}$ to possess a well-defined gauge at $\Gamma$ in the moir\'e BZ, which can be written as 
\begin{equation}
    \ket{\psi^{\text{TI}}_{+,-i} (\vec k)} = 
    \begin{pmatrix}
        i \cos \frac{\theta_\vec k}{2} \\
        \sin \frac{\theta_\vec k}{2} e^{i \phi_\vec k}
    \end{pmatrix}
    \quad
    \ket{\psi^\text{TI}_{-,-i} (\vec k)} = 
	\begin{pmatrix}
	-i \sin \frac{\theta_k}{2} e^{- i \phi_k }  \\ 
	\cos \frac{\theta_k}{2}  \\ 
	\end{pmatrix}  
\end{equation}
for $m>0$ with the eigen-energies $E^\text{TI}_\pm (\vec k) = \pm \sqrt{m^2 + v^2 k^2}$ and 
\begin{equation}
    \ket{\psi^\text{TI}_{+,-i} (\vec k)} = 
    \begin{pmatrix}
	i \sin \frac{\theta_k}{2}   \\ 
	-\cos \frac{\theta_k}{2} e^{i \phi_k } \\ 
	\end{pmatrix}   \quad
    \ket{\psi^{\text{TI}}_{-,-i} (\vec k)} = 
    \begin{pmatrix}
        i \cos \frac{\theta_\vec k}{2} e^{-i \phi_\vec k}\\
        \sin \frac{\theta_\vec k}{2} 
    \end{pmatrix} 
\end{equation}
for $m<0$ with the eigen-energies $E^\text{TI}_\pm (\vec k) = \mp \sqrt{m^2 + v^2 k^2}$, where $\cos \theta_k  = m / \sqrt{m^2 + v^2 k^2}$ and $k e^{i \phi_k} =k_x + i k_y$.
%\addCXL{some ambiguity in defining $\phi_k$ here}. The first lower-indices $I = \pm$ label the transformation property of the wave functions under inversion, 
%The gauges are picked to be well defined at $\Gamma$. They have parities 
\begin{equation}
	 \mathcal I \ket{\psi^{\text{TI}}_{I,-i} (\vec k)} =I \ket{\psi^{\text{TI}}_{I,-i} (-\vec k)},
\end{equation}
and the expression for the eigen-energy can be unified as 
\begin{equation}\label{eq:eig_eng_TI}
	 E^\text{TI}_I (\vec k) = {\sgn m} I \sqrt{m^2 + v^2 k^2},
\end{equation}
so the inversion parity $I$ also labels different eigen-energies of our model Hamiltonian. 
%\addCXL{CX: check if the above expression is correct or not.  }
The second lower-index $-i$ in the eigen-state labels the $\mathcal M_z$ eigen-values.
%\addCXL{CX: I'm not sure the purpose for the discussion below. It helps in evaluating H_{eff}^{CB} in Eq.S11}
% \begin{equation}
%     \bra{\psi^{\text{TI}}_\pm (\vec k)} {\psi^{\text{TI}}_\pm (- \vec k)} \rangle = \sgn m \cos \theta_\vec k.
% \end{equation}
This definition of the eigen-states $\ket{\psi^{\text {TI}}_{I,m_z}}$ is also used in the main text.  
%\addKJ{The TI surface states $\ket{\psi^{\text {TI}}_{\pm,\pm i}}$ are the same as those in the main text $\ket{\psi^{\text {TI}}_{I,m_z}}$}.
From the expression of the eigen-energies (\ref{eq:eig_eng_TI}), the higher energy state with $E^\text{TI}_{\text{CB}} (\vec k) = E^\text{TI}_{I=+ \sgn m} (\vec k)$ that corresponds to the conduction bands should be given by
\begin{equation}
	\ket{\psi^\text{TI}_{\text{CB},-i}(\vec k)}=\ket{\psi^\text{TI}_{+\sgn m,-i}(\vec k)}
\end{equation}
and the lower energy state  with $E^\text{TI}_{\text{VB}} (\vec k) = E^\text{TI}_{I=- \sgn m} (\vec k)$ for the valence bands should be
\begin{equation}
    \ket{\psi^\text{TI}_{\text{VB},-i}(\vec k)}=\ket{\psi^\text{TI}_{-\sgn m,-i}(\vec k)}. 
\end{equation}

For $m_z=+i$ subspace, we use TR symmetry operator, given by
\begin{equation}
\mathcal T = 
\begin{pmatrix}
    0 & 0 & -1 & 0 \\
    0 & 0 & 0 & 1 \\
    1 & 0 & 0 & 0 \\
    0 & -1 & 0 & 0 \\
\end{pmatrix}
\mathcal K
\end{equation}
in the basis \eqnref{eq:basis} with $\mathcal K$ for complex conjugate, to define 
%\addCXL{CX: definite the TR symmetry operator and $\psi$ for $+i$ subspace below. I think one may also define $\psi_{CB1, VB1}$ here and simplify the discussion in the later part. }
\begin{equation}
	 \ket{\psi^{\text{TI}}_{I,+i} (\vec k)} =  - i \mathcal T \ket{\psi^{\text{TI}}_{I,-i} (-\vec k)},
\end{equation}
and the commutation relation $[\mathcal T,\mathcal I]=0$ leads to the same inversion parity for two degenerate states $\ket{\psi^{\text{TI}}_{I,m_z=\pm i} (\Gamma_i)}$ at any TR-invariant momentum $\Gamma_i$.

The band gap and the inversion parity of mini-bands at $\Gamma$ of the moir\'e BZ are determined by the hybridization term $m$ in the limit $|\Delta_1| \ll  |m|$, for which the moir\'e potential does not play a role. Thus, we only need to consider the unperturbed Hamiltonian $H^\text{TI}_{m_z=-i}$ in Eq. (\ref{eq:HTI_mz}), which is diagonal, and the eigen-state $\ket{\psi^\text{TI}_{+,-i} (\Gamma)}=(1,0)^T$ has the eigen-energy $m$ and $\ket{\psi^{\text{TI}}_{-,-i} (\Gamma)}=(0,1)^T$ has the eigen-energy $-m$. 
The lower index $I$ directly gives the parity of the eigen-state at $\Gamma$, namely $\mathcal I \ket{\psi^{\text{TI}}_{I,-i} (\Gamma)} =I \ket{\psi^{\text{TI}}_{I,-i} (\Gamma)}$. 
The parity of the CB1 for the eigen-state $\ket{\psi^\text{TI}_{\text{CB},-i}(\Gamma)}$ is $\lambda_{\Gamma} = +\sgn m$ and that of the VB1 for the eigen-state $\ket{\psi^\text{TI}_{\text{VB},-i}(\Gamma)}$ is $\lambda_{\Gamma} = -\sgn m$, depending on the sign of $m$. 
%We start at the parity of states at $\Gamma$ for CB1 and VB1 as shown in \figref{fig:spectrum}(d) of the main text.  The gap is opened by $m$ and the parity is determined by the zero order $H^{\text TI}$ and $H_{eff}(\Gamma) = H^\text{TI}_{m_z}(\Gamma)$ by
% \begin{equation}
% 	H_{eff}(\Gamma) = 
% 	\begin{pmatrix}
% 		m & 0 \\ 
% 		0 & -m \\ 
% 	\end{pmatrix}.
% \end{equation}
% The eigenstates at $\Gamma$ are $\ket{\psi^\text{TI}_\pm(\Gamma)}$ with the energies $\pm m$ and parities $\pm$.
% Thus, the higher energy state (eigen-energy $+\vert m \vert$), which corresponds to the CB1, should be given by
% \begin{equation}
% 	\ket{\psi_{\text{CB1},-i}(\Gamma)}=\ket{\psi^\text{TI}_{+\sgn m,-i}(\Gamma)}
% \end{equation}
% with its parity $\lambda^\text{CB1}_{\Gamma} = +\sgn m$ depending on the sign of $m$, while the lower energy state (eigen-energy $-\vert m \vert$) for the VB1 should be
% \begin{equation}
%     \ket{\psi_{\text{VB1},-i}(\Gamma)}=\ket{\psi^\text{TI}_{-\sgn m,-i}(\Gamma)}
% \end{equation}
% with the parity $\lambda^\text{VB1}_{\Gamma} = -\sgn m$. 
Therefore, the parities of CB1 and VB1 at $\Gamma$ are opposite,
\begin{equation}\label{eq:parityGamma}
    \lambda^\text{CB1}_{\Gamma} = - \lambda^\text{VB1}_{\Gamma}.
\end{equation}

%For $\eta_{\pm}$, the lower index $\pm$ labels conduction or valence bands \addCXL{Is this correct? Should we label CB and VB instead of + and -? }. 

% The CB1 that has high the energy $+\vert m \vert$ and the eigenstates $\ket{\psi^\text{TI}_{+\sgn m}(\Gamma)}$ with the parity $\eta_+ = +\sgn m$.
% The VB1 has the energy $-\vert m \vert$ and the eigenstates $\ket{\psi^\text{TI}_{-\sgn m}(\Gamma)}$ with the parity $\eta_- = -\sgn m$.

Different from the $\Gamma$ point, the moir\'e potential is essential in determining the parities of the mini-bands at $M$ in the moir\'e BZ.  
Without moir\'e potential, the eigen-states $\ket{\psi^{\text{TI}}_{I,-i} (\vec k)}$ of $H^\text{TI}_{m_z=-i}$ at $\vec k=\vec M=\frac{1}{2} \vec b_1^\text M$ and $\vec k=\vec M- \vec b_1^\text M=-\vec M$ are degenerate, so the spectrum is gapless at M, even with a finite $m$. 
The moir\'e potential will couple these two states at $\vec M$ and $-\vec M$ as both belong to the same momentum in moir\'e BZ. 
By projecting the full Hamiltonian $H_0(\vec r)$ into the subspace spanned by these two states $\ket{\psi^\text{TI}_{\text {CB},-i}(\pm \vec M)}$, we find the effective Hamiltonian $H^\text{CB}_{eff}(\vec M)$ for CB1 and CB2 is given through the degenerate perturbation by
%Next, the parity of states for CB1, CB2, VB1 and VB2 at M are derived. 
%The unperturbed states at $M$ are $\ket{\psi^{\text{TI}}_{\eta_\pm} (\vec M)}$ and $\ket{\psi^{\text{TI}}_{\eta_\pm} (\vec M- \vec b_1^\text M)}=\ket{\psi^{\text{TI}}_{\eta_\pm} (-\vec M)}$. 
%The $H_{eff}(k_\text M)$ for CB1 and CB2 by the first order perturbation $H^{\text M_1}$ is 
\begin{equation}
\begin{split}
	H^\text{CB}_{eff}(\vec M) 
	&= 
    \begin{pmatrix}
        \bra{\psi^\text{TI}_{\text{CB},-i}(\vec M)} H^\text{TI} \ket{\psi^\text{TI}_{\text{CB},-i}(\vec M)} &
        \bra{\psi^\text{TI}_{\text{CB},-i}(\vec M)} H^\text{M} \ket{\psi^\text{TI}_{\text{CB},-i}(-\vec M)} \\
        \bra{\psi^\text{TI}_{\text{CB},-i}(-\vec M)} H^\text{M} \ket{\psi^\text{TI}_{\text{CB},-i}(\vec M)} &
        \bra{\psi^\text{TI}_{\text{CB},-i}(-\vec M)} H^\text{TI} \ket{\psi^\text{TI}_{\text{CB},-i}(-\vec M)}
    \end{pmatrix} \\
    & =
    \begin{pmatrix}
        E^\text{TI}_\text{CB}(\vec M) & \Delta_1 \vert \cos \theta_{k_\text M} \vert\\
        \Delta_1 \vert \cos \theta_{k_\text M} \vert & E^\text{TI}_\text{CB}(-\vec M)
    \end{pmatrix},
	% \begin{pmatrix}
	% 	\sqrt{m^2 + v^2 k_\text M^2} &  \Delta_1 \vert m \vert/ \sqrt{m^2 + v^2 k_\text M^2}  \\ 
	% 	\Delta_1 \vert m \vert / \sqrt{m^2 + v^2 k_\text M^2}   & \sqrt{m^2 + v^2 k_\text M^2} \\ 
	% \end{pmatrix},
\end{split}
\end{equation}
where $ E^\text{TI}_\text{CB}(\vec M) =  E^\text{TI}_\text{CB}(-\vec M) = \sqrt{m^2 + v^2 k_\text M^2}, \cos \theta_{k_\text M} = m / \sqrt{m^2 + v^2 k_\text M^2}$, and $k_\text M = \vert \vec M \vert $.
The eigen-state $ (\ket{\psi^\text{TI}_{\text{CB},-i} (\vec M)} + \ket{\psi^\text{TI}_{\text{CB},-i} (-\vec M)}) / \sqrt 2$ has eigen-energy $ E^\text{TI}_\text{CB}(\vec M) +  \Delta_1 \vert \cos \theta_\vec k \vert$ and the parity $+\sgn m$ while the eigen-state $ (\ket{\psi^\text{TI}_{\text{CB},-i} (\vec M)} - \ket{\psi^\text{TI}_{\text{CB},-i} (-\vec M)}) / \sqrt 2$ has eigen-energy $  E^\text{TI}_\text{CB}(\vec M) - \Delta_1 \vert \cos \theta_\vec k \vert$ and the parity $-\sgn m$.
The lower energy state (eigen-energy $ E^\text{TI}_\text{CB}(\vec M) - \vert \Delta_1 \vert \vert \cos \theta_\vec k \vert$), which corresponds to CB1, depends on the sign of $\Delta_1$ and is given by 
\begin{equation}
    \ket{\psi_{\text{CB1},-i}(\vec M)} = (\ket{\psi^\text{TI}_{\text{CB},-i} (\vec M)} - \sgn{\Delta_1} \ket{\psi^\text{TI}_{\text{CB},-i} (-\vec M)}) / \sqrt 2
\end{equation}
with the parity $\lambda^{\text{CB1}}_ \vec M = (+\sgn{m})(-\sgn{\Delta_1})$.
For VB1 and VB2, the effective Hamiltonian at $\vec M$ is given by
\begin{equation}
\begin{split}
	H^\text{VB}_{eff}(\vec M) 
	&= 
    \begin{pmatrix}
        \bra{\psi^\text{TI}_{\text{VB},-i}(\vec M)} H^\text{TI} \ket{\psi^\text{TI}_{\text{VB},-i}(\vec M)} &
        \bra{\psi^\text{TI}_{\text{VB},-i}(\vec M)} H^\text{M} \ket{\psi^\text{TI}_{\text{VB},-i}(-\vec M)} \\
        \bra{\psi^\text{TI}_{\text{VB},-i}(-\vec M)} H^\text{M} \ket{\psi^\text{TI}_{\text{VB},-i}(\vec M)} &
        \bra{\psi^\text{TI}_{\text{VB},-i}(-\vec M)} H^\text{TI} \ket{\psi^\text{TI}_{\text{VB},-i}(-\vec M)}
    \end{pmatrix} \\
    & =
    \begin{pmatrix}
        E^\text{TI}_\text{VB}(\vec M) & \Delta_1 \vert \cos \theta_{k_\text M} \vert\\
        \Delta_1 \vert \cos \theta_{k_\text M} \vert & E^\text{TI}_\text{VB}(-\vec M)
    \end{pmatrix},
\end{split}
\end{equation}
where $E^\text{TI}_\text{VB}(\vec M) = - \sqrt{m^2 + v^2 k_\text M^2}$.
% \begin{equation}
% 	H_{eff}^\text{VB}(\vec M) = 
% 	\begin{pmatrix}
% 		-\sqrt{m^2 + v^2 k_\text M^2} &  \Delta_1 \vert m \vert/ \sqrt{m^2 + v^2 k_\text M^2}  \\ 
% 		\Delta_1 \vert m \vert / \sqrt{m^2 + v^2 k_\text M^2}   & -\sqrt{m^2 + v^2 k_\text M^2} \\ 
% 	\end{pmatrix}.
% \end{equation}
%The eigen-state $ (\ket{\psi^{\text{TI}}_{-\sgn m,-i} (\vec M)} + \ket{\psi^{\text{TI}}_{-\sgn m,-i} (-\vec M)}) / \sqrt 2$ has eigen-energy $ -\sqrt{m^2 + v^2 k_\text M^2} +  \Delta_1 \vert m \vert / \sqrt{m^2 + v^2 k_\text M^2}$ and the parity $-\sgn m$ while the eigen-state $ (\ket{\psi^{\text{TI}}_{-\sgn m,-i} (\vec M)} - \ket{\psi^{\text{TI}}_{-\sgn m,-i} (-\vec M)}) / \sqrt 2$ has eigen-energy $ -\sqrt{m^2 + v^2 k_\text M^2} - \Delta_1 \vert m \vert / \sqrt{m^2 + v^2 k_\text M^2}$ and the parity $\sgn m$. 
The eigen-state $ (\ket{\psi^\text{TI}_{\text{VB},-i} (\vec M)} + \ket{\psi^\text{TI}_{\text{VB},-i} (-\vec M)}) / \sqrt 2$ has eigen-energy $ E^\text{TI}_\text{VB}(\vec M) +  \Delta_1 \vert \cos \theta_{k_\text M}\vert$ and the parity $-\sgn m$ while the eigen-state $ (\ket{\psi^\text{TI}_{\text{VB},-i} (\vec M)} - \ket{\psi^\text{TI}_{\text{VB},-i} (-\vec M)}) / \sqrt 2$ has eigen-energy $ E^\text{TI}_\text{VB}(\vec M) - \Delta_1 \vert \cos \theta_{k_\text M}\vert$ and the parity $\sgn m$. 
The higher energy state (eigen-energy $ E^\text{TI}_\text{VB}(\vec M) + \vert \Delta_1 \vert \vert \cos \theta_{k_\text M}\vert$), which corresponds to VB1, is given by 
% \begin{equation}
%     \ket{\psi_{\text{VB1}}(\vec M)} = (\ket{\psi^{\text{TI}}_{-\sgn m,-i} (\vec M)} + \sgn{\Delta_1} \ket{\psi^{\text{TI}}_{-\sgn m,-i} (-\vec M)}) / \sqrt 2
% \end{equation}
\begin{equation}
    \ket{\psi_{\text{VB1},-i}(\vec M)} = (\ket{\psi^\text{TI}_{\text{VB},-i} (\vec M)} + \sgn{\Delta_1} \ket{\psi^\text{TI}_{\text{VB},-i} (-\vec M)}) / \sqrt 2
\end{equation}
with the parity $\lambda^\text{VB1}_{ \vec M} = (-\sgn{m})(+\sgn{\Delta_1})$. 
Thus,
\begin{equation}\label{eq:parityM}
    \lambda^\text{CB1}_{ \vec M} = \lambda^\text{VB1}_{ \vec M} = -\sgn{m}\sgn{\Delta_1}.
\end{equation}
The parity at $\vec M$ for CB1 and VB1 are the same.
Because the $\mathbb Z_2$ invariant is $(-1)^\nu = \lambda_\Gamma (\lambda_\vec M)^3$, $(-1)^{\nu_\text{CB1}} = - \sgn{\Delta_1}$ and $(-1)^{\nu_\text{VB1}} = +\sgn{\Delta_1}$, so that
$\nu_\text{CB1}$ and $\nu_\text{VB1}$ are differed by 1. Thus, we conclude $\nu_{CB1}+\nu_{VB1}=1$ mod 2, namely one of CB1 and VB1 has nonzero $\mathbb Z_2$ invariant and the other has trivial $\mathbb Z_2$ invariant.

\begin{figure}
	\centering
	\includegraphics[width=0.6\columnwidth]{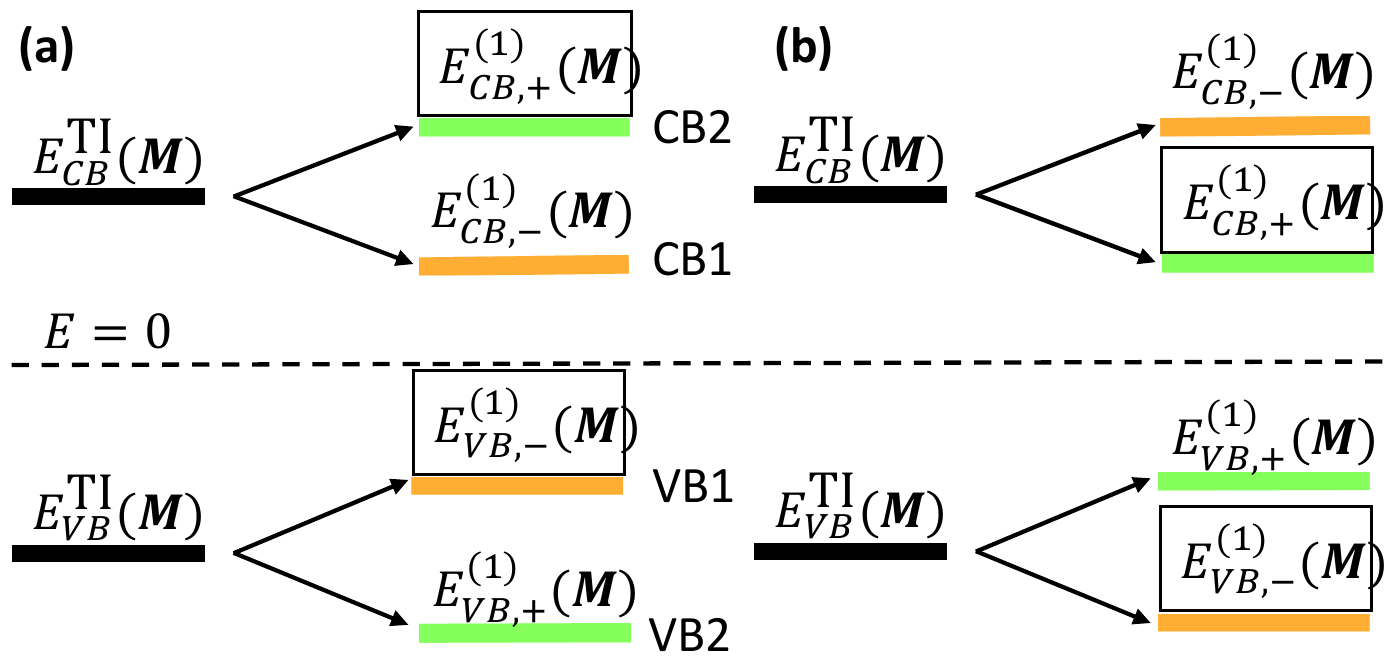}
	\caption{
	    Schematic figures of energies under the first order perturbation. The orange (green) lines are states with odd (even) parities. The framed/unframed energies have the first order energy perturbation related by chiral symmetries. (a) $\bra{ \Psi_{\text{CB}, +, m_z} (\vec M) } H^{\text{M}} \ket{ \Psi_{\text{CB}, +, m_z} (\vec M) } > \bra{ \Psi_{\text{CB}, -, m_z} (\vec M) } H^{\text{M}} \ket{ \Psi_{\text{CB}, -, m_z} (\vec M) }$.
	    (b) $\bra{ \Psi_{\text{CB}, +, m_z} (\vec M) } H^{\text{M}} \ket{ \Psi_{\text{CB}, +, m_z} (\vec M) } < \bra{ \Psi_{\text{CB}, -, m_z} (\vec M) } H^{\text{M}} \ket{ \Psi_{\text{CB}, -, m_z} (\vec M) }$ for $m_z=\pm i$. 
	    %\addCXL{Where is $m_z$ index here? }
	}
	\label{fig:parity by chiral symmetry}
\end{figure}

%\addCXL{CX: It seems you define the wave function for CB1 and VB1 states here, we should make the definition unified and consistent throughout the paper. }
The above conclusion of topology of CB1 and VB1 can also be understood from the chiral symmetry operator $\mathcal C$ of $H^{\text{TI}}$, defined by $\mathcal C = \tau _z s_z$, when the chemical potential is at the charge neutrality point, where $\tau $ acts on the top/bottom surface degrees of freedom and $s$ acts on spin. 
The emergence of the chiral symmetry requires dropping higher-order $k$ terms, e.g. $k^2$ terms, in $H^\text{TI}$, which are not important at the moir\'e energy scale. This operator has the commutation relations
\begin{equation}
\left\{\mathcal{C},H^{\text{TI}}\right\}=0, \left[\mathcal{C},H^{\text{M}}\right]=0, \left\{\mathcal{C}, \mathcal I\right\}=0, \left\{\mathcal C, \mathcal M_z \right\} = 0. \end{equation}
On the basis of \eqnref{eq:basis}, the form of chiral symmetry operator is transformed into 
\begin{equation}
\mathcal C = 
\begin{pmatrix}
    0 & 0 & 0 & 1 \\
    0 & 0 & -1 & 0 \\
    0 & -1 & 0 & 0 \\
    1 & 0 & 0 & 0 \\
\end{pmatrix},
\end{equation}
which mixes the eigen-states with opposite $\mathcal M_z$ eigen-values, namely 
\begin{equation}
   \ket{\psi^\text{TI}_{-I,+i}(\vec k)}=\mathcal C \ket{\psi^\text{TI}_{I,-i}(\vec k)}. 
\end{equation}
This implies
\begin{eqnarray}
\ket{\psi_{\text{VB},+i}(\vec k)}=\mathcal C \ket{\psi_{\text{CB},-i}(\vec k)}=\mathcal C \ket{\psi^\text{TI}_{+\sgn m,-i}(\vec k)}= \ket{\psi^\text{TI}_{-\sgn m,+i}(\vec k)}. 
\end{eqnarray}

At $\Gamma$, the opposite parities between $\ket{\psi_{\text{CB},-i}(\Gamma)}$ and $\ket{\psi_{\text{VB},+i}(\Gamma)}$ ($+\sgn m$ for $\ket{\psi_{\text{CB},-i}(\Gamma)}$ and $-\sgn m$ for $\ket{\psi_{\text{VB},+i}(\Gamma)}$ directly come from the anti-commutation relation $ \left\{\mathcal C, \mathcal I\right\} = 0$.

At $M$, the CB1 (VB1) and CB2 (VB2) are degenerate for $H^\text{TI}$, so we need to consider the first order perturbation from $H^{\text M}$. For the convenience of the discussion, we introduce the inversion adapted basis functions for CB1, CB2, VB1, and VB2 as 
%\addCXL{CX: this statement is very unclear. which states in the following definition are for CB1, CB2, VB1, VB2? Also does one have to introduce this basis to discuss the parity at $M$? }
% \begin{equation}
% \begin{split}
%     \ket{ \Psi_{I, +} (\vec M)} = \frac{1}{2} \left( \ket{\psi^\text{TI}_{I,-i}(\vec M)} + \ket{ \psi^\text{TI}_{I,-i}(-\vec M)} + \ket{\psi^\text{TI}_{I,+i}(\vec M)} + \ket{ \psi^\text{TI}_{I,+i}(-\vec M)} \right) \\
%     \ket{ \Psi_{I, -} (\vec M)} = \frac{1}{2} \left( \ket{\psi^\text{TI}_{I,-i}(\vec M)} - \ket{ \psi^\text{TI}_{I,-i}(-\vec M)} + \ket{\psi^\text{TI}_{I,+i}(\vec M)} - \ket{ \psi^\text{TI}_{I,+i}(-\vec M)} \right).
% \end{split}
% \end{equation}
% with $I=\pm$.
\begin{equation}
\begin{split}
    \ket{\Psi_{\text{CB},I,m_z}(\vec M)} & = \frac{1}{\sqrt 2} \left( \ket{\psi^\text{TI}_{\text{CB},m_z}(\vec M)} +  I \sgn m  \ket{\psi^\text{TI}_{\text{CB},m_z}(-\vec M)} \right) \\
    \ket{\Psi_{\text{VB},I,m_z}(\vec M)} & = \frac{1}{\sqrt 2} \left( \ket{\psi^\text{TI}_{\text{VB}, m_z}(\vec M)} - I \sgn m \ket{\psi^\text{TI}_{\text{VB},m_z}(-\vec M)} \right)
\end{split}
\end{equation}
with the parity 
\begin{equation}
\mathcal I \ket{\Psi_{\text{CB},I,m_z}} = I \ket{\Psi_{\text{CB},I,m_z}} \quad
\mathcal I \ket{\Psi_{\text{VB},I,m_z}} = I \ket{\Psi_{\text{VB},I,m_z}}. 
\end{equation}
They are related by chiral symmetry
\begin{equation}
    \ket{ \Psi_{\text{VB}, I, m_z} (\vec M) } = \mathcal{C} \ket{ \Psi_{\text{CB}, -I, -m_z} (\vec M) }.
\end{equation}
% They have parities
% \begin{equation}
%     \mathcal I \ket{ \Psi_{I, \pm} (\vec M) } = I \ket{ \Psi_{I, \pm} (\vec M) }
% \end{equation}
% which are opposite for $\ket{ \Psi_{+, \pm} (\vec M) }$ and $\ket{ \Psi_{-, \pm} (\vec M) }$ related by $\mathcal C$.
% At zero order, eigenstates $\ket{ \Psi_{+, \pm} (\vec M) }$ are degenerate determined by $H^\text{TI}$ with the energy $E^\text {TI}_+(\vec M)$, which is opposite to the energies $E^\text{TI}_-$ of $\ket{ \Psi_{-, \pm} (\vec M) }$ from chiral symmetry by
% \begin{equation}\label{eq:chiralsymmetryzeroorder}
%     E^\text {TI}_+(\vec M) = -E^\text {TI}_-(\vec M).
% \end{equation}
As $\left[\mathcal I, H^{\text M} \right] = 0$, the first order perturbation correction from $H^{\text M}$ is diagonal.
For CB1 and CB2 $\ket{ \Psi_{\text{CB}, I, m_z} (\vec M) }$, we find the perturbation Hamiltonian is
\begin{equation}
\begin{split}
&\begin{pmatrix}
    \bra{  \Psi_{\text{CB}, +, m_z} (\vec M) } H^{\text{M}} \ket{ \Psi_{\text{CB}, +, m_z} (\vec M) } & 0 \\
    0 & \bra{  \Psi_{\text{CB}, -, m_z} (\vec M) } H^{\text{M}} \ket{  \Psi_{\text{CB}, -, m_z} (\vec M) } \\
\end{pmatrix} \\ = 
&\begin{pmatrix}
    \Delta_1 \cos \theta_{k_\text M} & 0 \\
    0 & -\Delta_1  \cos \theta_{k_\text M} \\
\end{pmatrix}
\end{split} 
\end{equation}
with $ \cos \theta_{k_\text M} = m  /\sqrt{m^2 + v^2 k_\text M^2}$,
while for VB1 and VB2 $\ket{ \Psi_{\text{VB}, I , m_z} (\vec M) }$, the perturbation Hamiltonian is
\begin{equation}
\begin{split}
&\begin{pmatrix}
    \bra{  \Psi_{\text{VB}, +, m_z} (\vec M) } H^{\text{M}} \ket{ \Psi_{\text{VB}, +, m_z} (\vec M) } & 0 \\
    0 & \bra{  \Psi_{\text{VB}, -, m_z} (\vec M) } H^{\text{M}} \ket{  \Psi_{\text{VB}, -, m_z} (\vec M) } \\
\end{pmatrix} \\ = 
&\begin{pmatrix}
    -\Delta_1 \cos \theta_{k_\text M} & 0 \\
    0 & \Delta_1  \cos \theta_{k_\text M} \\
\end{pmatrix}
\end{split}. 
\end{equation}
% \begin{equation}
% \begin{pmatrix}
%     \bra{ \Psi_{-, +} (\vec M) } H^{\text{M}_1} \ket{ \Psi_{-, +} (\vec M) } & 0 \\
%     0 & \bra{ \Psi_{-, -} (\vec M) } H^{\text{M}_1} \ket{ \Psi_{-, -} (\vec M) } \\
% \end{pmatrix} = 
% \begin{pmatrix}
%     \Delta_1 \vert m \vert /\sqrt{m^2 + v^2 k_\text M^2} & 0 \\
%     0 & -\Delta_1 \vert m\vert /\sqrt{m^2 + v^2 k_\text M^2} \\
% \end{pmatrix}.
% \end{equation}
The eigen-energy of the system at $\vec M$ after taking into first order perturbation is
% \begin{equation}
% \begin{split}
%     E_{+,\pm}^{(1)} (\vec M) = E_+^\text{TI}(\vec M) + \bra{ \Psi_{+, \pm} (\vec M) } H^{\text{M}_1} \ket{ \Psi_{+, \pm} (\vec M) } \\
%     E_{-,\pm}^{(1)} (\vec M) = E_-^\text{TI}(\vec M) + \bra{ \Psi_{-, \pm} (\vec M) } H^{\text{M}_1} \ket{ \Psi_{-, \pm} (\vec M) }.
% \end{split}
% \end{equation}
\begin{equation}
\begin{split}
    E_{\text{CB},I}^{(1)} (\vec M) = E^\text{TI}_{\text{CB}}(\vec M) + \bra{  \Psi_{\text{CB}, I, m_z} (\vec M) } H^{\text{M}} \ket{ \Psi_{\text{CB},I, m_z} (\vec M) } \\
    E_{\text{VB},I}^{(1)} (\vec M) = E^\text{TI}_{\text{VB}}(\vec M) + \bra{  \Psi_{\text{VB}, I, m_z} (\vec M) } H^{\text{M}} \ket{ \Psi_{\text{VB},I, m_z} (\vec M) }.
\end{split}
\end{equation}
The two $m_z$ states are degenerate due to the $\TR \mathcal I$ symmetry so the index $m_z$ is dropped in the above labelling for the eigen-energy.
Chiral symmetry leads to
\begin{equation}
    \bra{ \Psi_{\text{VB}, I, m_z} (\vec M) } H^{\text{M}} \ket{ \Psi_{\text{VB}, I, m_z} (\vec M) } = \bra{ \Psi_{\text{CB}, -I, -m_z} (\vec M) } H^{\text{M}} \ket{  \Psi_{\text{CB}, -I, -m_z} (\vec M) }
\end{equation}
as $\left[ H^{\text M}, \mathcal C \right] = 0$.
If
\begin{equation}
\begin{split}
    \bra{ \Psi_{\text{CB}, +, m_z} (\vec M) } H^{\text{M}} \ket{  \Psi_{\text{CB}, +, m_z} (\vec M) } & > \bra{ \Psi_{\text{CB}, -, m_z} (\vec M) } H^{\text{M}} \ket{  \Psi_{\text{CB}, -, m_z} (\vec M) }, \\
    \bra{ \Psi_{\text{VB}, -, -m_z} (\vec M) } H^{\text{M}} \ket{ \Psi_{\text{VB}, -, -m_z} (\vec M) } & > \bra{ \Psi_{\text{VB}, +, -m_z} (\vec M) } H^{\text{M}} \ket{ \Psi_{\text{VB}, +, -m_z} (\vec M) }, \\
\end{split}
\end{equation}
which is equivalently
\begin{equation}
\bra{ \Psi_{\text{VB}, -, m_z} (\vec M) } H^{\text{M}} \ket{ \Psi_{\text{VB}, -, m_z} (\vec M) } > \bra{ \Psi_{\text{VB}, +, m_z} (\vec M) } H^{\text{M}} \ket{ \Psi_{\text{VB}, +,m_z} (\vec M) }.
\end{equation}
So, $E_{\text{CB},+}^{(1)} (\vec M) > E_{\text{CB},-}^{(1)} (\vec M)$ and $E_{\text{VB},-}^{(1)} (\vec M) > E_{\text{VB},+}^{(1)} (\vec M)$ as shown in \figref{fig:parity by chiral symmetry}(a).
CB1 has the eigenstate $\ket{ \Psi_{\text{CB},-} (\vec M) }$ with the energy $E_{\text{CB},-}^{(1)} (\vec M)$ while VB1 has the eigenstate $\ket{ \Psi_{\text{VB},-} (\vec M)}$ with the energy $E_{\text{VB}, -}^{(1)} (\vec M)$.
CB1 and VB1 has the same parity at $\vec M$. 
The other cases are shown in \figref{fig:parity by chiral symmetry}(b). 
CB1 and VB1 has the same parity as \eqnref{eq:parityM} for all cases. In the above analysis, the key is that $H^\text M$ commutes with $\mathcal C$ and leads to the same parity at $\vec M$, different from the case at $\Gamma$ where $H^\text{TI}$ anti-commutes with $\mathcal C$ and results in opposite parities. This leads to one of CB1 and VB1 to be topologically non-trivial while the other to be trivial.

\begin{figure}
	\centering
	\includegraphics[width=0.8\columnwidth]{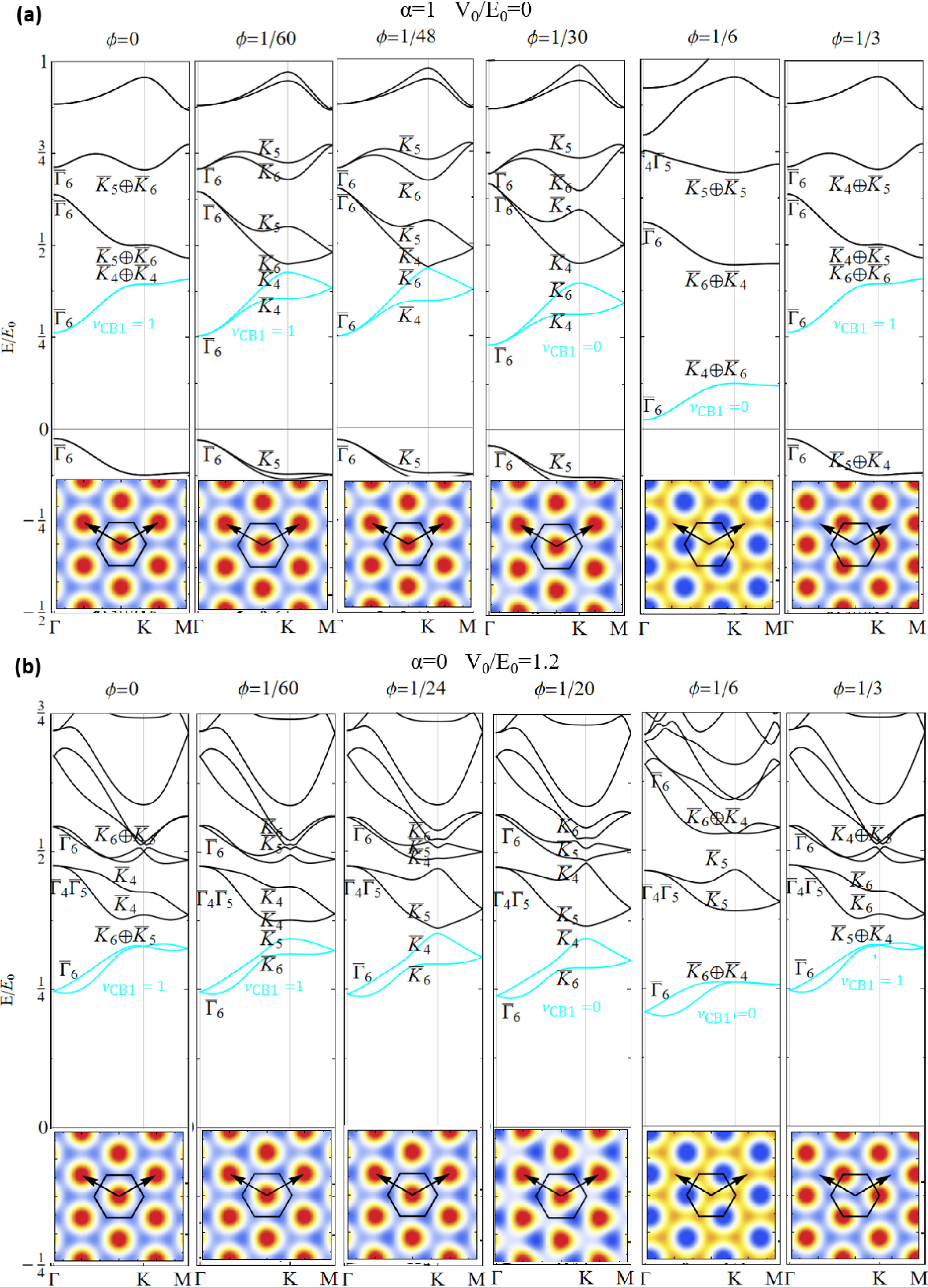}
	\caption{
        (a) Spectra for different $\phi$ in \figref{fig:spectrum}(a) of the main text.
        (b) Spectra for different $\phi$ in \figref{fig:spectrum}(b) of the main text.
        Insets are real space moir\'e potentials for different $\phi$. States are labelled with the irreducible representations by the little groups $C_{3v}$ at $\Gamma$ and $C_3$ at $K$. $\bar K_5$, $\bar K_6$, $\bar K_4$ represent the angular momentum states under $C_3$ with $J_z= 1/2, -1/2, 3/2$ respectively. $\nu$ is the $\mathbb Z_2$ invariant for the lowest conduction bands CB1.
	}
	\label{fig:topological phase transition for phi}
\end{figure}

\begin{table}[t]
\subfloat[$\Gamma$]{
    \begin{tabular}{|c|c|c|c|c|c|c|c|c|c|}
        \hline
        \ & $\mathcal C_3$ & $\mathcal M_y$ & $\mathcal T$ \\
        \hline
        $\bar{\Gamma}_4\bar{\Gamma}_5$ & 
        $\begin{pmatrix}
            -1 & 0 \\
            0 & -1 \\
        \end{pmatrix}$ & 
        $\begin{pmatrix}
            -i & 0 \\
            0 & i \\
        \end{pmatrix}$ & 
        $\begin{pmatrix}
            0 & -1 \\
            1 & 0 \\
        \end{pmatrix}$\\
        \hline
        $\bar{\Gamma}_6$ & 
        $\begin{pmatrix}
            e^{-i\pi/3} & 0 \\
            0 & e^{i\pi/3} \\
        \end{pmatrix}$ &
        $\begin{pmatrix}
            0 & -1 \\
            1 & 0 \\
        \end{pmatrix}$ &
        $\begin{pmatrix}
            0 & 1 \\
            -1 & 0 \\
        \end{pmatrix}$\\
        \hline
    \end{tabular}}
\qquad
\subfloat[$K$]{
    \begin{tabular}{|c|c|c|c|c|c|c|c|c|c|}
        \hline
        \ & $\mathcal C_3$ \\
        \hline
        $\bar{K}_4$ & $-1$\\
        \hline
        $\bar{K}_5$ & $e^{-i\pi/3}$\\
        \hline
        $\bar{K}_6$ & $e^{i\pi/3}$\\
        \hline
    \end{tabular}}
% \qquad
% \subfloat[EBR]{
%     \begin{tabular}{|c|c|c|c|c|c|c|c|c|c|}
%         \hline
%         \ & ${}^1\bar E {}^2\bar E@1a$ & $\bar E_1@1a$ & ${}^1\bar E {}^2\bar E@1b$ & $\bar E_1@1b$ & ${}^1\bar E {}^2\bar E@1c$ & $\bar E_1@1c$\\
%         \hline
%         $\Gamma$ & $\bar{\Gamma}_4\bar{\Gamma}_5$ & $\bar \Gamma_6$ & $\bar{\Gamma}_4\bar{\Gamma}_5$ & $\bar \Gamma_6$ & $\bar{\Gamma}_4\bar{\Gamma}_5$ & $\bar \Gamma_6$\\
%         \hline
%         $K$ & $2\bar K_4$ & $\bar K_5 \oplus \bar K_6$ & $2\bar K_5$ & $\bar K_4 \oplus \bar K_6$ & $2\bar K_6$ & $\bar K_4 \oplus \bar K_5$\\
%         \hline
%     \end{tabular}}
    \caption{(a)(b) symmetry operators in the irreducible representation at high symmetry momenta $\Gamma$ and $K$ for the double space group 156 $P3m1$ corresponding to the point group $C_{3v}$. %(c) EBR with the time-reversal symmetry of the double space group 156 $P3m1$ corresponding to the point group $C_{3v}$. $1a,1b,1c$ are Wyckoff positions shown in Fig.1c of the main text.
    }
    \label{tab:irrep and EBR under C3v}
\end{table}

\subsection{Topological phase transition when varying \(\phi\)}

%\addCXL{CX: can we include a table to summarize the irreducible representations for the Fig. S2? You can take it from Bilbao and give the relevant information, such as EBR. I think that's useful for the reader.}\addKJ{Here, we may have some confusion of topology from comparing irrep at high symmetry points of bands with those of EBR in Fig. S2e. CB1 in Fig. S2e has same irrep as  $\bar E_1@1a$ but it is topological. This may need some explanation to avoid confusion.}

%\addCXL{CX: do we refer to this section somewhere in the main text? Each section in Appendix should be referred somewhere in the main text. I previously had but we don't have now. We can cite this and the section D for Wanier functions of Normal insulators by adding a sentence : the cases with other $\phi$ are discussed in XXX at the begeinning of right column of page 3 if needed.} \addCXL{CX: please do that and clear the comment after it is done. }
In this section, we study the topological phase transition of our system when varying $\phi$ of the moir\'e potentials. For general $\phi$, there is no inversion symmetry. As shown in \figref{fig:topological phase transition for phi}(a), $\nu_{CB1}$ changes from $1$ to $0$ when $\phi$ varies from $0$ to $1/6$.
Between the two phases, there is a gap closing around $\phi \approx 1/48$ at $K$ and $K'$.
The gap closing at $K, K'$ happens between two states $\ket{u_{J_z=-1/2}(K)}$ and $\ket{u_{J_z = 3/2}(K)}$, belonging to $\bar K_6$ and $\bar K_4$ irreducible representations as summarized in Tab.\ref{tab:irrep and EBR under C3v}\cite{xu2020high,elcoro2021magnetic}, respectively, with different angular momenta $J_z = -1/2$ and $J_z =3/2 $ under three-fold rotation $C_3$, where $\ket{u_{J_z}(K)}$ are eigen-states of $H_0(K)$ as \eqnref{eq:eign equation for u}.
The effective Hamiltonian on the basis $\ket{u_{J_z = -1/2}(K)}$ and $\ket{u_{J_z = 3/2}(K)}$ has the Dirac fermion form $H_{eff}(\vec k) = v_K (k_x \sigma_x + k_y \sigma_y) + m_K \sigma_z$, up to the linear order, with $\sigma_{x,y,z}$ are Pauli matrices for the two band basis.
 The gap closing can be captured by one parameter, namely the Dirac mass $m_K$ that is controlled by $\phi$, corresponding to the co-dimension 1 case.
The $\TR$ symmetry guarantees the gap closing also occurring at $K'$, and the gap closings at $K$ and $K'$ lead to the change of $\mathbb Z_2$ number $\nu$ by $1$.
%\addCXL{I think you only shows half of the Hamiltonian here, right? It is Z2 change, so each band should be doubly degenerate. }
The normal insulator (NI) states are localized at moir\'e potential minima of the Wyckoff position $1b$ shown by the insets of spectrum for $\phi=1/6$ in \figref{fig:topological phase transition for phi}(a) (See SM Sec.\MakeUppercase{\romannumeral 1}.D).
From $\phi = 1/6$ to $\phi=1/3$ in \figref{fig:topological phase transition for phi}(a), another Dirac-type gap closing should happen at $K$ and $K'$ , and we find the system with $\phi = 1/3$ has $\nu_\text{CB1}=1$.

From the phase diagram in Fig. 2(a)(b) of the main text, we notice that the $\mathbb Z_2$ topological property of the system shows a periodicity when $\phi$ varies by $1/3$. Indeed, one can show that the moir\'e potential $\Delta(r)$ with $\phi$ and $\Delta'(r)$ with $\phi+1/3$ (with the same $\Delta_1$ parameter) are related by a constant shift as
\begin{equation}
\begin{split}
    \Delta'(\vec r) & = \Delta_1  e^{i 2\pi\phi} e^{i 2\pi/3} (e^{i \vec b_1^\text M. \vec r} + e^{i  (-\vec b_1^\text M+ \vec b_2^\text M). \vec r} + e^{i  (- \vec b_2^\text M). \vec r} )+ c.c. \\
    & = \Delta_1  e^{i 2\pi\phi} (e^{i \vec b_1^\text M. (\vec r + \vec a_1^\text M / 3 + 2 \vec a_2^\text M /3)} + e^{i  (-\vec b_1^\text M+ \vec b_2^\text M). (\vec r + \vec a_1^\text M /3 + 2\vec a_2^\text M /3)} + e^{i  (- \vec b_2^\text M). (\vec r + \vec a_1^\text M /3 + 2 \vec a_2^\text M /3} )+ c.c.\\
    & = \Delta (\vec r + \vec a_1^\text M /3 + 2 \vec a_2^\text M /3 ).
\end{split}    
\end{equation}
As a constant shift of potential term cannot change the band topology of the system, $\nu$ must keep the same for $\phi$ and $\phi+1/3$ 
while keeping other parameters. For NI phase, the Wyckoff position of Wannier orbitals should also shift accordingly by $ \vec a_1^\text M /3 + 2 \vec a_2^\text M /3$, as shown in \figref{fig:Wannier functions}.

Similar topological phase transitions happen for $\alpha = 0 $ and $V_0 /E_0 = 1.2$ by a Dirac-type gap closing at $K$ and $K'$ between two states with different angular momenta when $\phi$ varies from $0$ to $1/6$ to $1/3$, as shown by \figref{fig:topological phase transition for phi}(b). 

\begin{figure}
	\centering
	\includegraphics[width=\columnwidth]{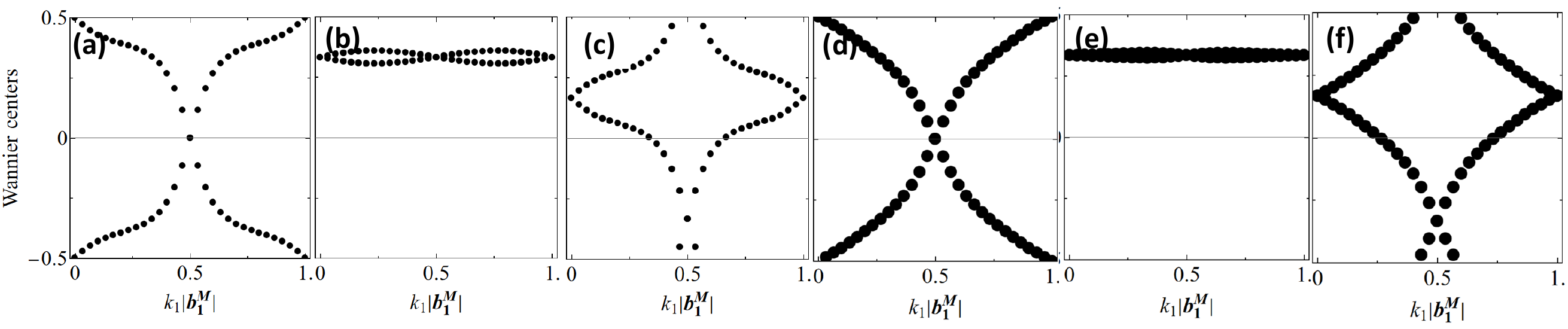}
	\caption{
        (a)(b)(c) The Wannier center flows for CB1 with $\phi = 0, 1/6, 1/3$, corresponding to  \figref{fig:topological phase transition for phi}(a)(c)(d), respectively.
        (d)(e)(f) The Wannier center flows for CB1 with $\phi = 0, 1/6, 1/3$, corresponding to  \figref{fig:topological phase transition for phi}(e)(g)(h), respectively.
	}
	\label{fig:Wannier center flow}
\end{figure}

%\addCXL{should it make more sense to show the Wannier center flow for e-h, as we cannot determine topological property from parity in this case? }
The Wannier centers flows for CB1 with $\phi=0,1/6,1/3$ in \figref{fig:topological phase transition for phi}(a) is shown in \figref{fig:Wannier center flow}(a)-(c).
CB1 with $\phi = 0, 1/3$ has nontrivial $\mathbb Z_2$ topology as analyzed in the main text. For the case with $\phi = 1/6$, CB1 are topologically trivial.
Similarly, the Wannier centers flows for CB1 with $\phi=0,1/6,1/3$ in \figref{fig:topological phase transition for phi}(b) is shown in \figref{fig:Wannier center flow}(d)-(f).
The $\mathbb Z_2$ number of CB1 is $\nu_{CB1}=1$ for $\phi = 0, 1/3$ and $\nu_{CB1}=0$ for $\phi=1/6$.

\begin{figure}
	\centering
	\includegraphics[width=0.8\columnwidth]{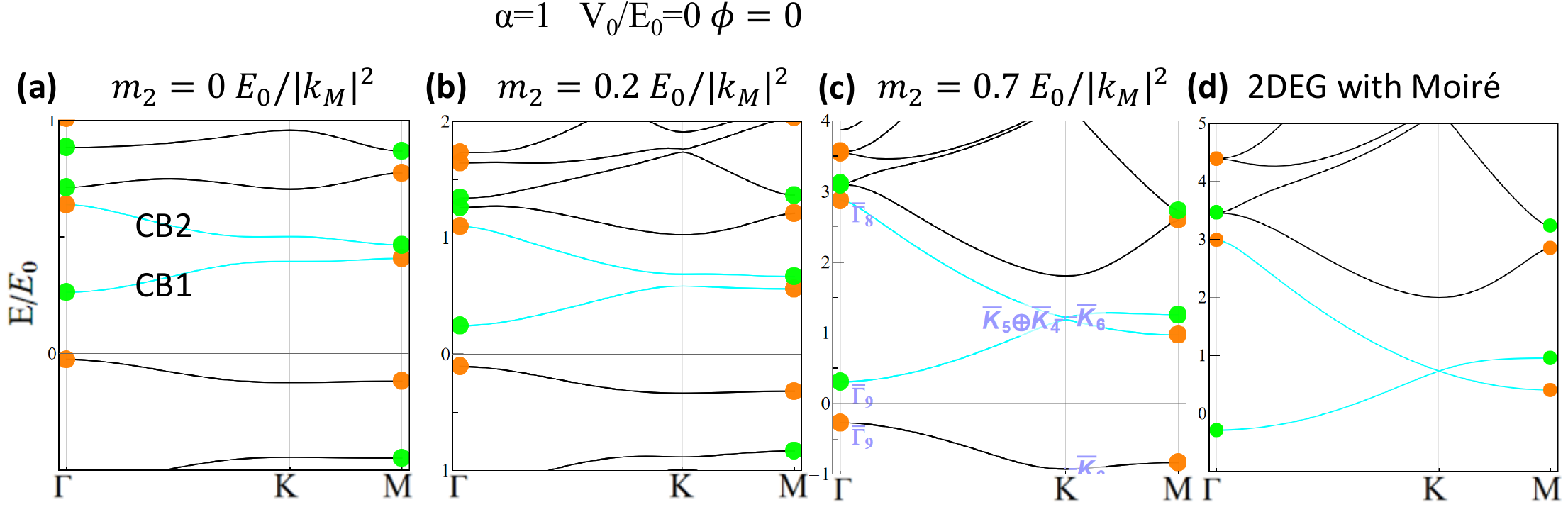}
	\caption{
        (a)(b)(c) Spectra with increasing $m_2$ for $\alpha = 1, V_0/E_0=0, \phi =0$ of Fig. 2(d) in the main text. Green (Orange) dots denote even (odd) parities at $\Gamma$ and $M$.
        (d) Spectrum of 2DEG on the moir\'e potential with $\phi=0$ shown in \figref{fig:system}(c) of the main text. Spectrum in (c) is labelled with irreps by the little group $C_{6v}$ at $\Gamma$ and $C_{3v}$ at $K$. 
        % \addCXL{Since you have energy scale on each plot, I do not think it is necessary to make them the same scale. }
	}
	\label{fig:atomic limits by large m2}
\end{figure}

\begin{figure}
	\centering
	\includegraphics[width=\columnwidth]{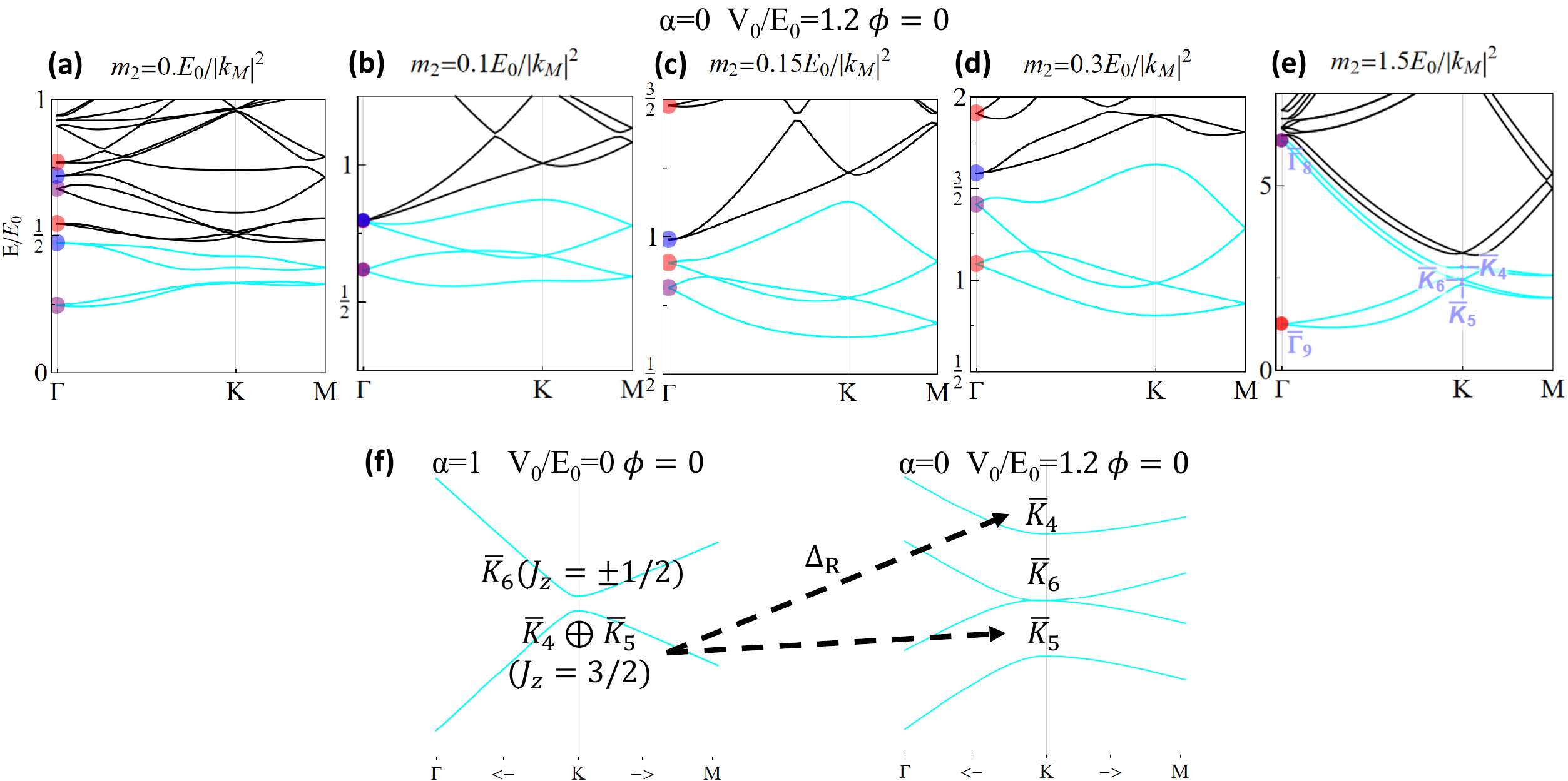}
	\caption{
        (a)-(e) Spectra with increasing $m_2$ for $\alpha = 0, V_0/E_0=1.2, \phi =0$ of Fig. 2(e) in the main text. Different colorful dots $\Gamma$ represents different irreps of the little group $C_{6v}$. 
        Spectrum in (c) is labelled with irreps by the little group $C_{6v}$ at $\Gamma$ and $C_{3v}$ at $K$.
        (f) spectrum around $K$ before and after breaking the inversion symmetry. $J_z$ is the angular momentum of the state at $K$ under $C_3$.
	}
	\label{fig:atomic limits by large m2 for asymmetric case}
\end{figure}

\begin{figure}
	\centering
	\includegraphics[width=0.8\columnwidth]{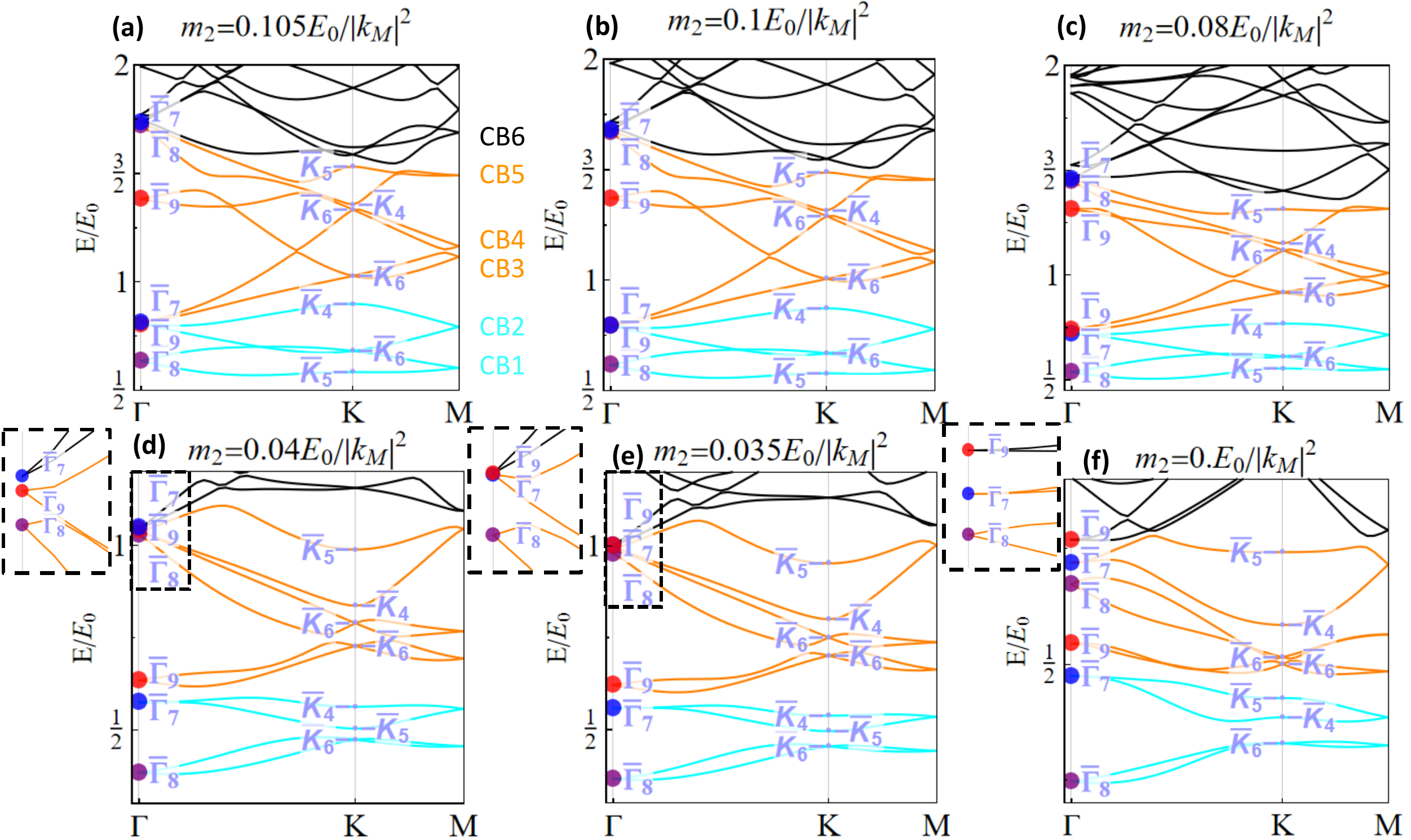}
	\caption{
        (a)-(f) Spectra with reducing $m_2$ for $\alpha = 0, V_0/E_0=1.2, \phi =0$ of Fig. 2(e) in the main text. Different colorful dots $\Gamma$ represents different irreps of the little group $C_{6v}$ shown in Tab.\ref{tab:irrep and EBR under C6v}. Cyan bands are CB1 and CB2. Orange ones are Cb3-5. Black ones are CB6 and higher energy bands. Insets in (d)(f) are enlargement of spectra around $\Gamma$ in the dashed boxes.
	}
	\label{fig:the other topological band for asymmetric case}
\end{figure}

\begin{table}[t]
\subfloat[$\Gamma$]{
    \begin{tabular}{|c|c|c|c|c|c|c|c|c|c|}
        \hline
        \ & $\mathcal C_6$ & $\mathcal M_y$ & $\mathcal T$ \\
        \hline
        $\bar{\Gamma}_7$ & 
        $\begin{pmatrix}
            -i & 0 \\
            0 & i \\
        \end{pmatrix}$ & 
        $\begin{pmatrix}
            0 & -1 \\
            1 & 0 \\
        \end{pmatrix}$ & 
        $\begin{pmatrix}
            0 & 1 \\
            -1 & 0 \\
        \end{pmatrix}$\\
        \hline
        $\bar{\Gamma}_8$ & 
        $\begin{pmatrix}
            e^{i5\pi/6} & 0 \\
            0 & e^{-i5\pi/6}
        \end{pmatrix}$ &
        $\begin{pmatrix}
            0 & -1 \\
            1 & 0 \\
        \end{pmatrix}$ &
        $\begin{pmatrix}
            0 & 1 \\
            -1 & 0 \\
        \end{pmatrix}$\\
        \hline
        $\bar{\Gamma}_9$ & 
        $\begin{pmatrix}
            e^{i\pi/6} & 0 \\
            0 & e^{-i\pi/6}
        \end{pmatrix}$ &
        $\begin{pmatrix}
            0 & -1 \\
            1 & 0 \\
        \end{pmatrix}$ &
        $\begin{pmatrix}
            0 & 1 \\
            -1 & 0 \\
        \end{pmatrix}$\\
        \hline
    \end{tabular}}
\qquad
\subfloat[$K$]{
    \begin{tabular}{|c|c|c|c|c|c|c|c|c|c|}
        \hline
        \ & $\mathcal C_3$ & $\mathcal M_x$\\
        \hline
        $\bar{K}_4$ & $-1$ & $-i$\\
        \hline
        $\bar{K}_5$ & $-1$ & $i$\\
        \hline
        $\bar{K}_6$ &
        $\begin{pmatrix}
            e^{-i\pi/3} & 0 \\
            0 & e^{i\pi/3}
        \end{pmatrix}$ & 
        $\begin{pmatrix}
            0 & -1 \\
            1 & 0
        \end{pmatrix}$\\
        \hline
    \end{tabular}}
    \caption{(a)(b) symmetry operators in the irreducible representation at high symmetry momenta $\Gamma$ and $K$ for the double space group 183 $P6mm$ corresponding to the point group $C_{6v}$.
    }
    \label{tab:irrep and EBR under C6v}
\end{table}

\subsection{Atomic limits at $m_2 \rightarrow \infty$}
%\addCXL{CX: Todo 1. Separate Fig. S3 into two: one for $\alpha=0$ and one for $\alpha=1$ and discuss them separately; 2. Discuss the effective model expanded around K for a large $m_2$ and compare it with the Kane-Mele model, also check the Rashba term;   }
In this section, we will provide theoretical understanding  of the non-trivial mori\'e mini-bands from the atomic limits of the CB1 and CB2 with a large $m_2$ term (the quadratic term of the inter-surface coupling $m$), and discuss how the realistic models with a small $m_2$ are connected to this atomic limit.

For $\alpha = 1, V_0/E_0=0, \phi =0$ with the inversion symmetry in Fig. 2(d) of the main text, the energy spectra for increasing $m_2$ are shown in \figref{fig:atomic limits by large m2}(a)-(c).
We focus on CB1 and CB2 as a whole for atomic limits because they together have $\nu_{CB1} + \nu_{CB2} = 0$ and are topologically trivial.
When increasing $m_2$, we do not find any gap closing between CB1, CB2 and other valence bands or higher conduction bands. Thus, the topological properties of CB1 and CB2 remain the same, and the CB1 and CB2 are adiabatically connected to those corresponding bands in the large $m_2$ limit.
When the $m_2$ term dominates in $H_0$, for $\phi=0$, we may consider the Hamiltonian in the $m_z=\pm i$ basis, Eq. (\ref{eq:hammoire}), and drop the linear term $\pm i v \hat k_{\pm}$ in the off-diagonal component first. Then, the remaining part of the Hamiltonian just describes the 2D electron gas (2DEG) with a simple parabolic dispersion on a hexagonal potential, 
\begin{equation}
    H^\text{2DEG} = \pm (m_0 - m_2 \nabla^2 ) + \Delta(\vec r), 
\end{equation}
with $\Delta(\vec r)$ the hexagonal potential, as shown in Fig.1(c) in the main text. The corresponding conduction band dispersion with $m_z = -i$ is shown \figref{fig:atomic limits by large m2}(d), while the $m_z=+i$ conduction bands are degenerate with $m_z=-i$ bands.
%\addCXL{Is the figure correct? }. 
% the conduction band spectra in \figref{fig:atomic limits by large m2}(c) almost resembles the energy spectrum of the 2D electron gas (2DEG) with a simple parabolic dispersion on a hexagonal potential, described by the Hamiltonian
% \begin{equation}
%     H^\text{2DEG} = - t \partial_\vec r^2 / 2 + \Delta(\vec r), 
% \end{equation}
% with $\Delta(\vec r)$ the hexagonal potential, as shown in \figref{fig:atomic limits by large m2}(d)
The lowest two conduction bands of the Hamiltonian $H^\text{2DEG} $ can be viewed as coming from two s-wave atomic orbitals localized at the moir\'e hexagonal potential minima of the Wyckoff positions $1b$ and $1c$ for the point group $D_{6h}$ and give rise to a Dirac cone at K and K', similar to the case of graphene.
The off-diagonal linear term in Eq. (\ref{eq:hammoire}) represents the strong spin-orbit-coupling (SOC) of TI thin films, which gives rise to a small gap opening for the dispersion in \figref{fig:atomic limits by large m2}(c) and can be treated perturbatively. 
%We notice a small gap opening for the dispersion in \figref{fig:atomic limits by large m2}(c), which should come from the linear momentum term in $H^\text{TI}$ that originates from strong spin-orbit-coupling (SOC) of TI films. 
We perform a ${\bf k\cdot p}$ type of perturbation expansion of the full Hamiltonian $H_0(\vec k)$ around $K$. 
%Adding SOC to 2DEG moir\'e systems can be described by the Kane-Mele model\cite{kane2005quantum} constructed under the same point group. We check whether Kane-Mele model can be applied to TI moir\'e system by comparing it with the $kp$ Hamiltonian $H^{kp}(\vec K+ \vec k)$ around $K$.
The basis wave functions are chosen to be the eigen-states of $H_0(\vec K)$ in Eq. 1 of the main eigenstates without SOC ($v_f=0$)
% \addCXL{Is this the correct form of the Hamiltonian? Please give the explicit form of the basis below. If you take any approximation for the eigen-state below, you need to make it clear. }
% \addCXL{Do you choose the basis of $H_0$ defined in the main text? Or the basis for the Hamiltonian without linear term? } 
\begin{equation}\label{eq:kp basis}
    \ket{\tilde u_{J_z, m_z}(\vec K) } = \ket{\tilde u_{-1/2, -i}(\vec K)},\ket{\tilde u_{ 3/2, -i}(\vec K)},\ket{\tilde u_{3/2, i}(\vec K)},\ket{\tilde u_{+1/2, i}(\vec K)}
\end{equation}
for CB1 and CB2 with the irreps $\bar K_6$ for $ \ket{u_{ 1/2, + i}(\vec K)},  \ket{u_{-1/2, -i}(\vec K)}$ and $\bar K_4$,$\bar K_5$ for $ \ket{u_{3/2, +i}(\vec K)},\ket{u_{3/2, -i}(\vec K)}$ (\figref{fig:atomic limits by large m2}(c)), the detailed forms of which can be numerically evaluated. 
The relevant symmetry operators are
\begin{equation}
    \mathcal M_z = -i \sigma_z  \tau_0 \quad \mathcal T \mathcal I =  i \sigma_y \tau_x \mathcal K
\end{equation}
with $\sigma$ acts on the different $m_z$,
%\addCXL{Cx: can we use $\tau$ instead? } 
$\tau$ acts on different $J_z$ in one $m_z$, and $\mathcal K$ is the complex conjugate.
The SOC couples $\ket{\tilde u_{J_z, m_z}(\vec K) }$ and valence bands and contributes a k-independent term from the first order L\"owdin perturbation\cite{winkler2003spin} by
\begin{equation}
    H_\text{SOC} = C_0' \sigma_0 \tau_0 + \Delta_\text{KM}\sigma_z \tau_z.
\end{equation}
The effective Hamiltonian $H_{eff}$ around $K$ to the first order in $\vec k$ with $m_2 = 0.7 E_0 / \vert k_M^2 \vert$ is
\begin{equation}\label{eq:Kane Mele model}
    H_{eff}(\vec k)\approx H_0(\vec K) + \left(\frac{\partial H_0 (\vec k)}{\partial \vec k}\right)_{\vec k=\vec K} \cdot {\vec k} + H_\text{SOC}= C_0 \sigma_0 \tau_0  + v_f ( k_x \sigma_0 \tau_x +  k_y \sigma_0 \tau_y) + \Delta_\text{KM}\sigma_z \tau_z,
\end{equation}
where $C_0, \Delta_\text{KM}, v_f$ are material dependent parameters and can be obtained numerically from the perturbation expansion. 
The above effective Hamiltonian $H_{eff}(\vec k)$ resembles the Kane-Mele model \cite{kane2005quantum} with the SOC term $\Delta_\text{KM}\sigma_z \tau_z$, 
which provides another understanding of the non-trivial $\mathbb Z_2$ topology of the CB1 and CB2 in our moir\'e system. 

%\addCXL{CX: I think it should be $V_/E_0=1.2$ here, right? }
For $\alpha = 0, V_0/E_0=1.2, \phi = 0$, similar procedure can be applied to find the atomic limits of CB1 and CB2 at a large $m_2$. The point group in this case is $C_{6v}$ group. For $m_2 = 1.5 E_0 / \vert k_M \vert^2$ in \figref{fig:atomic limits by large m2 for asymmetric case}(e), the effective Hamiltonian on the same basis as \eqnref{eq:kp basis} is given by
\begin{equation}
    H_{eff}(\vec k) = C_0 \sigma_0 \tau_0 + v_f (k_x \sigma_0 \tau_x + k_y \sigma_0 \tau_y) + \Delta_\text{KM} \sigma_z \tau_z + \Delta_\text R (\sigma_x \tau_y - \sigma_y \tau_x).
\end{equation}
Besides Kane-Mele SOC term $\Delta_\text{KM}$, there is another Rashba type of SOC term $\Delta_\text R (\sigma_x \tau_y - \sigma_y \tau_x)$ as the inversion symmetry is broken for $\alpha = 0$ \cite{kane2005quantum}.
The Rashba term couples two basis functions $\ket{\tilde u_{3/2, \pm i}(\vec K)}$ ($\bar K_4$ and $\bar K_5$ irreps) and opens the gap between these two states, as schematically shown in \figref{fig:atomic limits by large m2 for asymmetric case}(f). The other two states $\ket{\tilde u_{ 1/2, i}(\vec K)}, \ket{\tilde u_{ 1/2, i}(\vec K)}$ ($\bar K_6$ irrep) remain degenerate and form a 2D irrep under the little group $C_{3v}$ at $K$.
When this energy splitting $\Delta_\text R$ is larger than the Kane-Mele SOC gap $\Delta_\text{KM}$, the degenerate states with the 2D irrep $\bar K_6$ lies between the $\bar K_4$ and $\bar K_5$ state,
%the higher energy state of the $\bar K_4$ and $\bar K_5$ bands is above the doubly degenerate $\bar K_6$ bands, \addCXL{CX: what does this mean? } 
leading to the band touching between CB1 and CB2 bands at K for $m_2 = 1.5 E_0 / \vert k_M \vert^2$ in \figref{fig:atomic limits by large m2 for asymmetric case}(e). 
In this limit,  the topology of the CB1 and CB2 is $\nu_\text{CB1} + \nu_\text{CB2} = 0$, as the CB1 and CB2 together form an atomic limit. With decreasing $m_2$ to $m_2 = 0.1 E_0 / \vert k_M \vert^2$, we notice the nodes at K between CB1 and CB2 remains, but there is another band crossing between CB2 and higher conduction bands at $\Gamma$ in \figref{fig:atomic limits by large m2 for asymmetric case}(b). 
This band crossing at $\Gamma$ changes the overall $\mathbb Z_2$ topology of CB1 and CB2 to $\nu_\text{CB1}+\nu_\text{CB2}=1$ for a smaller $m_2$. In \figref{fig:the other topological band for asymmetric case}, we also show the band dispersion and the irreducible representations at high symmetry momenta for other higher-energy mini-bands (labelled by CB3, CB4, CB5 and CB6). We find the mini-bands of CB3, CB4 and CB5 are always touching each other and their total $\mathbb Z_2$ number is $\nu_\text{CB3}+\nu_\text{CB4}+\nu_\text{CB5}=1$ for $0.04 E_0 / \vert k_M \vert^2< m_2 <0.1 E_0 / \vert k_M \vert^2$. Another transition between CB5 and CB6 occurs at $m_2 \approx 0.035 E_0 / \vert k_M \vert^2$ (See \figref{fig:the other topological band for asymmetric case}e), and after this transition, $\nu_\text{CB3}+\nu_\text{CB4}+\nu_\text{CB5}$ becomes zero while the other non-trivial $\mathbb Z_2$ number is moved to even higher energy mini-bands. For $m_2 <0.1 E_0 / \vert k_M \vert^2$, these additional transitions only occur for higher-energy mini-bands, while the $\mathbb Z_2$ topology of CB1 and CB2 remains the same ($\nu_\text{CB1}+\nu_\text{CB2}=1$). For $m_2 < 0.04 E_0 / \vert k_M \vert^2$, we find a gap between CB1 and CB2 opens at K due to the interchange between the $\bar K_6$ and $\bar K_5$ mini-bands. Thus, the isolated CB1 with $\nu_\text{CB1} = 1$ and CB2 with $\nu_\text{CB2} = 0$ states can be found in \figref{fig:atomic limits by large m2 for asymmetric case}(a) for $m_2=0$. 

% When lowering $m_2$, there is a band inversion happens at $\Gamma$ between CB1, CB2 and higher conduction bands as shown in \figref{fig:atomic limits by large m2 for asymmetric case}(a)-(c).
% It changes the topology of CB1 and CB2 to $\nu_\text{CB1}+\nu_\text{CB2}=1$ for $m_2 = 0  E_0/\vert k_M \vert^2$ in \figref{fig:atomic limits by large m2 for asymmetric case}(a).
% When the gap is opened between CB1 and CB2, it happens $\nu_\text{CB1} = 1$ and $\nu_\text{CB2} = 0$.

% For $\alpha = 0, V_0/E_0=0, \phi = 1$, CB1 has $\nu_{CB1} = 1$ and CB2 $\nu_{CB1} = 0$. CB1 and CB2 combined are nontrivial.
% When increasing $m_2$, there is a band inversion at $\Gamma$ between CB2 and higher conduction bands for $m_2$ between \figref{fig:atomic limits by large m2}(e) and \ref{fig:atomic limits by large m2}(f).
% The band inversion changes the topology of CB1 and CB2 from nontrivial to trivial.
% For $m_2$ from \figref{fig:atomic limits by large m2}(f) and \ref{fig:atomic limits by large m2}(g), there is band inversion between CB1 and CB2 that does not change their topology.
% The CB1 and CB2 in \figref{fig:atomic limits by large m2}(g) are adiabatically to the lowest conduction bands also come from 2DEG on a moir\'e potential with Rashba SOC from the external electrical fields at large $m_2$ shown in \figref{fig:atomic limits by large m2}(h).

\begin{figure}
	\centering
	\includegraphics[width=0.8\columnwidth]{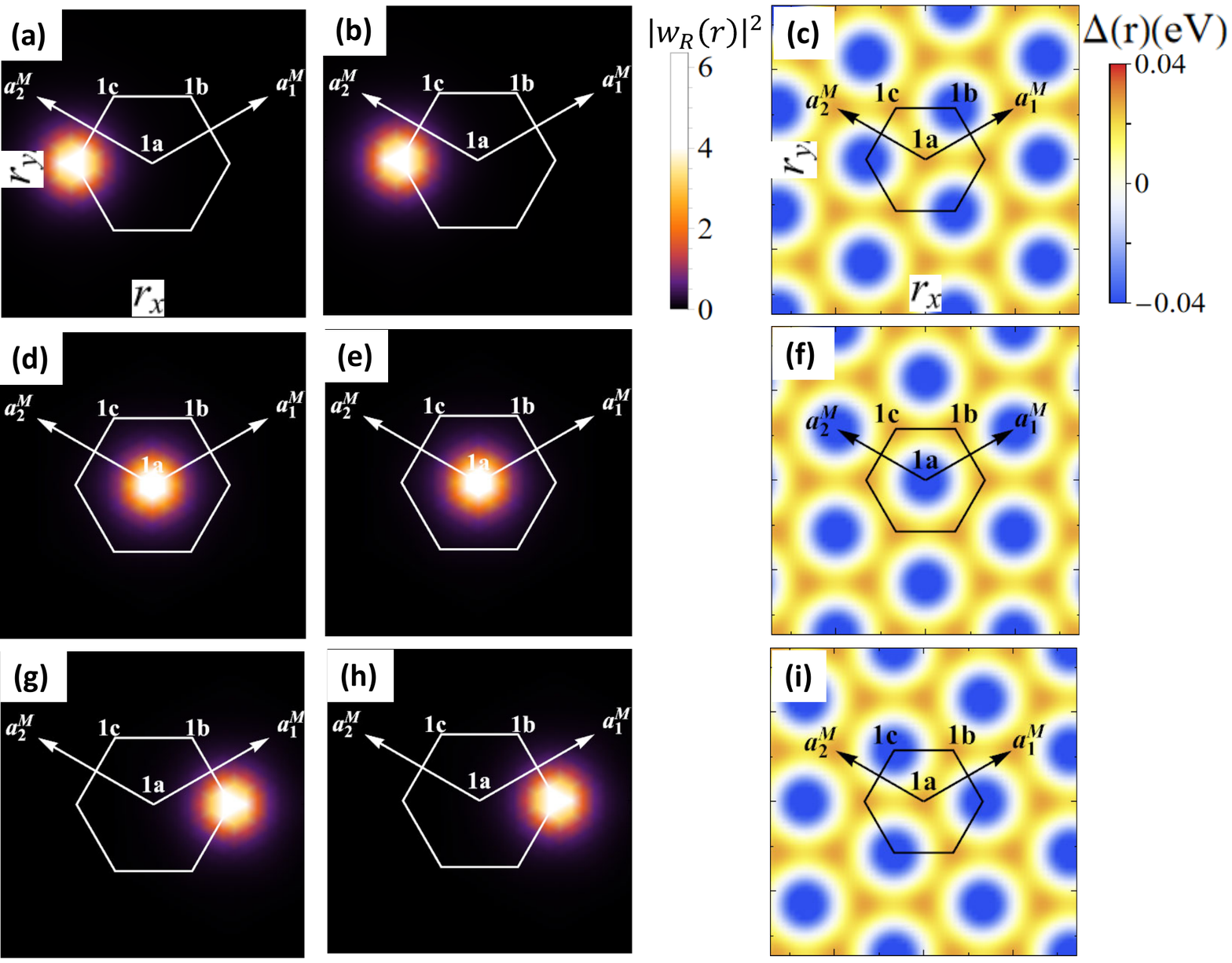}
	\caption{
        (a)(b) The real-space maximally localized Wannier functions $w_\vec R (\vec r)$ for the lowest conduction bands with $\phi = 1/6$ corresponding to \figref{fig:spectrum}(a) of the main text.
        (c) The real space moir\'e potentials with $\phi = 1/6$.
        (d)(e)(f) Those for $\phi = 1/2$ and (g)(h)(i) Those for $\phi = 5/6$.
	}
	\label{fig:Wannier functions}
\end{figure}

\subsection{Normal insulator phases of atomic limits}
We construct the maximally localized Wannier functions \cite{Pizzi2020} for the topologically trivial region for the CB1 as shown in \figref{fig:Wannier functions}.
The locations of Wannier functions show the NI phase of CB1 has localized orbitals at Wyckoff positions 1b for $\phi = 1/6$, 1a for $\phi = 1/2$, 1c for $\phi = 5/6$, as indicated in the phase diagram \figref{fig:spectrum}(a)(b) of the main text.
Comparing the Wannier functions with the moir\'e potentials, they are located at minima of moir\'e potentials and correspond to the lowest conduction bands as expected.
Since the minima of potentials change from one to another when tuning $\phi$, the localized orbitals shift from one location to the other.
The phase transition between two NI phases with orbitals at different Wyckoff positions has gap closing\cite{xu2021filling}, shown as the semi-metal phase in \figref{fig:spectrum}(a)(b) of the main text, as they belong to different atomic limits.

\section{Hartree Fock methods for Coulomb interaction}

%\addCXL{CX: 1. references need to be added. 2. We need to keep the formula applicable to both two band and four-band model. 3. More numerical results for self-consistent calculation should be added. }

\subsection{Eigenbasis projection}

In this section, we project the Coulomb interaction into the eigenbasis of the non interacting Hamiltonian $H_0(\vec k)$\cite{zhang2020correlated,lian2021twisted}.
The non-interacting moir{\' e} Hamiltonian in the second quantization form is
\begin{equation}
    H_0(\vec k)=\sum _{\vec G, \vec G',\alpha ,\alpha '} f_{\alpha }^{\dagger }(\vec k+ \vec G)\left(H^\text{TI}_{\alpha ,\alpha '}(\vec k+\vec G)\delta _{\vec G,\vec G'}+H^\text M (\vec G- \vec G')\delta _{\alpha ,\alpha'}\right)f_{\alpha '}(\vec k+\vec G'),
\end{equation}
where $\alpha=1,...,4$ labels both spin and layer index, 
\(f_{\alpha }^{\dagger }(\vec k+\vec G)\) is a fermion creation operator, $\vec k$ is within the first moir\'e BZ and \(\vec G\) is Moir{\' e} reciprocal lattice vectors.
The creation operators for eigenstates of \(H_0(\vec k)\) are are defined as
\begin{equation}
    c_n^{\dagger }(\vec k)=\sum _{\vec G,\alpha } u_{\vec G,\alpha }^n(\vec k)f_{\alpha }^{\dagger }(\vec k+\vec G),
\end{equation}
where \(u_{\vec G,\alpha }^n(\vec k)\) satisfies the eigen equation 
\begin{equation}\label{eq:eign equation for u}
    \sum _{\vec G',\alpha '} \left({H^\text{TI}}_{\alpha ,\alpha '}(\vec k+\vec G)\delta _{\vec G,\vec G'}+H^\text M (\vec G- \vec G')\delta _{\alpha ,\alpha'}\right)  u_{\vec G',\alpha' }^n(\vec k)
    = E^n_0(\vec k)  u_{\vec G,\alpha }^n(\vec k) 
\end{equation}
for \(H_0(\vec k)\) with energies \(E_0^n(\vec k)\).
By replacing $\vec G $ with $\vec G + \vec G_0$ and $\vec G'$ with $\vec G' + \vec G_0$ in \eqnref{eq:eign equation for u}, we obtain
\begin{equation}
    \sum _{\vec G',\alpha '} \left({H^\text{TI}}_{\alpha ,\alpha '}(\vec k+ \vec G_0 +\vec G)\delta _{\vec G,\vec G'}+H^\text M (\vec G- \vec G')\delta _{\alpha ,\alpha'}\right)  u_{\vec G' + \vec G_0,\alpha' }^n(\vec k)
    = E^n_0(\vec k)  u_{\vec G + \vec G_0,\alpha }^n(\vec k),
\end{equation}
which can be viewed as the eigen equations for $ u_{\vec G,\alpha }^n(\vec k + \vec G_0) $ by replacing $\vec k$ with $\vec k + \vec G_0$ in \eqnref{eq:eign equation for u}, 
\begin{equation}
        \sum _{\vec G',\alpha '} \left({H^\text{TI}}_{\alpha ,\alpha '}(\vec k+ \vec G_0 +\vec G)\delta _{\vec G,\vec G'}+H^\text M (\vec G- \vec G')\delta _{\alpha ,\alpha'}\right)  u_{\vec G',\alpha' }^n(\vec k + \vec G_0)
    = E^n_0(\vec k+\vec G_0)  u_{\vec G,\alpha }^n(\vec k + \vec G_0).
\end{equation}
Thus, we can fix the periodic gauge for the eigen-state as
\begin{equation}
    u_{\vec G + \vec G_0,\alpha }^n(\vec k) =  u_{\vec G,\alpha }^n(\vec k + \vec G_0). 
\end{equation}
As \(u_{G,\alpha }^n(\vec k)\) is a set of orthonormal basis, we can take the inverse of the above expansion as 
%$c_n^{\dagger }(k)$ has the properties
\begin{equation}
f_{\alpha }^{\dagger }(\vec k+\vec G)=\sum_n u_{\vec G,\alpha }^{n*}(\vec k)c_n^{\dagger}(\vec k)
\end{equation}
and
\begin{equation}
    \left\{c_n(\vec k),c_{n'}^{\dagger }(\vec k')\right\}=\sum _{\vec G,\alpha } u_{G,\alpha }^{n*}(\vec k)\sum _{\vec G',\alpha '} u_{\vec G',\alpha '}^{n'}(\vec k')\left\{f_{\alpha }(\vec k+\vec G),f_{\alpha'}^{\dagger }(\vec k'+\vec G')\right\}=\delta _{n,n'}\delta(\vec k - \vec k').
\end{equation}
% \(u_{G,\alpha }^n(k)\) is chosen to satisfy the periodic gauge \addCXL{CX: I think this is something one needs to choose, right? }
% \begin{equation}
% \begin{split}
% c_n^{\dagger }(k)&=c_n^{\dagger }(k+G')=\sum _{G,\alpha } u_{G,\alpha }^n(k+G')f_{\alpha }^{\dagger }(k+G'+G)=\sum _{G,\alpha } u_{G-G',\alpha }^n(k+G')f_{\alpha
% }^{\dagger }(k+G)\\
% &\Longrightarrow u_{G-G',\alpha }^n(k)=u_{G,\alpha }^n(k-G')
% \end{split}
% \end{equation}
% for any moir\'e reciprocal lattice vector $G'$. 
% \addCXL{Is the current Hamiltonian $H_0$ real? If so, we just need to say $C_{2z}T$ symmetry is a complex conjugate for the current Hamiltonian $H_0$ and that's enough. We do not need to make it so complicated. current $H_0$ is not real for $k_\pm$ in $H^\text{TI}$. And $C_2\TR$ are the symmetry used to take eigen states in HF}
To improve the efficiency of the numerical calculations, we need to further fix the gauge freedom of eigenstates.
An important step is to choose the real gauge for the Hamiltonian and eigenbasis due to the space-time inversion symmetry \(C_{2z}\mathcal{T}\) in 2D for moir\'e potential with $\phi=0$. 
Take \(C_{2z}\mathcal{T}=U_{\text{CT}}\mathcal{K}\) with \(\mathcal{K}\) as complex conjugate.
\(U_{\text{CT}}\) is unitary and satisfies the \(U_{\text{CT}}{}^*U_{\text{CT}}=1\) from \(\left(C_{2z}\mathcal{T}\right){}^2=1\).
Under the basis transformation \(U_{\text{CT}}{}^{1/2}\),
\begin{equation}
\left(U_{\text{CT}}{}^{1/2}\right){}^{\dagger }C_{2z}\mathcal{T} U_{\text{CT}}{}^{1/2}=U_{\text{CT}}{}^{-1/2}U_{\text{CT}}\left(U_{\text{CT}}{}^{1/2}\right){}^*\mathcal{K}=\mathcal{K},
\end{equation}
and the corresponding Hamiltonian and eigenbasis can be chosen to be real. There is still a $SO(2)$ gauge freedom left for eigenstates for Fig. 2(d) in the main text with inversion and $\pm$ gauge freedom for Fig. 2(e) in the main text without inversion.

In the eigenbasis, the non-interacting Hamiltonian is 
\begin{equation}
H_0(\vec k)=\sum _n c_n^{\dagger }(\vec k)E_0^n(k)c_n(\vec k).
\end{equation}
%We currently keep only the topologically non trivial lowest conduction bands to study its interaction effects.
The dual-gated Coulomb interaction potential is \cite{zhang2020correlated,bernevig2021twisted}
%\addCXL{CX: it is not good to use some words in the formula. Where did you take this form? We should cite it. I look at this paper Bernevig B A, Song Z D, Regnault N, et al. Twisted bilayer graphene. III. Interacting Hamiltonian and exact symmetries[J]. Physical Review B, 2021, 103(20): 205413. }
\begin{equation}
V(\vec q)=\frac{e^2\tanh \vert \vec q\vert d}{2\epsilon _0\epsilon _r  \vert \vec q\vert }\frac{1}{S},
\end{equation}
where $S$ is the area, \(d\) is the dual-gate distance, \(\epsilon _0\epsilon _r\) are permittivity, \(e\) is electron charge.
The Coulomb interaction Hamiltonian in second quantization form is
\begin{equation}
\begin{split}
H_\text I & =\frac{1}{2}\sum _{\vec k_1, \vec k_2,\vec q,\vec G}  \sum _{\vec G_1,\alpha _1,\vec G_2,\alpha _2} V(\vec q+\vec G) \\
    & f_{\alpha _1}^{\dagger}\left(\vec k_1+\vec G_1+\vec q+\vec G\right)f_{\alpha_2}^{\dagger}\left(\vec k_2+\vec G_2-\vec q-\vec G\right)f_{\alpha_2}\left(\vec k_2+\vec G_2\right)f_{\alpha _1}\left(\vec k_1+\vec G_1\right)\\
    &=\frac{1}{2}\sum _{\vec k_1,\vec k_2,\vec q,\vec G}  \sum _{m_1,n_1,m_2,n_2} V(\vec q+\vec G)\Lambda_{m_1n_1}\left(\vec k_1+\vec q+\vec G,\vec k_1\right)\Lambda _{m_2n_2}\left(\vec k_2-\vec q-\vec G,\vec k_2\right)\\
    & c_{m_1}^{\dagger}\left(\vec k_1+\vec q\right)c_{m_2}^{\dagger}\left(\vec k_2-\vec q\right)c_{n_2}\left(\vec k_2\right)c_{n_1}\left(\vec k_1\right)
\end{split}
\end{equation}
with the form factor
\begin{equation}
\Lambda _{m_1n_1}\left(\vec k_1+\vec G,\vec k_2\right)=\sum _{\vec G',\alpha '} u_{\vec G',\alpha '}^{m_1*}\left(\vec k_1+\vec G\right)u_{\vec G',\alpha '}^{n_1}\left(\vec k_2\right)=\left\langle u^{m_1}\left(\vec k_1+\vec G\right)|u^{n_1}\left(\vec k_2\right)\right\rangle.
\end{equation}
The form factor satisfies
\begin{equation}
\Lambda _{m_1n_1}\left(\vec k_1+\vec G,\vec k_2\right)=\Lambda _{n_1m_1}^*\left(\vec k_2,\vec k_1+\vec G\right)=\Lambda _{m_1n_1}\left(\vec k_1,\vec k_2-\vec G\right).
\end{equation}
In the real eigenbasis, the form factors are all real.

\subsection{Self-consistent Hartree-Fock mean field Theory}

In this section, we treat the Coulomb interaction under the Hartree-Fock (HF) approximation\cite{zhang2020correlated}.
The basic idea is the decoupling of four-fermion operators by
\begin{equation}
\begin{split}
c_1^{\dagger }c_1c_2^{\dagger }c_2
& =\left(\left\langle c_1^{\dagger }c_1\right\rangle +c_1^{\dagger }c_1-\left\langle c_1^{\dagger }c_1\right\rangle
\right)\left(\left\langle c_2^{\dagger }c_2\right\rangle +c_2^{\dagger }c_2-\left\langle c_2^{\dagger }c_2\right\rangle \right)\\
&\approx \left\langle c_1^{\dagger }c_1\right\rangle \left\langle c_2^{\dagger }c_2\right\rangle +\left\langle c_1^{\dagger }c_1\right\rangle \left(c_2^{\dagger
}c_2-\left\langle c_2^{\dagger }c_2\right\rangle \right)+\left(c_1^{\dagger }c_1-\left\langle c_1^{\dagger }c_1\right\rangle \right)\left\langle
c_2^{\dagger }c_2\right\rangle \\
&=\left\langle c_1^{\dagger }c_1\right\rangle c_2^{\dagger }c_2+c_1^{\dagger }c_1\left\langle c_2^{\dagger }c_2\right\rangle -\left\langle c_1^{\dagger
}c_1\right\rangle \left\langle c_2^{\dagger }c_2\right\rangle .
\end{split}
\end{equation}
The expectation value of the two-fermion operator is the density matrix
\begin{equation}\label{eq:density matrix}
\rho _{mn}(\vec k)= \left\langle c_m^{\dagger }(\vec k)c_n(\vec k)\right\rangle =\sum _j \psi^\text{HF *} _{j,m}(\vec k)\psi^\text{HF} _{j,n}(\vec k)n_F\left(E_j^\text{HF}(\vec k)\right)
\end{equation}
determined by
with \(n_F\) as the Fermi distribution function and \( \psi^\text{HF} _{j,m}(\vec k) \),\(E_j^\text{HF}(\vec k)\) as the $j$-th eigenstates and eigen-energies of Hartree-Fock Hamiltonian 
\begin{equation}\label{eq:hartree fock wavefunctions}
    \sum_m H^\text{HF}_{nm}[\rho](\vec k) \psi^\text{HF} _{j,m}(\vec k) = E_j^\text{HF}(\vec k) \psi^\text{HF} _{j,n}(\vec k),
\end{equation}
where $ H^\text{HF}_{nm}[\rho](\vec k)$ is defined in Eq. \eqnref{eq:hartree fock hamiltonian} below. We always choose $E_{j=1}^\text{HF}(\vec k)<E_{j=2}^\text{HF}(\vec k)<....$, so the mean field ground state is given by the eigen-state $\psi^\text{HF} _{j=1}(\vec k)$.
% \addCXL{CX: so this is different from the eigen-state and eigen-energy of $H_0$ right? we need to write down the eigen-equation to make it clear. }.
% \addCXL{CX: which step we choose the uniform order parameter? We should illustrate all the approximation we have made. We donot choose uniform density matrix. i dont understand the question. $\rho (\vec k)$ always depend on $\vec k$. I think you get this impression by $H_\text I[\rho]$. So I change to $H_\text I[\rho(\vec k)]$}
%\addCXL{CX: I mean the uniform order parameter in real space, not momentum space, which correspond to the charge or spin density wave order parameter. That's the question Ribu asked. That should correspond to the order parameter form $\rho _{mn}(\vec k, \vec q)= \left\langle c_m^{\dagger }(\vec k)c_n(\vec k+q)\right\rangle$ with non-zero $\vec q$.   }
Here, we do not consider non-uniform order parameters in real space with the form $\left\langle c_m^{\dagger }(\vec k)c_n(\vec k+\vec q)\right\rangle$ for $\vec q \neq 0$.

The Coulomb interaction under Hartree-Fock approximation is
\begin{equation}
\begin{split}
    H_\text I[\rho (\vec k)] & =\frac{1}{2}\sum _{\vec k_1,\vec k_2,\vec q,\vec G} V(\vec q+\vec G) \sum _{m_1,n_1,m_2,n_2} \Lambda_{m_1n_1}\left(\vec k_1+\vec q+\vec G,\vec k_1\right)\Lambda _{m_2n_2}\left(\vec k_2-\vec q-\vec G,\vec k_2\right) \\ 
    & \quad ( \delta_{\vec q=0}\left(\rho _{m_1n_1}\left(\vec k_1\right)c_{m_2}^{\dagger }\left(\vec k_2\right)c_{n_2}\left(\vec k_2\right)+c_{m_1}^{\dagger }\left(\vec k_1\right)c_{n_1}\left(\vec k_1\right)\rho_{m_2n_2}\left(\vec k_2\right) -\rho _{m_1n_1}\left(\vec k_1\right)\rho _{m_2n_2}\left(\vec k_2\right) \right)\\
    & -\delta _{\vec q=\vec k_2-\vec k_1}\left(\rho _{m_1n_2}\left(\vec k_2\right)c_{m_2}^{\dagger}\left(\vec k_1\right)c_{n_1}\left(\vec k_1\right)+c_{m_1}^{\dagger }\left(\vec k_2\right)c_{n_2}\left(\vec k_2\right)\rho _{m_2n_1}\left(\vec k_1\right)-\rho _{m_1n_2}\left(\vec k_2\right)\rho_{m_2n_1}\left(\vec k_1\right)\right) )\\
    & =\sum _{\vec k_1} C^{\dagger }\left(\vec k_1\right)\left(H_\text I^\text H[\rho ]\left(\vec k_1\right)-H_\text I^\text F[\rho]\left(\vec k_1\right)\right)C\left(\vec k_1\right)-E_C[\rho].
\end{split}
\end{equation}
with the Hartree term $H_\text I^\text H[\rho]\left(\vec k_1\right)$, Fock term $H_\text I^\text F[\rho]\left(\vec k_1\right)$, condensation energy $E_C[\rho]$ defined as
\begin{equation}
\begin{split}
    H_\text I^\text H[\rho]\left(\vec k_1\right) & = \sum _{\vec k_2,\vec G} V(\vec G)\Lambda \left(\vec k_1-\vec G,\vec k_1\right)\text{Tr} \left( \rho \left(\vec k_2\right)\Lambda ^*\left(\vec k_2-\vec G,\vec k_2\right)\right)\\
    H_\text I^\text F[\rho]\left(\vec k_1\right)& = \sum _{\vec k_2,\vec G} V\left(\vec k_2-\vec k_1+\vec G\right)\Lambda \left(\vec k_1-\vec G,\vec k_2\right)\rho ^T\left(\vec k_2\right)\Lambda ^{\dagger }\left(\vec k_1-\vec G,\vec k_2\right)\\
    E_C[\rho] & = \frac{1}{2}\sum _{\vec k_1,\vec k_2,\vec G} V(\vec G) \text{Tr}\left(\rho \left(\vec k_1\right)\Lambda ^T\left(\vec k_1-\vec G,\vec k_1\right)\right) \text{Tr}\left(\rho \left(\vec k_2\right)\Lambda^*\left(\vec k_2-\vec G,\vec k_2\right)\right)\\
    & -\frac{1}{2}\sum _{\vec k_1,\vec k_2,\vec G} V\left(\vec k_2-\vec k_1+\vec G\right)\text{Tr} \left(\rho ^T\left(\vec k_1\right)\Lambda \left(\vec k_1-\vec G,\vec k_2\right)\rho^T\left(\vec k_2\right)\Lambda ^{\dagger }\left(\vec k_1-\vec G,\vec k_2\right)\right).\\
\end{split}
\end{equation}
%\addCXL{CX: the explicit form of three terms should be written down. It is not clear which term corresponds to which term in the current form. }
\(C^{\dagger }(k)=(c_1^{\dagger }(\vec k),c_2^{\dagger}(\vec k)),\dots,c_n^{\dagger}(\vec k))\) with $n$ as the number of bands projected.
Since the \(H_0(\vec k)\) comes from DFT with Hartree-Fock interaction, the non-interacting states \(\psi^\text{HF}_{j,m}(k)=\delta _{j,m}\) or \(\rho_0(\vec k)\) would be a solution to the Hartree-Fock mean-field Hamiltonian.
To achieve this, the \(H_\text I \left[\rho_0\right]\) is subtracted from \(H_\text I \left[\rho\right]\)\addKJ{\cite{zhang2020correlated,liu2021nematic}}. 
%\addCXL{cite the references for this treatment. }
We define the Hartree-Fock Hamiltonian to be
\begin{equation}\label{eq:hartree fock hamiltonian}
\begin{split}
H^{\text{HF}}[\rho](\vec k)
&=H_0(\vec k)+H^\text H_\text I[\rho](\vec k)-H^\text F_\text I[\rho](\vec k)-\left(H^\text H_\text I\left[\rho _0\right](\vec k)-H^\text F_\text I[\rho_0](\vec k) \right) \\
&=H_0(\vec k)+H^\text H_\text I[\rho-\rho_0](\vec k)-H^\text F_\text I[\rho - \rho_0](\vec k).
\end{split}
\end{equation}

We solve \( H^{\text{HF}}[\rho](\vec k) \) self-consistently in the following standard procedures. 
We first choose an initial guess of the density matrix, denoted as $\rho_{ini}(\vec k)$, as the order parameter for the filling of one band (half filling in two-band model and one quarter filling for four-band model).
% The first one is \(\rho \propto \sigma _y\), which breaks the \(\mathcal{T}\) symmetry and \(C_{2z}\mathcal{T}\) symmetry.
% The other one is \(\rho \propto \sigma _{x,z}\) that preserves the \(C_{2z}\mathcal{T}\) symmetry but breaks \(U(1)\) symmetry  \addCXL{CX: what U(1) symmetry do you mean here? spin U(1)? }.
% Here $\sigma_{x,y,z} $ acts on the lowest two conduction band. 
% These two types of order parameters are decoupled in the self-consistent calculations because \(H_H^I\left[\sigma _y\right]-H_F^I\left[\sigma
% _y\right]\) is purely imaginary \addCXL{I do not understand here. Hamiltonian needs to be hermitian, right? So do you mean here the interaction Hamiltonian is purely off-diagonal and imaginary? } and \(H_H^I\left[\sigma _{x,z}\right]-H_F^I\left[\sigma _{x,z}\right]\) is real in the real eigenbasis.
Based on $\rho_{ini}(\vec k)$, we can construct \(H^{\text{HF}}[\rho_{ini}](\vec k)\) from \eqnref{eq:hartree fock hamiltonian} and calculate
the corresponding new eigenstates that allow us to construct the new density matrix, denoted as $\rho_{new}(\vec k)$. 
% \addCXL{CX: what's the criterion for the convergence? }
We reset $\rho_{ini}(\vec k)=\rho_{new}(\vec k)$ and continue the iterative process until the convergence is achieved. 
The criterion for the convergence is taken as the spectra $\tilde E_{j}^\text{HF}(\vec k)$ of $H^\text{HF}[\rho_{ini}]$ and $E_{j}^\text{HF}(\vec k)$ of $H^\text{HF}[\rho_{new}]$ satisfy
\begin{equation}
\max_{j,\vec k} \vert \tilde E_{j}^\text{HF}(\vec k) -    E_{j}^\text{HF}(\vec k) \vert < 10^{-5}E_0
\end{equation}
with $\max$ taken for all bands $j$ in $H^\text{HF}$ and $\vec k$ on the high symmetry lines $\Gamma-K-M$ as shown in \figref{fig:HF for CB1 with inversion rhoy}(a). \addKJ{ $E_0 = v \vert \vec b_1^\text M\vert$.}

The final self-consistent solution for the density matrix is denoted as $\rho^{HF}(\vec k)$ which is determined by the eigen wavefunctions $\ket{\psi^\text{HF} _{j}(\vec k)}$ by \eqnref{eq:density matrix} and (\ref{eq:hartree fock wavefunctions}).
% \addCXL{Define $j$ early here.  }
The energy per particles for each self-consistent solution to the mean-field Hamiltonian is
\begin{equation}
E_\text I [\rho]=\frac{1}{N}\sum _\vec k \text{Tr$\rho $}^T(\vec k)H^{\text{HF}}[\rho](\vec k) - \left(E_C[\rho]-E_C\left[\rho_0\right]\right).
\end{equation}
with $N$ as the number of electrons for the filling.
% \addCXL{When we get the self-consistent density matrix, do you still need to minimize $E$ here? }
% The ground state is determined by minimizing $E_\text I [\rho]$.

\subsection{Two-band model of CB1}
%\addCXL{CX: not clear here. This part needs to be integrated into the numerical results. All the figures for the analysis in your talk should be included into the appendix. }
%\addCXL{CX: the logic in this paragraph is not so clear. Before going into any details, you need to have several sentences before each paragraph to let readers know what you want to discuss and why you want to do that. }
In this section, we discuss the self-consistent solutions of $H^\text{HF}(\vec k)$ at half filling of two-band model for CB1. Below we will discuss both the inversion-symmetric and asymmetric cases.

We first describe our gauge choice of the non-interacting eigen-states for the case $\alpha = 1, V_0/E_0 = 0, \phi = 0$ with inversion symmetry, which is important to simplify the numerical calculations. 
The non-interacting states are $\ket{u^\text{CB1}_{\pm i}(\vec k)}$ with the $m_z$ eigenvalues $\pm i$. 
%\addCXL{ $\ket{u^\text{CB1}_{\pm i}(\vec k)}$ here is the same as $\ket{\psi_{\text{CB1},-i}(\vec k)}$ in the early part, or not? If it is the same, we should unify the notation.  }
% For the system with inversion symmetry \( I\), the bands are doubly degenerate for \(\mathcal{T} I\). 
%\(\text{im}_z=\text{$ I$C}_{2z}\) block diagonalize \(H^0\) and mirror Chern number can be defined.
The mirror Chern number\cite{fu2011topological} $C = \pm 1$ can be defined for $\ket{u^\text{CB1}_{\pm i}(\vec k)}$.
$C_{2z}\mathcal{T}$ relates two $m_z$-eigen states by
\begin{equation}
    C_{2z}\mathcal{T} \ket{u^\text{CB1}_{\pm i}(\vec k)} =\ket{u^\text{CB1}_{\mp i}(\vec k)}.
\end{equation}
It turns out that the Hartree-Fock calculations can be simplified by taking the real gauge due to the $C_{2z}\mathcal{T}$ symmetry and thus we transform the basis wavefunctions into the real-gauge form
\begin{equation}\label{eq:real gauge}
\begin{split}
\ket{u^{\text R,\text{CB1}}_+(\vec k)} = \frac{1}{\sqrt{2}}\left(e^{\text{i$\varphi $}_\vec k}\ket{u^\text{CB1}_{- i}(\vec k)} +e^{-\text{i$\varphi $}_\vec k}\ket{u^\text{CB1}_{+ i}(\vec k)} \right) \\
\ket{u^{\text R,\text{CB1}}_-(\vec k)} = \frac{1}{\sqrt{2}i}\left(e^{\text{i$\varphi$}_\vec k}\ket{u^\text{CB1}_{- i}(\vec k)} -e^{-\text{i$\varphi $}_\vec k}\ket{u^\text{CB1}_{+ i}(\vec k)}\right), 
\end{split}
\end{equation}
% \addCXL{CX: is $\varphi_k$ obtained from the self-consistent calculation, or is it arbitrary? Any value of $\varphi_k$ can satisifies the self-consistent equation? }
where $\varphi_k$ is the remaining relative $U(1)$ phase between eigen-states opposite $m_z$ (spin $U(1)$ symmetry).
The real eigen-states with different $\varphi_k$ can be related by a $SO(2)$ transformation %\addCXL{CX: is this a U(1) or SO(2) transformation? }
\begin{equation}\label{eq:U(1)}
    R (\tilde \varphi_\vec k)=
    \begin{pmatrix}
        \cos \tilde \varphi_\vec k & - \sin \tilde \varphi_\vec k \\
        \sin \tilde \varphi_\vec k & \cos \tilde \varphi_\vec k \\
    \end{pmatrix},
\end{equation}
which shifts $\varphi_\vec k$ to $\varphi_\vec k+ \tilde \varphi _\vec k $.
The other symmetry operators can be taken as
\begin{equation}
    \TR = i \sigma_y \mathcal K;  \quad
    C_{2z} = - i \sigma_y;  \quad
    C_{2z} \TR =  \mathcal K; \quad
    \mathcal M_z = - i \sigma_y,
\end{equation}
with Pauli matrices $\sigma$ redefined under the basis $\ket{u^{\text R,\text{CB1}}_\pm(\vec k)}$.
$\ket{u^{\text R,\text{CB1}}_\pm(\vec k)}$ are taken as the eigenstates projected for the self-consistent Hartree-Fock calculations, which are related to the basis $\ket{u^\text{CB1}_{m_z = \pm i}}$ used in the main text by \eqnref{eq:real gauge}.
%\addCXL{CX: In Eq. S65, there is a $\varphi$ phase, but in $U^R$ below, there is no $\varphi$ phase. Please make it consistent. }
The density matrices in the main text, denoted as $[\rho]_{\alpha\beta}=\langle u^\text{CB1}_{\alpha}| \hat{\rho} |u^\text{CB1}_{\beta}\rangle $ with $\alpha,\beta=\pm i $, are related to the density matrices $\rho^\text R$  in the real basis discussed below, denoted as $[\rho^\text R]_{\alpha\beta}=\langle u^\text{R,CB1}_{\alpha}| \hat{\rho} |u^\text{R, CB1}_{\beta}\rangle $ with $\alpha,\beta=\pm $, by
\begin{equation}
    \rho^\text R = U^{\text R \dagger} \rho U^\text R
\end{equation}
and
%\addKJ{The two basis and the density matrices written in these two basis are related by a gauge transformation}
%The density matrices in this section and in Fig. 3(b) of the main text are related by a gauge transformation %\addCXL{ Need to check.  $U^R \sigma_{x,y,z} U^R\dagger = \sigma_{z,x,y}$}
% \addCXL{CX: we do not use $\ket{u^\text{CB1}_{\pm i}(\vec k)}$ in the main text. Do you mean that in the early part of the SM?  }
\begin{equation}
    U^\text R(\vec k) = \frac{1}{\sqrt 2}
    \begin{pmatrix}
        e^{i\varphi_\vec k} & -i e^{i\varphi_\vec k} \\
        e^{-i\varphi_\vec k} & i e^{-i\varphi_\vec k} \\
    \end{pmatrix},
\end{equation}
which transforms Pauli matrices $\sigma$ as $U^{\text R \dagger} \sigma_y U^\text R =\sigma_x  \cos{2\varphi_\vec k} - \sigma_z \sin{2\varphi_\vec k}, U^{\text R \dagger} \sigma_x U^\text R = \sigma_x  \sin{2\varphi_\vec k} + \sigma_z \cos{2\varphi_\vec k}, U^{\text R \dagger}\sigma_z U^\text R = \sigma_y$.
%\addCXL{CX: Give the expressions how related. If we want to talk about density matrix, we should also give how density matrices are related.  }

%\addCXL{KJ: Here the gauge is picked by convenience of presentation. In the numerical calculation, the gauge between states at $\pm \vec k$ for $\TR,C_{2z}$ is not fixed. The question is how to fix the $U(1)$ gauge numerically. How can we find a smooth $U(1)$ gauge for every $\vec k$ giving the symmetry operators above relating only $\pm \vec k$?}
%\addCXL{CX: Is your question for me? What did you do with your calculation? Do we really need to fix gauge for the HF calculation? }

\begin{figure}
	\centering
	\includegraphics[width=\columnwidth]{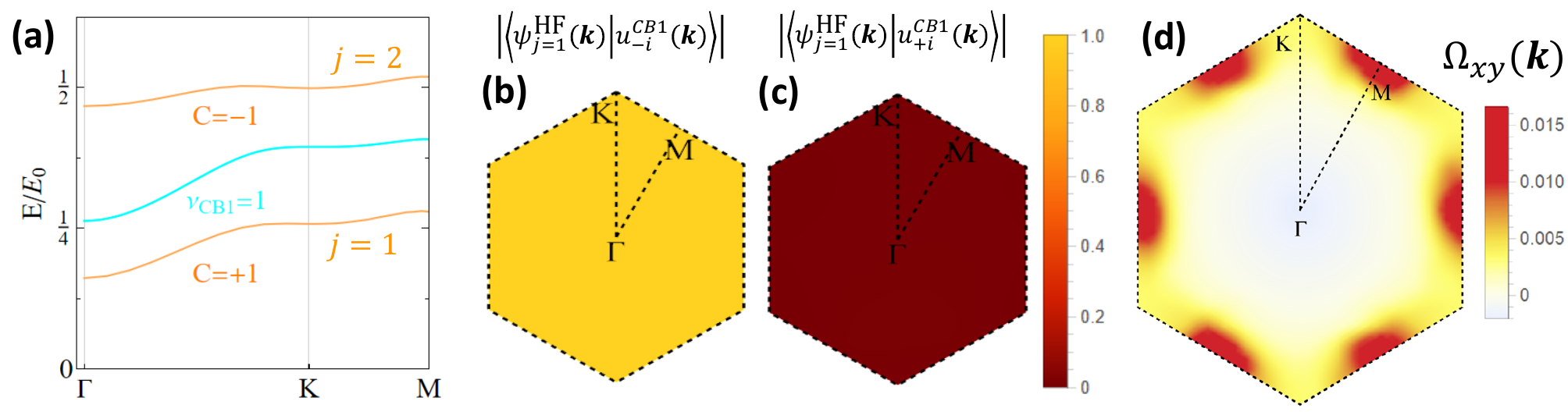}
	\caption{
	    (a) The spectrum of $H^\text{HF}(\vec k)$ for mirror polarized states with $\rho^\text R_y(\vec k)$ (orange lines) . 
        (b)(c) The overlap between Hartree-Fock states $\ket{\psi^{\text HF}_{j=1} (\vec k)}$ in (a) and the non-interacting mirror polarized basis wavefunction $\ket{u^\text{CB1}_{\pm i}}(\vec k)$. %\addCXL{What $j$ is for the two plots? }
        (d) The Berry curvature $\Omega_{xy} (\vec k)$ over the moir\'e BZ for the filled band. Here the calculation is for the two-band model with the parameters $\phi=0, \alpha=1, V_0 /E_0 = 0$. %CX: please add the parameters for this figure.
	}
	\label{fig:HF for CB1 with inversion rhoy}
\end{figure}

%\addCXL{CX: Change the notation in the table to $\rho^{HF}$ for the ground state.  }
% \addCXL{CX: one needs a paragraph to summarize what order parameters we are considering and why?  }
We performed the self-consistent calculations on the basis $\ket{u^{\text R,\text{CB1}}_\pm(\vec k)}$ and generally consider the following two types of order parameters: $\rho^\text R_y (\vec k) = f_{0y}(\vec k) \sigma_0 +  f_y(\vec k) \sigma_y$ and $\rho^\text R_{zx} (\vec k) = f_{0zx}(\vec k) \sigma_0 + f_x(\vec k) \sigma_x + f_z(\vec k) \sigma_z$ with $\sigma$ acting on the basis $\ket{u^{\text R,\text{CB1}}_\pm(\vec k)}$ in \eqnref{eq:real gauge}. These two types of order parameters possess different symmetry properties as summarized in Tab.\ref{tab:symmetry operators for interacting states}. For $\rho^\text R_y (\vec k)$, the density matrix breaks the $C_{2z} \mathcal T$ symmetry with complex $f_y(\vec k) \sigma_y$ and preserves $SO(2)$ symmetry in \eqnref{eq:U(1)} by $[R(\tilde \varphi _ \vec k), \rho^\text R_y(\vec k)]=0$. $\rho^\text R_y (\vec k)$ also preserves the z-directional mirror symmetry, $[\rho^\text R_y(\vec k), \mathcal M_z] = 0$, as $\mathcal M_z = -i\sigma_y$ in the $\ket{u^{\text R,\text{CB1}}_\pm(\vec k)}$ basis, which is the generator of the $SO(2)$ symmetry.
%\addCXL{CX: do you mean mirror z is the generator of SO(2)? Is that just the mirror U(1) symmetry?  }
It represents the many-body states polarized to one of the mirror states $\ket{u^\text{CB1}_{m_z}}$, dubbed as mirror-polarized states.
For $\rho^\text R_{zx} (\vec k)$, the density matrix is real and preserves the $C_{2z} \mathcal T$ symmetry ($[C_{2z} \mathcal T, \rho^\text R_{zx}(\vec k)]=0$) but breaks $SO(2)$ symmetry.
It represents the many-body states with superposition of both mirror states $\ket{u^\text{CB1}_{m_z}}$, dubbed as mirror-coherent states.
The identity matrix $\sigma_0$ appears in both order parameters and mainly determines the filling of states $\ket{\psi^\text{HF}_j(\vec k)}$ at different momenta $\vec k$.
% XXX \addCXL{CX: $\sigma_0$ appear in both order parameter. Does it just shift the chemical potential? }
\begin{table}[t]
    \centering
    \begin{tabular}{|c|c|c|c|c|c|c|c|c|c|}
        \hline
        \ & $\mathcal T$ & $C_{2z}$ & $C_{2z}\mathcal T$ & $\mathcal M_z$ & $\mathcal I$ & $C_{6z}$ & $C_{3z}$ & $\mathcal M_x$ & $\mathcal M_y$ \\
        \hline
        \ &  $i \tau_0 s_y \mathcal K$ & $ -i \tau_0 s_z$ & $ i\tau_0 s_x \mathcal K $ & $ -i \tau_x s_z$ & $ \tau_x s_0$ & $\exp(-i \pi \tau_0 s_z /6)$ & $\exp(-i \pi \tau_0 s_z / 3)$ & $-i \tau_0 s_x $  & $-i \tau_0 s_y $\\
        \hline
        \ & $i \sigma_y \mathcal K$ & $-i \sigma_y$  & $\mathcal K$ & $ - i \sigma_y $ & $\sigma_0$ & $\exp(-i\pi \sigma_y / 6)$ & $\exp(-i\pi \sigma_y / 3)$ & $-i \sigma_z$ & $-i \sigma_x$ \\
        \hline
        $\rho_y^\text{R,HF}(\vec k)$  & $\times$ & \checkmark & $\times$ & \checkmark & \checkmark & \checkmark & \checkmark & $\times$ & $\times$ \\
        \hline 
        $\rho_{zx}^\text{R,HF}(\vec k)$ & $\times$ & $\times$ & \checkmark & $\times$ & \checkmark & $\times$ & $\times$  & \checkmark & $\times$\\
        \hline
    \end{tabular}
    \caption{A summary of symmetries preserved (\checkmark) or broken ($\times$) by the mirror polarized states with $\rho_y^\text{R,HF}(\vec k)$ and the mirror coherent states with $\rho_{zx}^\text{R,HF}(\vec k)$. $\rho^\text{R,HF}(\vec k)$ are the self-consistent solutions from the mean-field Hamiltonian $H^\text{HF}(\vec k)$. The symmetry operators are written in two basis. $\tau,s$ are Pauli matrices for the surface and spin basis as Eq.1 in the main text. $\sigma$ are the Pauli matrices for the real basis $\ket{ u^\text{R,CB1}_\pm(\vec k) }$.}
    \label{tab:symmetry operators for interacting states}
\end{table}

Different symmetry properties of $\rho^\text R_y (\vec k)$ and $\rho^\text R_{zx} (\vec k)$ under the $C_{2z}\TR$ and $SO(2)$ symmetry guarantee that they will not mix with each other. We may start from the initial density matrix $\rho^\text R_{ini}(\vec k)=\rho^\text R_y (\vec k)$ with certain forms of $f_{0y}$ and $f_y$, which preserves the $R(\tilde \varphi _\vec k)$ symmetry, $[\rho^\text R_{ini}(\vec k),R(\tilde \varphi _\vec k)]=0$. As the Hartree-Fock Hamiltonian $H^{\text{HF}}[\rho^\text R_{ini}](\vec k)$ is constructed from $\rho^\text R_{ini}$, direct calculation shows that $[H^{\text{HF}}[\rho^\text R_{ini}](\vec k),R(\tilde \varphi _\vec k)]=0$ for any $\varphi _\vec k$. From Eq. (\ref{eq:U(1)}) of $R(\tilde \varphi _\vec k)$, the Hamiltonian has to take the form 
\begin{eqnarray}
H^{\text{HF}}[\rho^\text R_{ini}](\vec k)= h_0(\vec k) \sigma_0 + h_y(\vec k) \sigma_y,
\end{eqnarray}
where $h_0(\vec k), h_y(\vec k)$ are some functions of $\vec k$ which can be determined numerically. From the above form of the Hamiltonian, the new density matrix can be evaluated as 
\begin{equation}
    \rho^\text R_{new}(\vec k) = \sum_{j=\pm} n_F \left( h_0(\vec k) + j  h_y(\vec k)  \right) \frac{1}{2} \left(\sigma_0 - j \sigma_y \right),
\end{equation}
which still satisfies $[\rho^\text R_{new}(\vec k),R(\tilde \varphi _\vec k)]=0$. $n_F(E)$ is the Fermi distribution function. Thus, the $R(\tilde \varphi _\vec k)$ symmetry is preserved in the self-consistent calculation process and thus the Pauli matrices $\sigma_x$ and $\sigma_z$ cannot be generated in the final $\rho^\text{R,HF}_y(\vec k)$. 

Similar argument can be applied to the initial density matrix $\rho^\text R_{ini}(\vec k)=\rho^\text R_{zx} (\vec k)$ with certain forms of $f_{0zx}, f_z, f_x$. The $C_{2z}\mathcal T$ symmetry is preserved for $\rho^\text R_{ini}(\vec k)$ and $H^{\text{HF}}[\rho^\text R_{ini}](\vec k)$. As a result, the Hamiltonian form has to be 
\begin{eqnarray}
H^\text{HF}= h_0(\vec k) \sigma_0 + h_x(\vec k) \sigma_x + h_z(\vec k) \sigma_z,
\end{eqnarray}
and the new density matrix is 
\begin{equation}
    \rho^\text R_{new} (\vec k) = \sum_{j=\pm} n_F \left( h_0(\vec k) + j \sqrt{h_x^2(\vec k) + h_z^2(\vec k)} \right) \frac{1}{2} \left(\sigma_0 + j \frac{h_x(\vec k)}{\sqrt{h_x^2(\vec k)+ h_z^2(\vec k)}}\sigma_x  + j \frac{h_z(\vec k)}{\sqrt{h_x^2(\vec k)+ h_z^2(\vec k)}}\sigma_z \right),
\end{equation}
which has the $C_{2z}\mathcal T$ symmetry, $[\rho^\text R_{new}(\vec k),C_{2z}\mathcal T]=0$. So the Pauli matrix $\sigma_y$ cannot be mixed into the density matrix $\rho^\text{R,HF}_{zx}(\vec k)$ in the above procedure. Based on this symmetry argument, we can discuss the self-consistent solutions for the density matrix form $\rho^\text R_y (\vec k)$ and $\rho^\text R_{zx} (\vec k)$, separately, below.

For $\rho^\text R_y(\vec k)$, we choose the initial density matrix as
\begin{equation}
    \rho^\text R_{ini}(\vec k) =  \frac{1}{2}(\sigma_0 - \sigma_y),
\end{equation}
which can be obtained from the states $\ket{u^\text{CB1}_{- i}(\vec k)}$.
% $\rho_i(\vec k)$ breaks the $C_{2z}\mathcal T$ by breaking the $\mathcal T$ and preserving $C_{2z}$.
% Here, we does not consider the initial density matrices breaking $C_{2z}$ and preserving $\mathcal T$.
% The HF ground state may break the $\TR$ symmetry. Because $\TR \ket{u^\text{CB1}_{\pm i}(\vec k)} = \pm i \ket{u^\text{CB1}_{\mp i}(\vec k)}$, a initial density matrix $\rho _i(\vec k)$ can be the mirror polarized states for half-filling.
% Take $\ket{u^\text{CB1}_{- i}(\vec k)}$ as an example. \addCXL{CX: is it confusing that you transform forth and back between two different basis? I do not understand your purpose. }
% \begin{equation}
%     \ket{u^\text{CB1}_{- i}(\vec k)} = e^{- i \varphi_\vec k} \frac{1}{\sqrt{2}} \left( \ket{u^{\text R,\text{CB1}}_+(\vec k)} + i \ket{u^{\text R,\text{CB1}}_-(\vec k)}  \right)
% \end{equation}
% with the initial density matrix \addCXL{Why this matrix for mirror polarized states? }
% \begin{equation}
%     \rho_{i}= \frac{1}{2}(\sigma_0 - \sigma_y).
% \end{equation}
% ,which breaks the $\TR$, $C_{2z}\TR$ but preserves $C_{2z}$.
% The $\rho_{i} $ is invariant under $U(1)$ gauge transformation in \eqnref{eq:U(1)} and $U(1)$ symmetry is preserved.
% Because $C_{2z}\TR = \mathcal K$ is broken, the density matrix is complex and has $\sigma_y$ terms.
% The self-consistent solutions is denoted as $\rho_y (\vec k)$ with a term $f_y(\vec k)\sigma_y$ breaking $C_{2z}\TR$ symmetry with $f_y(\vec k)$ as a real function capturing $\vec k$ dependence.
Although the initial density matrix $\rho^\text R_{ini}$ is independent of $\vec k$, the $H^\text{HF}(\vec k)$ in \eqnref{eq:hartree fock hamiltonian} depends on $\vec k$ and the self-consistent density matrix should in principle depend on $\vec k$.
The self-consistent solutions are shown in \figref{fig:HF for CB1 with inversion rhoy}, in which we evaluate the overlap 
%It can be seen that the Hartree-Fock solutions \addCXL{ground state? } $\ket{\psi^\text{HF} _{j}(\vec k)}$ are mirror polarized states from 
\begin{equation}
\vert \langle \psi^\text{HF} _{j=1}(\vec k)\vert u^\text{CB1}_{-i}(\vec k) \rangle \vert = 1 \quad \vert \langle \psi^\text{HF} _{j=1}(\vec k) \vert u^\text{CB1}_{+i}(\vec k)\rangle \vert = 0
\end{equation}
in \figref{fig:HF for CB1 with inversion rhoy}(b)(c) with $\ket{\psi^\text{HF} _{j=1}(\vec k)} = \sum _{m=\pm} \psi^\text{HF} _{1,m}(\vec k) \ket{u^{\text R,\text CB1}_m(\vec k)}$ in \figref{fig:HF for CB1 with inversion rhoy}(a) for the filled bands at half-filling. 
% \addCXL{Can you change the notation related to $j$ above and below, also in the figure? }
%\addCXL{specify what values $j$ can be and what $j$ you take here? }.
Furthermore, the Chern number for the band $j$ can be evaluated by 
\begin{equation}
    C = \frac{1}{2\pi} \int d^2 \vec k \ \Omega_{xy}(\vec k),
\end{equation}
where the Berry curvature is calculated by \cite{fukui2005chern} 
\begin{equation}
\begin{split}
    \Omega_{xy}(\vec k) =- \arg \Big(  &\bra{\psi^\text{HF}_{j}(\vec k)} \psi^\text{HF}_{j}(\vec k + \delta k_x ) \rangle  \bra{\psi^\text{HF}_{j}(\vec k  + \delta k_x)} \psi^\text{HF}_{j}(\vec k +\delta k_x + \delta k_y )\\ 
    & \bra{\psi^\text{HF}_{j}(\vec k  + \delta k_y)} \psi^\text{HF}_{j}(\vec k +\delta k_x + \delta k_y )\rangle^{-1} \bra{\psi^\text{HF}_{j}(\vec k )} \psi^\text{HF}_{j}(\vec k + \delta k_y )\rangle^{-1} \Big)
\end{split}
\end{equation}
with $\delta k_x, \delta k_y$ as the momenta connecting neighboring momentum grid points in the $x,y$ direction. 
Our calculation shows $C=+1$ for the filled band $\ket{\psi^\text{HF} _{j=1}(\vec k)}$ with the Berry curvature distribution shown in \figref{fig:HF for CB1 with inversion rhoy}(d). 

%carried by $\vert u^\text{CB1}_{-i}(\vec k) \rangle$, which is checked with Berry curvature shown in \figref{fig:HF for CB1 with inversion rhoy}(c).

\begin{figure}
	\centering
	\includegraphics[width=\columnwidth]{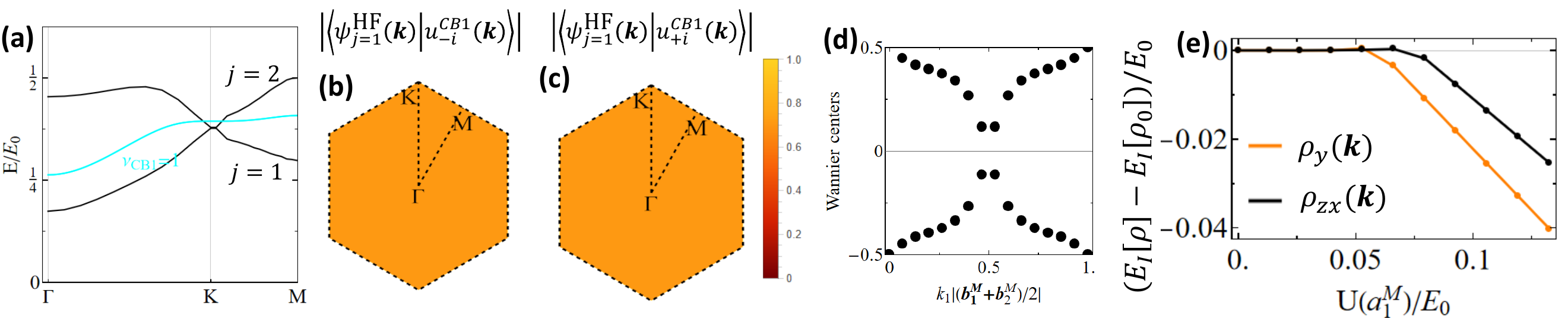}
	\caption{
	    (a) The spectrum of $H^\text{HF}(\vec k)$ for the mirror coherent states with $\rho^\text R_{zx}(\vec k)$ (black lines) .
        (b)(c) The overlap between mirror coherent states $\ket{\psi^{\text HF}_{j=1} (\vec k)}$ for the filled band in (a) and the non interacting states $\ket{u^\text{CB1}_{\pm i}}(\vec k)$.
        (d) The Wannier center flow for both eigen-states of $H^\text{HF}(\vec k)$.
        (e) The energy per particle $E_\text{I}[\rho^\text R]$ with non-interacting energy $E_\text{I}[\rho_0]$ subtracted for $\rho^\text R_{y}(\vec k)$ and $\rho^\text R_{zx}(\vec k)$. Here the calculation is for the two-band model with the parameters $\phi=0, \alpha=1, V_0 /E_0 = 0$. %CX: please add the parameters for this figure. 
	}
	\label{fig:HF for CB1 with inversion rhozx}
\end{figure}

For $\rho^\text R_{zx}(\vec k)$, the initial density matrices are taken as 
\begin{equation}
    \rho^\text R_{ini} = \frac{1}{2}(\sigma_0 + \sigma_z \cos 2 \tilde \varphi - \sigma_x \sin 2 \tilde \varphi )
\end{equation}
for a certain uniform value of $\tilde \varphi$, which corresponds to states $\cos \tilde \varphi \ket{u^{\text R,\text{CB1}}_+(\vec k)} - \sin \tilde \varphi  \ket{u^{\text R,\text{CB1}}_-(\vec k)} $.
% The other symmetry breaking states break the $U(1)$ symmetry in \eqnref{eq:U(1)}, denoted as mirror coherent states.
% Take $\cos \tilde \varphi \ket{u^{\text R,\text{CB1}}_+(\vec k)} - \sin \tilde \varphi  \ket{u^{\text R,\text{CB1}}_-(\vec k)} $ for a fixed and uniform $\tilde \varphi$ as an example. \addCXL{CX: what's the value of $\varphi$? Just some random number? } The initial density matrix is
% \begin{equation}
%     \rho_{i} = \frac{1}{2}(\sigma_0 + \sigma_z \cos 2 \tilde \varphi - \sigma_x \sin 2 \tilde \varphi ),
% \end{equation}
% which breaks the $U(1)$ symmetry and preserves the $C_{2z}\TR$ symmetry.
% Because $C_{2z}\TR = \mathcal K$, the density matrix is real.
%The self-consistent solutions is denoted as $\rho_{zx} (\vec k)$ with terms $f_z(\vec k)\sigma_z, f_x(\vec k)\sigma_x$ preserving $C_{2z}\TR$ symmetry with $f_z(\vec k)$,$f_x(\vec k)$ as real functions of $\vec k$.
The HF energy spectrum from this initial $\rho^\text R_{ini}$ in \figref{fig:HF for CB1 with inversion rhozx} shows nodes at $K,K'$.
These nodes can be understood from nonzero Euler number, denoted as $\mu$, a topological invariant defined for a two-band model with the $C_{2z}\TR$ symmetry \cite{bouhon2020non,yu2022euler,ahn2019failure}.
The non-interacting eigen-state of CB1 has non-trivial $\mathbb Z_2$ number $\nu_{\text{CB1}} = 1$, and the Euler number can be related to the $\mathbb Z_2$ number by $\nu_{CB1} = \mu \mod 2$\cite{ahn2019failure}. Thus, when $\nu_{\text{CB1}} = 1$, $\mu$ has to be an odd number, which gives rise to $2\mu$ of gapless Dirac nodes in the spectrum. Because $C_{2z} \TR$ is preserved for the initial density matrix $\rho^\text R_{ini}$, this symmetry remains throughout the whole self-consistent calculation process, so Euler class is still well-defined for the final self-consistent Hatree-Fock ground state. We evaluate the Wannier center flow for the final Hatree-Fock ground state, which is shown in \figref{fig:HF for CB1 with inversion rhozx}(d). 
%\addCXL{CX: is it possible to show the Wannier function flow for HF spectrum? the HF states are a linear combination of non-interacting states, which does not change the Wilson loop.}
The nonzero Euler class with $\mu = 1$ from the Wannier center flow guarantees the existence of $2$ Dirac nodes in the Hartree-Fock spectrum.
\figref{fig:HF for CB1 with inversion rhozx}(b)(c) shows that the Hartree-Fock solutions $\ket{\psi^\text{HF} _{j=1}(\vec k)}$ shown in \figref{fig:HF for CB1 with inversion rhozx}(a) are superposition of two $m_z$ states with the same probability
\begin{equation}
\vert \langle \psi^\text{HF} _{j=1}(\vec k)\vert u^\text{CB1}_{-i}(\vec k) \rangle \vert = 1/\sqrt{2} \quad \vert \langle \psi^\text{HF} _{j=1}(\vec k) \vert u^\text{CB1}_{+i}(\vec k)\rangle \vert = 1/\sqrt{2},
\end{equation}
which are denoted as mirror coherent states. %\addCXL{clarify $j$ issue here.}

% These two initial density matrices exhaust all three Pauli matrices for two bands and are all possible initial uniform density matrices one can take.
The true ground state of the system is obtained by comparing the energies $E_\text{I}[\rho^\text R]$ of two self-consistent density matrices %\addCXL{CX: can we use some different labelling for these two self-consistent ground state? That will make the discussion here clear.  Or shall we just use the density matrix notation to distinguish these two states? } 
in \figref{fig:HF for CB1 with inversion rhozx}(e). Above the critical interaction value around $0.05E_0 \approx 2$ meV, our calculation shows that the mirror polarized state with $\rho^\text R_y(\vec k)$ has lower energies than the non-interacting ground state and the mirror coherent states with $\rho^\text R_{zx}(\vec k)$. 
This is because non-interacting ground state and mirror coherent state have gapless excitations in their spectrum, while the mirror polarized states are fully gapped. Thus, we conclude that the true ground state is a mirror polarized Chern insulator.

\begin{figure}
	\centering
	\includegraphics[width=\columnwidth]{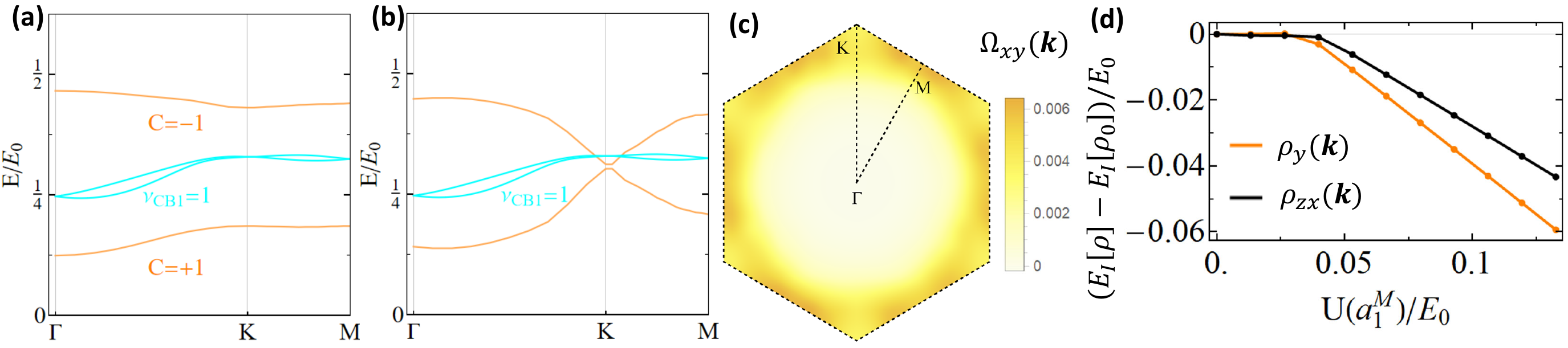}
	\caption{
	    (a)(b) The spectra (orange) of $H^\text{HF}(\vec k)$ with $\rho^\text R_y(\vec k)$ for (a) and $\rho^\text R_{zx}(\vec k)$ for (b). The cyan lines are non-interacting spectrum.
        (c) Berry curvature for the lower band of $H^\text{HF}(\vec k)$ in (a).
        (c) The energy per particle $E_\text{I}[\rho^\text R]$ with non-interacting energy $E_\text{I}[\rho_0]$ subtracted for $\rho^\text R_{y}(\vec k)$ and $\rho^\text R_{zx}(\vec k)$. Here the calculation is for the two-band model with the parameters $\phi=0, \alpha=0, V_0 /E_0 = 1.2$.
	}
	\label{fig:HF for CB1 without inversion}
\end{figure}

For the case with $\alpha = 0, V_0 / E_0 = 1.2, \phi = 0$ without inversion, the mirror symmetry $\mathcal M_z$ is broken so we cannot characterize the non-interacting eigen-state with mirror eigen-values and mirror Chern number. However, the $C_{2z}\mathcal T$ symmetry remains, so we can still choose the real gauge for non-interacting states as $\ket{u^\text{R,CB1}_{1}(\vec k)}, \ket{u^\text{R,CB1}_{2}(\vec k)}$, which satisfies
\begin{equation}
    C_{2z}\TR \ket{u^\text{R,CB1}_{n}(\vec k)} = \ket{u^\text{R,CB1}_{n}(\vec k)},
\end{equation}
where $n=1,2$ labels two spin-split bands for the Kramers' pair of CB1.
Consequently, two types of order parameters, $\rho^\text R_y(\vec k)$ that breaks $C_{2z} \TR$ and $\rho^\text R_{zx}(\vec k)$ that breaks spin $U(1)$ symmetry, do not mix with each other. The self-consistent solutions with two types of order parameters are shown in \figref{fig:HF for CB1 without inversion}. $\rho^\text R_y(\vec k)$ breaks $\TR$ symmetry and one band of CB1 with nonzero Chern number is gapped from the other band.
$\rho^\text R_{zx}(\vec k)$ has Dirac nodes in spectra at $K,K'$ with energies $E_\text I[\rho^\text R_{zx}(\vec k)]$ higher than the other case.
The ground state is an interaction-driven Chern insulator, same as the inversion symmetric case.

% The density matrix of the mirror-polarized state contains the \(\sigma _y\) part.
% The other density matrix \(\rho \propto \sigma _{x,z}\) fixes
% the \(\phi _k\) for each \(k\) and breaks the \(U(1)\) symmetry, which is the $m_z$ coherent state.

\subsection{Four-band model with CB1 and CB2}
In the main text, we have discussed the important role of the band mixing between CB1 and CB2 induced by the Coulomb interaction, which can result in the interacting ground state varying from the QAH state to a trivial insulator state for the realistic Coulomb interaction strength for the inversion symmetric case ($V_0=0$), while the QAH state remains for the realistic Coulomb interaction when a large asymmetric potential $V_0$ is applied. The difference between  the inversion symmetric and asymmetric cases is that both CB1 and CB2 carry non-trivial $\mathbb{Z}_2$ number,  $\nu_{CB1}=\nu_{CB2}=1$, for inversion symmetric case, while a strong asymmetric potential $V_0$ gives a trivial insulator phase for CB2, $\nu_{CB1}=1$ and $\nu_{CB2}=0$, for inversion asymmetric case. This effect can only be taken into account when considering both CB1 and CB2, and thus it is important to go beyond the two-band model discussed above and consider a four-band model with both CB1 and CB2. In this section, we will provide more details of our numerical self-consistent calculations of the interacting ground state within the HF approximations for the four-band model. Below we always assume the $1/4$ filling of four bands, which corresponds to the $1/2$ filling of CB1.  

%As the mini-gap between CB1 and CB2 is of the same order as Coulomb interaction strength, one needs to understand if the inter-band mixing between CB1 and CB2 can change the HF solutions (QAH state) of the two-band model. Thus, in this section, we discuss the self-consistent HF solutions for the four-band model of CB1 and CB2 at the $1/4$ filling of four bands (corresponding to $1/2$ filling for CB1). We will show that a dramatic difference of the interacting ground states occurs between the inversion-symmetric and inversion-asymmetric cases. 

\begin{figure}
	\centering
	\includegraphics[width=0.8\columnwidth]{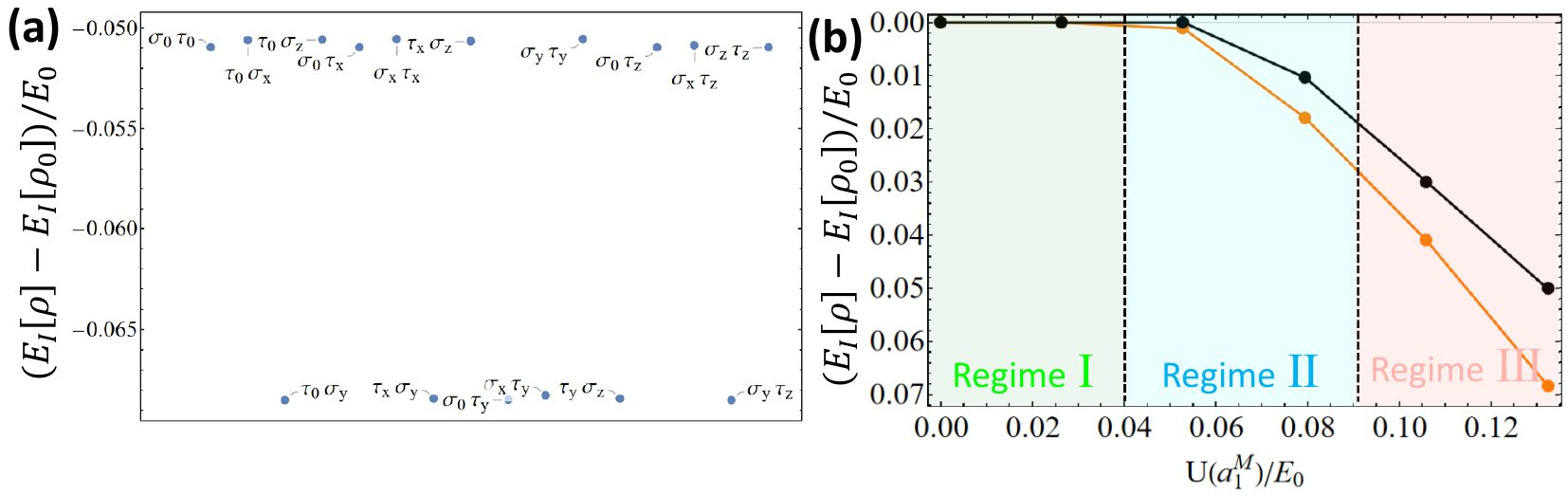}
	\caption{
	    (a) The energy per particle for the self-consistent Hartree-Fock solutions labelled by initial density matrix $\rho_{ini}$ for $U(a_1^\text M) = 0.13 E_0$.
	    (b) The energy per particle for the self-consistent Hartree-Fock solutions under different interaction strength. Orange (Black) lines are $C_{2z}\TR$ symmetry breaking (preserving) states. Here the calculation is for the four-band model with the parameters $\phi=0, \alpha=1, V_0 /E_0 = 0$}. 
	\label{fig:HF energy for CB1 CB2 with inversion}
\end{figure}

\begin{figure}
	\centering
	\includegraphics[width=\columnwidth]{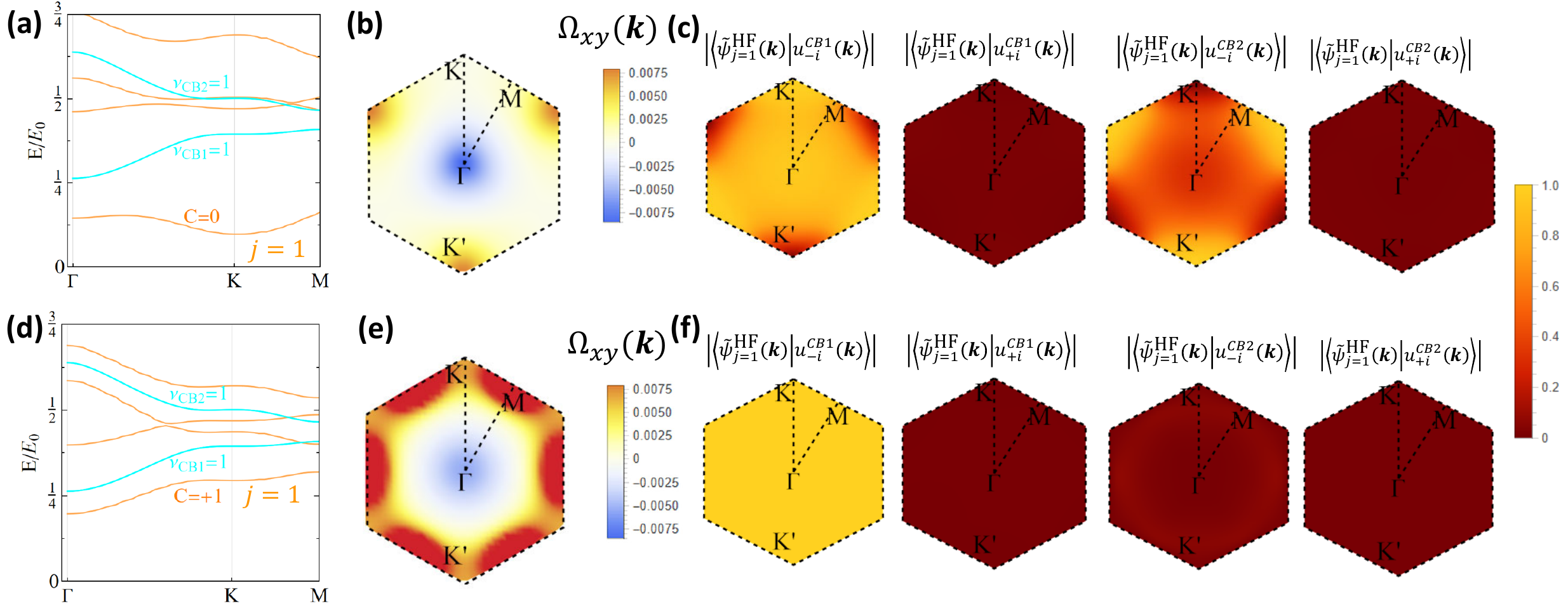}
	\caption{
	    (a) The spectra (orange) of $H^\text{HF}(\vec k)$ with $\rho^\text R_{ini}(\vec k) = \tau_0 \sigma_y$ and $U(a_1^\text M) = 0.13 E_0$. The cyan lines are non-interacting spectrum.
        (b) Berry curvature for the lowest band $j=1$ of $H^\text{HF}(\vec k)$ in (a).
        (c) The overlap between the ground states $\ket{\tilde \psi^\text{HF}_{j=1} (\vec k)}$  in (a) and the non-interacting states $\ket{u^\text{R,CB1/CB2}_{\pm i}(\vec k)}$.
        (d)(e)(f) are same as (a)(b)(c), respectively, for $U(a_1^\text M) = 0.08 E_0$. Here the calculation is for the four-band model with the parameters $\phi=0, \alpha=1, V_0 /E_0 = 0$. 
	}
	\label{fig:HF for CB1 CB2 with inversion}
\end{figure}

For the case $\alpha = 1, V_0/E_0 =0, \phi = 0$ with inversion symmetry, the non-interacting states now have $\ket{u^\text{CB1}_{\pm i}(\vec k)}$ with $C = \pm 1$ and $\ket{u^\text{CB2}_{\pm i}(\vec k)}$ with $C=\mp 1$, where $C$ denotes the Chern number of the mini-bands in the $m_z=-i$ subspace (mirror Chern number).
%$\ket{u^\text{CB1}_{- i}(\vec k)}$ and $\ket{u^\text{CB2}_{- i}(\vec k)}$ have opposite Chern numbers as the single particle Hamiltonian in the $m_z=-i$ subspace can be described by the Haldane model \cite{haldane1988model}. 
%because they are adiabatically connected to one copy of Kane-Mele model in \eqnref{eq:Kane Mele model} and can be described by Haldane model\cite{haldane1988model} with opposite Chern number for two bands.
For the convenience of the calculations, we choose the real gauge by applying the transformation given in \eqnref{eq:real gauge} to the basis wave-functions for both CB1 and CB2, denoted as $\ket{u^\text{ R,CB1}_\pm},\ket{u^\text{ R,CB2}_\pm}$.
The initial density matrices $\rho^\text R_{ini}$ are taken as one of $\tau_i \sigma_j$
with $i,j= 0,x,y,z$ and $\tau$ acting on CB1,CB2 and $\sigma$ acts on two real basis in one Kramer pair of bands, which are all possible $4\times 4$ uniform density matrices.
From $E_I[\rho^\text R(\vec k)]$ for different $\rho^\text R_{ini}$ in \figref{fig:HF energy for CB1 CB2 with inversion}, the self-consistent solutions can also be divided into two groups: one with complex density matrices breaking $C_{2z}\TR$ (e.g. $\tau_i\sigma_y$ and $\tau_y\sigma_i$ with $i=0,x,z$)
%$\tau_0\sigma_y, \tau_x \sigma_y, \tau_y\sigma_0, \tau_y \sigma_x,  \tau_y \sigma_z, \tau_z \sigma_y$)
and the other with real density matrices preserving breaking $C_{2z}\TR$ (e.g. $\tau_i\sigma_j$ with $i,j=0,x,z$ and $\tau_y\sigma_y$), as the $C_{2z}\TR$ symmetry is preserved at the single-particle Hamiltonian level for $\phi=0$. 
%$\tau_0\sigma_0, \tau_0\sigma_x. \tau_0 \sigma_z, \tau_x \sigma_0, \tau_x\sigma_x, \tau_x \sigma_z, \tau_y\sigma_y, \tau_z\sigma_0, \tau_z\sigma_x, \tau_z\sigma_z$).
We generally find that the self-consistent solutions with the initial complex density matrices have lower energies, as shown in \figref{fig:HF energy for CB1 CB2 with inversion}(a). Although the initial density matrices $\rho^\text R_{ini}$ are different, we numerically find the self-consistent density matrices are all mirror polarized states related by $C_{2z}$ or $\TR$.
%\addCXL{Is this easy to see? }.
From Fig.4(a) in the main text or \figref{fig:HF energy for CB1 CB2 with inversion}(b) reproduced here,  %\addCXL{Is this the same as that in the main text? }
the self-consistent solutions with complex density matrices becomes the ground states when the Coulomb interaction exceeds $0.04 E_0$. The inter-band mixing between CB1 and CB2 is negligible for interaction strength in regime \MakeUppercase{\romannumeral 2} with $0.04 E_0 < U(a^M_1) < 0.09E_0$ but a strong band mixing is found in regime \MakeUppercase{\romannumeral 3} with larger Coulomb interaction $0.09 E_0 < U(a^M_1)$.
\figref{fig:HF for CB1 CB2 with inversion} shows the self-consistent solutions for the initial density matrix $\rho^\text R_{ini} = \tau_0 \sigma_y$ as an example. Here \figref{fig:HF for CB1 CB2 with inversion}(a)-(c) are for $U(a_1^\text M) = 0.13 E_0$ (regime III) while (d)-(f) are for $U(a_1^\text M) = 0.08 E_0$ (regime II). 
% \addCXL{you did not discuss Fig. S10, b.  }
% \addCXL{KJ: These complex density matrices are not the same. I am not sure whether they are different density matrices or they have not converged over iteration. For smaller coulomb interaction, the complex density matrices are all the same.}
The filled band $\ket{\tilde \psi^\text{HF}_{j=1} (\vec k)}$ in \figref{fig:HF for CB1 CB2 with inversion}(a) for regime \MakeUppercase{\romannumeral 3} has the Chern number $C=0$ and that in \figref{fig:HF for CB1 CB2 with inversion}(d) for regime \MakeUppercase{\romannumeral 2} has $C=+1$. \figref{fig:HF for CB1 CB2 with inversion}(b) and (e) show the distribution of Berry curvature $\Omega_{xy}$ in the moir\'e BZ for regime \MakeUppercase{\romannumeral 3} and regime \MakeUppercase{\romannumeral 2}, respectively. 
\figref{fig:HF for CB1 CB2 with inversion}(c) and (f) show the projection of $\ket{\tilde \psi^\text{HF}_{j=1} (\vec k)}$ into non-interacting states $\ket{u^\text{CB1}_{\pm i}(\vec k)}$ and $\ket{u^\text{CB2}_{\pm i}(\vec k)}$ in regime \MakeUppercase{\romannumeral 3} and regime \MakeUppercase{\romannumeral 2}, respectively. One can see that the interacting ground state $\ket{\tilde \psi^\text{HF}_{j=1}(\vec k)}$ has a strong component from $\ket{u^\text{CB2}_{\pm i}(\vec k)}$, in addition to $\ket{u^\text{CB1}_{\pm i}(\vec k)}$, due to the strong band mixing in regime \MakeUppercase{\romannumeral 3}, while only the $\ket{u^\text{CB1}_{\pm i}(\vec k)}$ part dominates the interacting ground state in regime \MakeUppercase{\romannumeral 2}. Thus, from \figref{fig:HF for CB1 CB2 with inversion}, we show that the Coulomb interaction can drive the interacting ground state into a trivial Mott insulator phase \cite{sorella1992semi,neto2009electronic} via band mixing between CB1 and CB2 when there is inversion symmetry.

%the ground state $\ket{\tilde \psi^\text{HF}_{j=1}(\vec k)}$ is mainly $\ket{u^\text{CB1}_{- i}(\vec k)}$ with mixture of $\ket{u^\text{CB2}_{- i}(\vec k)}$ around $K'$.
% A band inversion happens at $K'$ and changes the Chern number of the ground state to $C=0$ like the Haldane model. \addCXL{What do you mean by Haldane model? }
%With smaller $U(a_1^\text M) = 0.08 E_0$ for regime \MakeUppercase{\romannumeral 2}, the filled band $\ket{\tilde \psi^\text{HF}_{j=1} (\vec k)}$ has $C= + 1$ as shown in \figref{fig:HF for CB1 CB2 with inversion}(d).
%The ground states $\ket{\tilde \psi^\text{HF}_{j=1}(\vec k)}$ are $\ket{u^\text{CB1}_{-i}}(\vec k)$ as shown in \figref{fig:HF for CB1 CB2 with inversion}(f), same as two-band model.

%Thus, a topological phase transition happens for the ground states when increasing the Coulomb interaction strength through inter band mixture. The topologically trivial phase can be understood in the large Coulomb interaction limit. With a large Coulomb interaction, the system can be described as a Hubbard model on a honeycomb lattice\cite{sorella1992semi,neto2009electronic}.The ground states are antiferromagnetic with $C=0$.
%\addCXL{KJ: One may need to think about an explanation for this transition. It may come from the competition between $H_0(\vec k)$ and $H^\text{F}[\rho]$.}
%\addCXL{CX: add my argument in the email. }

\begin{figure}
	\centering
	\includegraphics[width=\columnwidth]{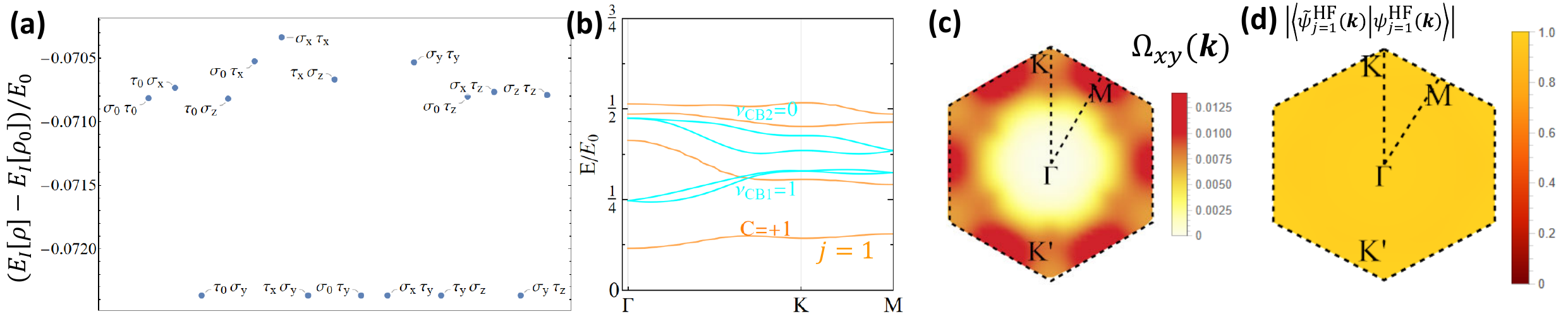}
	\caption{
	    (a) The energies of the self-consistent Hartree Fock states with different initial density matrices $\rho^\text R _{ini} (\vec k)$.
	    (b) The spectra (orange) of $H^\text{HF}(\vec k)$ with $\rho^\text R_{ini}(\vec k) = \sigma_y \tau_0$ and $U(a_1^\text M) = 0.13 E_0$. The cyan lines are non-interacting spectrum.
        (c) Berry curvature for the lowest band of $H^\text{HF}(\vec k)$ in (b).
        (d) The overlap between the ground states $\ket{\tilde \psi^\text{HF}_{j=1} (\vec k)}$ for four-band Hartree-Fock calculations and the ground states $\ket{\psi^\text{HF}_{j=1} (\vec k)}$ for two-band Hartree-Fock calculations. Here the calculation is for the four-band model with the parameters $\phi=0, \alpha=0, V_0 /E_0 = 1.2$.
	}
	\label{fig:HF for CB1 CB2 without inversion}
\end{figure}

For the case $\alpha = 0, V_0/E_0 = 1.2, \phi = 0$ without inversion symmetry, the self-consistent solutions are summarized in \figref{fig:HF for CB1 CB2 without inversion}. As shown in the phase diagram of Fig. 4(b) in the main text, the system stays in the QAH state with $C=+1$ for the realistic Coulomb interaction strength $U(a_1^\text M) = U_0 \approx 0.13 E_0$, which is quite different from the inversion symmetric case. Here we show more details of this calculation in \figref{fig:HF for CB1 CB2 without inversion} for $U(a_1^\text M) = 0.13 E_0$. \figref{fig:HF for CB1 CB2 without inversion}(a) shows that the self-consistent solutions with the initial complex density matrices that break the $C_{2z}\TR$ still have lower energy. We consider $\sigma_0 \tau_y$ as an example and show the energy dispersion of HF bands in \figref{fig:HF for CB1 CB2 without inversion}(b). The distribution of the Berry curvature $\Omega_{xy}$ in the moir\'e BZ is shown in \figref{fig:HF for CB1 CB2 without inversion}(c). We further project the interacting ground state of the four-band model, denoted as $\ket{\tilde \psi^\text{HF}_{j=1}(\vec k)}$, into that of the two-band model, denoted as $\ket{\psi^\text{HF}_{j=1}(\vec k)}$, and find their overlap is almost 1 in the whole moir\'e BZ, as shown in \figref{fig:HF for CB1 CB2 without inversion}(d). Thus, the inter-band mixing is negligible in the inversion asymmetric case for $U(a_1^\text M) = U_0 \approx 0.13 E_0$.

%the non-interacting states are $\ket{u^\text{R,CB1}_{n}(\vec k)}$ and $\ket{u^\text{R,CB2}_{n}(\vec k)}$ with $n=1,2$.
%The density matrices are also grouped into two with the ground states breaking the $C_{2z}\TR$ as shown in \figref{fig:HF for CB1 CB2 without inversion}(a).
%Take $\sigma_0 \tau_y$ as an example.
%The ground states is gapped from other higher energy bands with $C=+1$ for $U(a_1^\text M) = 0.13 E_0$ as shown in \figref{fig:HF for CB1 CB2 without inversion}(b)(c).
%It is exactly the ground state from the two-band Hartree-Fock calculations as shown in \figref{fig:HF for CB1 CB2 without inversion}(d).
% Thus, the QAH ground states are more robust under Coulomb interaction than the previous case.

\begin{figure}
	\centering
	\includegraphics[width=0.6\columnwidth]{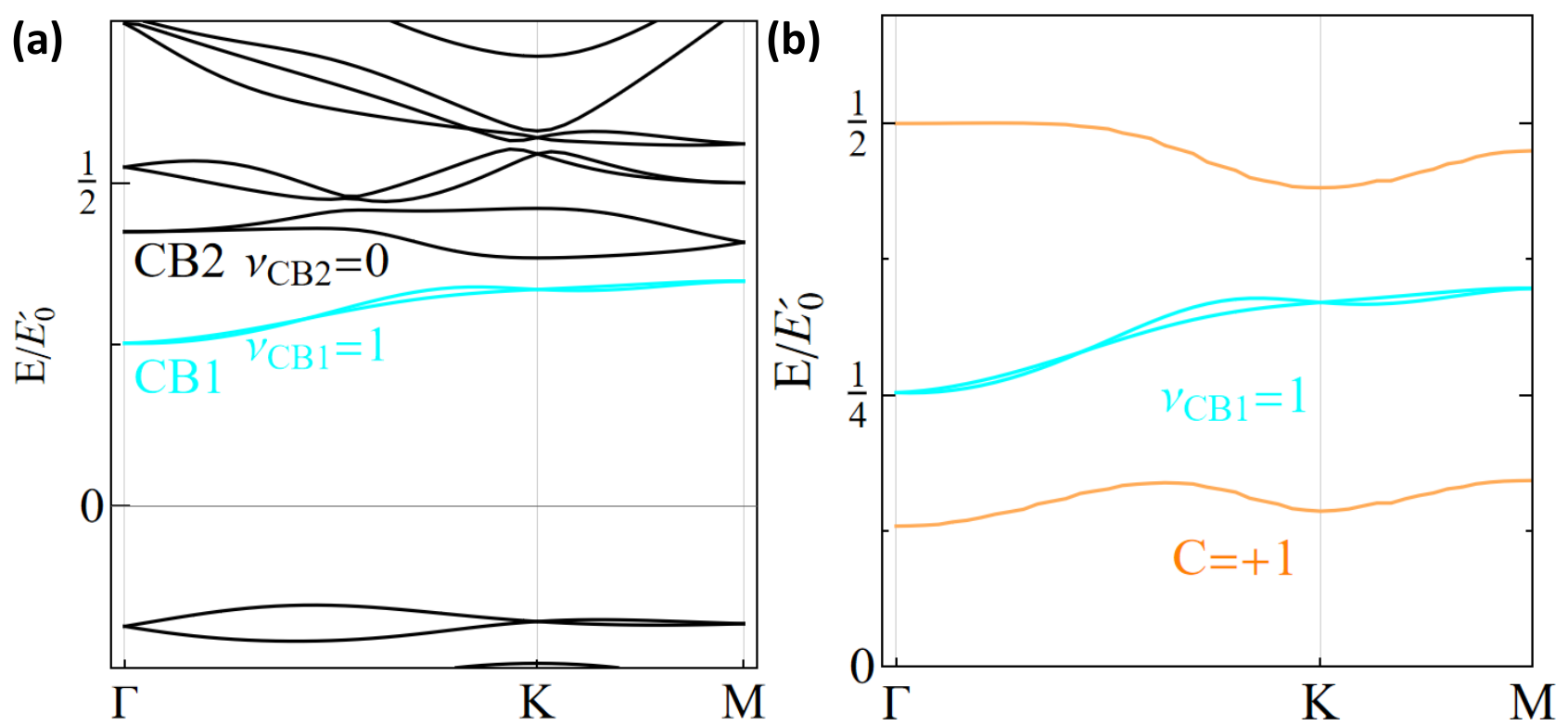}
	\caption{
	    (a) Spectrum for the moir\'e system for smaller moir\'e unit cells.
        (b) The spectra (orange) of $H^\text{HF}$ with $\rho^\text R_y(\vec k)$ for mirror polarized states.  Here the calculation is for the two-band model with the parameters $\phi=1/3, \alpha=0.16, V_0 /E_0' = 1.1$.
	}
	\label{fig:HF for CB1 of small moire unit cell}
\end{figure}

%\addKJ{
\subsection{Coulomb interaction for smaller moir\'e unit cells}

%\addCXL{Update the calculations for $\alpha=0$ here. }

In this section, we discuss the moir\'e systems with a smaller moir\'e lattice constant, $\vert \vec a_1^\text M\vert = 14$nm, for a twist angle $\theta = 1.0^\circ$. The Coulomb interaction scales inversely with the moir\'e unit cell length and its strength can be estimated as $U(\vec a_1^\text M) = 10$ meV.

The spectrum is shown in \figref{fig:HF for CB1 of small moire unit cell}(a). The parameters for the spectrum are $\Delta_1=14$meV, $\phi=1/3$, and $m_0 = 30$meV.
%, fitted from DFT as in Sec.\ref{Sec:DFT}.
The energy scale is $E_0' = v \vert \vec b_1^\text M\vert = 77$meV. 
The bandwidth of CB1 is $7.4$meV and the direct gap between CB1 and CB2 is $4$meV. 
The ratio between $U(\vec a_1^\text M)$ and bandwidth is smaller for the smaller $\vert \vec a_1^\text M\vert$.
When the Coulomb interaction is considered for CB1 with the density matrix $\rho^\text R_y(\vec k)$ as shown in  \figref{fig:HF for CB1 of small moire unit cell}(b), the mirror polarized states with $C=+1$ can be induced. 
%}

\section{Computational Methods for DFT calculations and moir\'e lattice}
\label{Sec:DFT}
The DFT calculations were performed with the Vienna \text{Ab initio} Simulation Package (VASP)$^{\cite{kresse1996efficient}}$. The exchange-correlation functional was chosen as the Perdew-Burke-Ernzerhof type generalized-gradient approximation$^{\cite{perdew1996generalized}}$ and the projector-augmented-wave method was used for the core-electron potentials$^{\cite{blochl1994projector,kresse1999ultrasoft}}$. The energy cutoff was set as 340 eV for all calculations. The convergence criterion was set as 10$^{-5}$ eV for self-consistent electronic calculations and the $k$-point meshes were set as $13 \times 13 \times 1$ to sample the Brillouin zone. We used the DFT-D3 method$^{\cite{grimme2010consistent}}$ to correctly describe the van der Waals interactions.

%As shown in \eqnref{eq:Morie potential} in the main text, the moir\'e potential $\Delta(\vec r)$ in such twisted Sb$_2$/Sb$_2$Te$_3$ heterostructures could be obtained via the Fourier transform of $\Delta_{\vec G}$ in each specific stacking. In this section and the next section, therefore, we calculated the band structures for the heterostructure models with different stacking and fitted $\Delta_{\vec G}$ correspondingly.

\begin{figure}
	\centering
	\includegraphics[width=0.8\columnwidth]{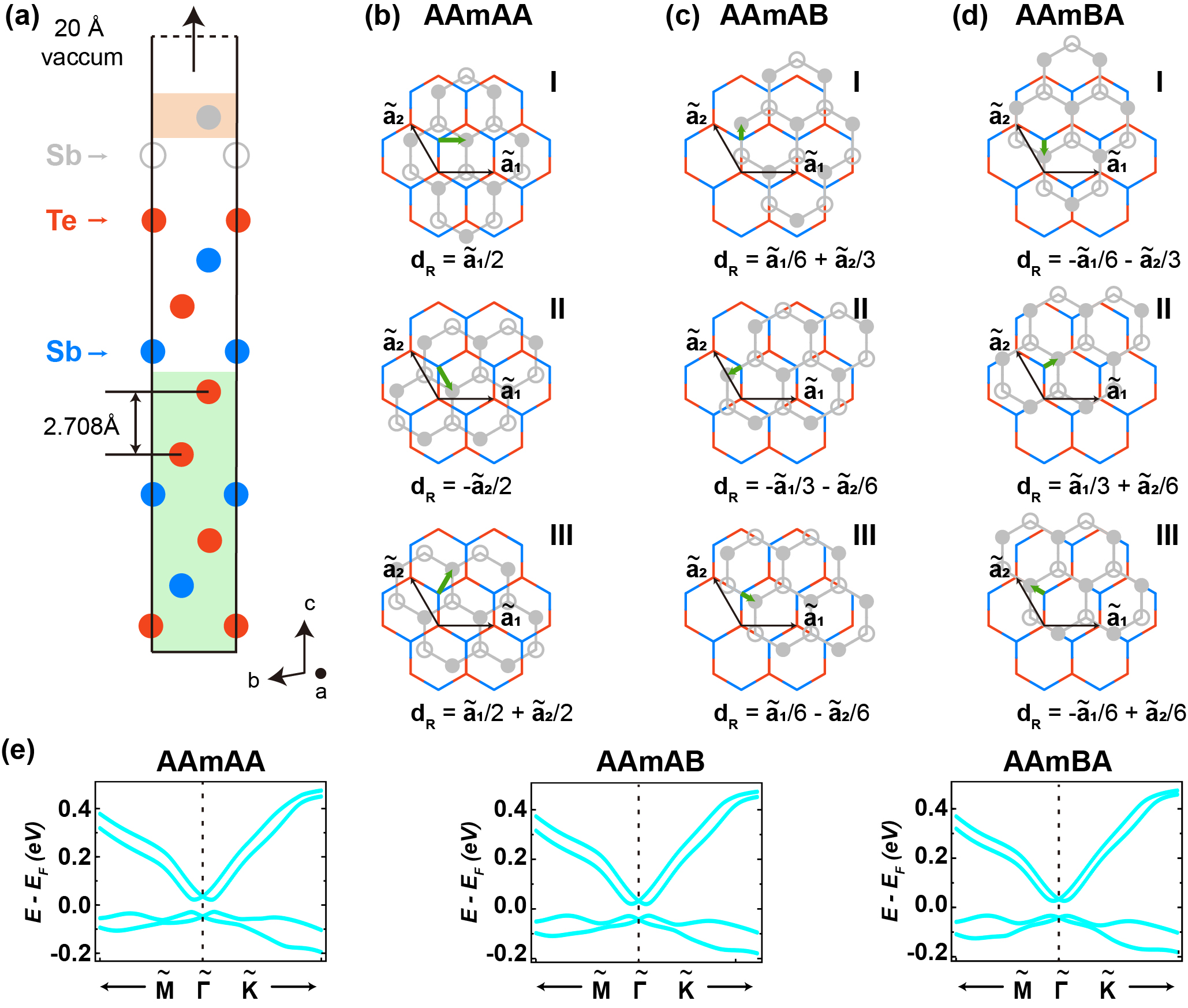}
	\caption{ %\addCXL{CX: there is no (b) in this figure? }
 (a) The side view of the Sb$_2$/Sb$_2$Te$_3$ heterostructure with AA stacking. The solid black lines mark the unitcell used in DFT calculations. Sb atoms in Sb$_2$ monolayer are marked as gray, Sb atoms and Te atoms in Sb$_2$Te$_3$ films are marked as red and blue. The atoms in the region with the green background are frozen when we relax the lattice structures, the $x$ and $y$ coordinates of Sb atoms in the region with the yellow background are fixed. The value of the van der Waals gap inside 2QL Sb$_2$Te$_3$ films is 2.078\AA. (b-d) The top view for heterostructures with the stacking of AAmAA-X, AAmAB-X, and AAmBA-X (X = I, II, III). Corresponding $\vec d_R$ is shown by the green arrows. The black arrows show the lattice vector of Sb$_2$ monolayer. \addKJ{(e)} Calculated band structures of the heterostructure with the stacking of AAmAA, AAmAB, and AAmBA, respectively. The Fermi levels are set as zero.}
	\label{fig:DFT-SI}
\end{figure}

As discussed in the main text and the next section \ref{sec:Moire_DFT}, the moir\'e potential $\Delta(\vec r)$ in the twisted Sb$_2$/Sb$_2$Te$_3$ hetero-structures could be obtained from the uniform potential $\tilde{\Delta}(\vec d_{\vec R})$ for the hetero-structure with a uniform shift $\vec d_{\vec R}$ between the Sb$_2$ layer and Sb$_2$Te$_3$ layer. Therefore, we will describe below our first principles calculations of the band structures with different stacking configurations that correspond to different shifting vectors $\vec d_{\vec R}$. As shown in \figref{fig:DFT-SI}(a), the heterostructure model with the specific stacking contains $1\times1$ Sb$_2$ monolayer and $1\times1$ 2QL Sb$_2$Te$_3$ thin film. A vacuum layer with 20 \AA\ was added along the \textit{z} direction to avoid the interaction between adjacent slabs. Because the uniform potential $\tilde{\Delta}(\vec d_{\vec R})$ is 
%we only care about $\Delta_{\vec G}$ 
induced by coupling between Sb$_2$ monolayer and Sb$_2$Te$_3$ thin films, we fixed the value of the van der Waals gap inside 2QL Sb$_2$Te$_3$ thin film as the bulk value (2.708 \AA), and then let Sb$_2$ monolayer and its neighboring atoms fully relax until the calculated forces are smaller than 0.01 eV/\AA. In \figref{fig:dft} of the main text, we plot the relaxed lattices and corresponding band structures for heterostructures with AA, AB, and BA stacking, whose interlayer distances between Sb$_2$ monolayer and 2QL Sb$_2$Te$_3$ thin film are 3.92 \AA, 2.81 \AA, and 2.91 \AA~respectively.

%we built the heterostructure models with AA, AB and BA stacking in a supercell (see \figref{fig:dft}(a)) which contains $1\times1$ Sb$_2$ monolayer and $1\times1$ 2QL Sb$_2$Te$_3$ thin film. In the relaxation of the crystal structure of the heterostructures, we fixed atoms in the bottom surface of 2QL Sb$_2$Te$_3$ thin film, which are marked by the black dashed lines in \figref{fig:DFT-SI}(a), and kept the interlayer distance between QLs in Sb$_2$Te$_3$ thin film as the bulk value of 2.708 \AA. Then we let others atoms in the heterostructure fully relax until forces were smaller than 0.01 eV/\AA. In these relaxed structures, the interlayer distances between Sb$_2$ monolayer and 2QL Sb$_2$Te$_3$ thin film are 3.92 \AA, 2.81 \AA, and 2.91 \AA~for AA, AB, and BA stacking, respectively.

Besides the stacking configurations shown in the main text, other stacked heterostructures can be also found in the moir\'e pattern (see \figref{fig:dft} in the main text). In order to calculate moir\'e potential $\Delta(\vec r)$ accurately (see Section \ref{sec:Moire_DFT} below), we consider extra nine stacked configurations, as shown in \figref{fig:DFT-SI}(b-d), which are located in the intermediate regions among AA, AB, and BA stackings, named AAmAA-X (\figref{fig:DFT-SI}(b)), AAmAB-X (\figref{fig:DFT-SI}(c)), and AAmBA-X (\figref{fig:DFT-SI}(d)) with X = I, II, III. We take the heterostructure models with AAmAA stackings as an example to show their lattices in details and the other intermediate stacked configurations could be obtained by using the same method. There are three types of AAmAA stacked structures in the moir\'e pattern, corresponding to the shifting vector $\vec d_R$ as $ \Tilde{\vec a}_1 /2+ 0 \Tilde{\vec a}_2$, $0 \Tilde{\vec a}_1 - \Tilde{\vec a}_2 /2$, and $\Tilde{\vec a}_1 /2+  \Tilde{\vec a}_2 /2$ for AAmAA-I, AAmAA-II, and AAmAA-III respectively. %\addCXL{CX: the shifting vector $\vec d_R$ seems not coincide with the green arrows in the figure. Or do I misunderstanding anything here? PZT: We correct these errors}. 
These three configurations are related by C$_{3z}$ rotation, and thus we only need to calculate the electronic structure for one of them. On the other hand, because these intermediate stacked structures are not in the local minimum of the potential energy surface, we only relax the $z$ direction coordinate while fix the $x$ and $y$ coordinates of the Sb$_2$ monolayer (see \figref{fig:DFT-SI}(a)). The corrugation effect, which is crucial for predicting the correct band structure in twisted bilayer graphene\cite{lucignano2019crucial,uchida2014atomic,koshino2018maximally}, is taken into account after this lattice relaxation process.
%thus if we fully relax these lattices, specific stacked configurations that we focus on will be destroyed. In order to overcome this issue, besides the setting in the relaxation process mentioned for AA stacked configuration, we also fixed $x$ and $y$ coordinates of Sb atoms in Sb$_2$ monolayer and only let them relax along the $z$ direction (see \figref{fig:DFT-SI}(a)). %\addCXL{CX: did people do the same thing in the twisted bilayer graphene? If so, we may mention that and cite the references. PZT: No, we have no need to do this thing for TBG, but for thick film, we need to fix the bulk value.}. 
Using the same method, we obtained the relaxed intermediate stacked heterostructures for AAmAB-X and AAmBA-X (X = I, II, III). The interlayer distances between Sb$_2$ monolayer and 2QL Sb$_2$Te$_3$ thin film are 2.94 \AA, 3.38 \AA, and 3.42 \AA\ for AAmAA, AAmAB, and AAmBA stacked heterostructures respectively and the related band structures are shown in \figref{fig:DFT-SI}(e) correspondingly.

\begin{figure}
	\centering
	\includegraphics[width=0.8\columnwidth]{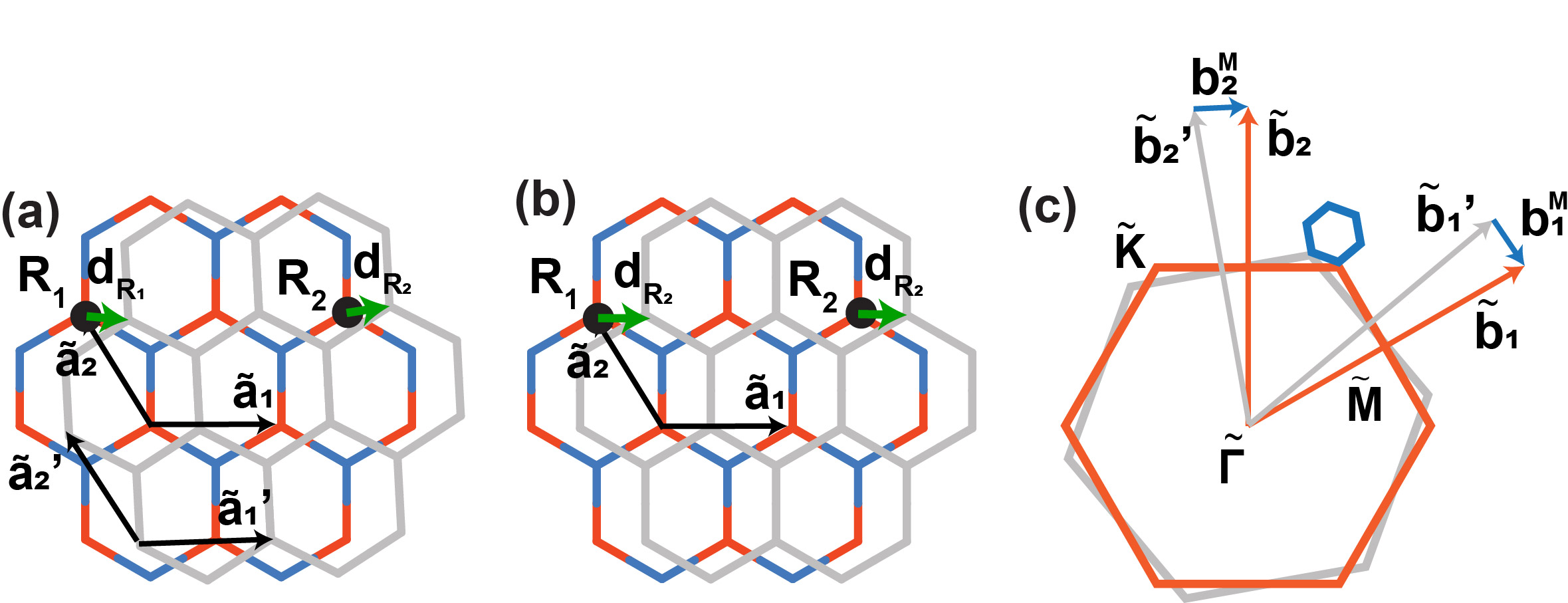}
	\caption{
	    (a) The lattice structures for the twisted Sb$_2$ (gray) on top of Sb$_2$Te$_3$ (blue and orange) at a location same as Fig.4(c) of the main text. $\tilde{\vec a}_{1,2}$ are primitive lattice vectors for the Sb$_2$Te$_3$ layer and $\tilde{\vec a}_{1,2}'$ are primitive lattice vectors for Sb$_2$ layer. $\vec d_{\vec R_{1,2}}$ are the local lattice shifts at sites $\vec R_{1,2}$.
	    (b) Commensurate Sb$_2$ (gray) on top of Sb$_2$Te$_3$ with a constant shift $\vec d_{\vec R_2}$ as an approximation for the local lattice structures in (a).
	    (c) BZs for Sb$_2$Te$_3$ (orange), Sb$_2$ (gray), and moir\'e superlattice (blue). $\tilde{\vec b}_{1,2}$, $\tilde{\vec b}_{1,2}'$, $\tilde{\vec b}^\text M_{1,2}$ are reciprocal lattice vectors for for Sb$_2$Te$_3$, Sb$_2$, moir\'e superlattice, respectively.
	    }
	\label{fig:DFT lattice structures}
\end{figure}

\section{Moir\'e potentials from the fitting to the DFT band structure}\label{sec:Moire_DFT}
%\addCXL{CX: we should ask Peizhe and his student to help with this part. }
%\addKJ{This section is rewritten.}
In this section, we discuss the method to obtain Moire potential $\Delta(\vec r)$ in Eq. 1 of the main text from the above DFT calculation in Sec. \ref{Sec:DFT} \cite{jung2014ab}. 

% \addCXL{CX: as we discuss the DFT calculation in the above, it is better to discuss $H^{DFT}$ from the DFT band structure first, and then discuss how to get $H_0$ from $H_{DFT}$. The current logic is reserved. }

We first consider the effective Hamiltonian $H^\text{DFT}$ in Eq. (\ref{eq:HDFT}) of the main text. Compared to the original Hamiltonian $H_0$ in Eq. \ref{eq:Hamiltonian}, the spatially dependent moir\'e potential term $\Delta(\vec r)$ is changed to a uniform potential term $\tilde \Delta(\vec d_R)$ for a fixed $\vec d_R$ that describes the relative shift between Sb$_2$ and Sb$_2$Te$_3$ layers. 
%for surface states of Sb$_2$Te$_3$ films with its dispersion fitted to the DFT-calculated spectra. 
%\addKJ{In addition of the original Hamiltonian, we obtain the moir\'e potential $\Delta(\vec r)$ by a potential $\tilde \Delta(\vec d_R)$ for a constant shift $\vec d_R$ between $Sb_2$ and Sb$_2$Te$_3$. 
$H^\text{DFT}$ describes the effective model for the hetero-structure with a uniform shift between two atomic layers, and
different values of the shifting vector $\vec d_R$ describe different stacking configurations. Thus, we can use the energy dispersion of $H^\text{DFT}$ to fit to that from the DFT calculations. 

%\addCXL{CX: Is this Hamiltonian already given in the main text? Or any difference. We should not repeat the definition. }
The effective Hamiltonian $\hat H^\text{DFT}(\vec d_R)$ is then given by
\begin{equation}\label{eq:DFT ham}
    \bra {\vec k_1,\beta_1}\hat H^\text{DFT}(\vec d_R)\ket{\vec k_2,\beta_2} = \delta(\vec k_1 - \vec k_2)
    (\begin{pmatrix}
        h_D^t(\vec k_1) & m s_0 \\
        m s_0 & h_D^b(\vec k_1)
    \end{pmatrix} + 
    \begin{pmatrix}
        \tilde \Delta(\vec d_R)s_0 & 0 \\
        0 & \alpha \tilde \Delta(\vec d_R) s_0
    \end{pmatrix}) =  \delta(\vec k_1 - \vec k_2) H^\text{DFT}(\vec k_1,\vec d_\vec R),
\end{equation}
where $H^\text{DFT}(\vec k,\vec d_\vec R)$ is just Eq.(4) in the main text, 
%\addCXL{We use $\alpha$ here and also use $\alpha_{1,2}$ as the indices. That's confusing. Can we change $\alpha_{1,2}$ to other labelling and define it? }
$h_D^{t/b}(\vec k)$ are the top/bottom Dirac surface states same as Eq. (1) of the main text, $s_0$ are the identical matrix in spin space, $m$ is the tunnelling between two surfaces, and $\alpha$ captures the difference in the potentials on two surfaces created by the Sb$_2$ layer. $\ket{\vec k, \beta}$ is the atomic Bloch states for the Sb$_2$Te$_3$ and Sb$_2$ lattice with a constant shift.
$\beta_{1,2}=1,...,4$ represents both the spin and layer degrees of freedom.
The spectra of this model are given by
\begin{equation}
    E^\text{DFT}_{\eta, \xi}(\vec k,\vec d) = \frac{1+\alpha}{2} \tilde \Delta(\vec d) + \eta \sqrt{m^2  + \left(\frac{1-\alpha}{2} \tilde \Delta(\vec d) +\xi v^2 k^2)\right)^2}.
    \label{eq:spectrum fitted to DFT}
\end{equation}
with $\eta=\pm, \xi=\pm$. 
By fitting $E^\text{DFT}_{\eta, \xi}(\vec k,\vec d)$ to the spectrum calculated from DFT in Fig.4(b) of the main text and \figref{fig:DFT-SI}(f), 
the parameters $\alpha$, $\tilde \Delta(\vec d)$, $v$ and $m$ in the model Hamiltonian can be obtained. 
%$\frac{1-\alpha}{2} \tilde \Delta(\vec d), v, m$ can be obtained.
% $\frac{1+\alpha}{2} \tilde \Delta(\vec d)$ as the chemical potential shift can be obtained from fitting to \addCXL{XXX}. \addCXL{Complete the above sentence. }\addKJ{XXX is the energy difference between CB1 at Gamma and the vacuum level which need some figures for illustration in Fig.S14 from DFT.}
%$\tilde \Delta(\vec d)$ can be obtained as shown in Fig.4(c) of the main text and $\alpha$ is also obtained.
From the three-fold rotation symmetry $C_3$ of the underlying lattice, one can obtain
\begin{equation}
    \bra {\vec k,\beta_1}\hat H^\text{DFT}(\vec d)\ket{\vec k,\beta_2} = \bra {C_3 \vec k,\beta_1}\hat H^\text{DFT}(C_3 \vec d)\ket{C_3 \vec k,\beta_2}
\end{equation}
and 
\begin{equation}
    E^\text{DFT}_{\eta,\xi}(\vec k, \vec d) = E^\text{DFT}_{\eta,\xi}(C_3 \vec k, C_3\vec d) = E^\text{DFT}_{\eta,\xi}(\vec k, C_3\vec d), 
\end{equation}
where the last line following from $E^\text{DFT}_{\eta,\xi}$ depends on $\vert \vec k\vert$ in \eqnref{eq:spectrum fitted to DFT}.
So, only one of three stacking related by $C_3$ shown in \figref{fig:DFT-SI}(b)-(d) needs to be fitted.

Next we will establish the relation between the moir\'e Hamiltonian $H_0$ in Eq.1 of the main text and the Hamiltonian $H^\text{DFT}$ 
%\addCXL{CX: what do you mean by $H^{DFT}$ here? we have not used it in the above, so it is confusing. DFT calculation only gives the band structure, not the Hamiltonian. } 
constructed from the fitting to the DFT energy spectrum.
We start from the momentum-space moir\'e Hamiltonian $\bra{\vec k_1, \beta_1} \hat H_0 \ket{\vec k_2, \beta_2}$ in Eq.1 of the main text, where $\beta_{1,2} = 1,\cdots, 4$ label both layer and spin indices and $\ket{\vec k,\beta}$ are atomic Bloch states for each layer underlying the moir\'e superlattice. As the moir\'e Hamiltonian $H_0$ does not preserve atomic lattice translation, the crystal momentum $\vec k$ is not a good quantum number and $H_0$ can mix different $\vec k$ states. The atomic Bloch wave function $\ket{\vec k,\beta}$ is related to atomic Wannier function $\ket{\vec R, \beta}$ by 
\begin{eqnarray}
    \ket{\vec k,\beta}=\sum_{\vec R} e^{i \vec k \cdot \vec R} \ket{\vec R, \beta},
\end{eqnarray}
so the moir\'e Hamiltonian is transformed into the form on the atomic Wannier function basis as
%Transforming this Hamiltonian into the atomic Wannier basis $\ket{\vec R, \beta}$ is
\begin{equation}\label{eq:DFT Wannier basis}
	\bra{\vec k_1, \beta_1} \hat H_0 \ket{\vec k_2, \beta_2} = \sum_{\vec R_1 \vec R_2} e^{- i \vec k_1 \cdot \vec R_1} \bra{\vec R_1, \beta_1} \hat H_0 \ket{\vec R_2 ,\beta_2} e^{i \vec k_2 \cdot \vec R_2}.
\end{equation}
%$\vec R$ is the atomic lattice vector on the accordingly layer for the atomic Wannier basis $\ket{\vec R, \beta}$.
Here $\bra{\vec R_1 ,\beta_1} \hat H_0 \ket{\vec R_2, \beta_2}$ describes the Hamiltonian matrix element between atomic Wannier functions located at $\vec R_{1}$ and $\vec R_{2}$ in the superlattice shown in \figref{fig:DFT lattice structures}(a).
As the overlap between atomic Wannier functions decays quickly as the distance increases, we only consider the local hopping within the length scale $\vert \vec R_2 - \vec R_1 \vert \sim \mathcal O(\vert \tilde{\vec a}_1 \vert)$,
%When $\vec R_{1,2}$ are close by $\vert \vec R_2 - \vec R_1 \vert \sim \mathcal O(\vert \tilde{\vec a}_1 \vert)$, the hopping are significant, 
where $\tilde{\vec a}_1$ is the atomic primitive lattice vector for the Sb$_2$Te$_3$ layer.
In this atomic length scale, the Hamiltonian matrix element between two Wannier orbitals near $\vec R$ on the superlattice structure with twist angle $\theta$ in \figref{fig:DFT lattice structures}(a) can be approximated locally by the Hamiltonian matrix element for two atomic layers with a constant shift $\vec d_\vec R$ in \figref{fig:DFT lattice structures}(b)\cite{jung2014ab}, where 
\begin{equation}
    \vec d_\vec R = \mathcal R(\theta) \vec R - \vec R
\end{equation}
and $\mathcal R(\theta)$ as the rotation operator for the Sb$_2$ layer with the rotating angle $\theta$.
This approximation is valid for a small twist angle $\theta$ because the local shift vector $\vec d_\vec R$ is almost uniform at the atomic length scale, 
%that is much smaller than moir\'e super-cell by
\begin{equation}
    \vec d_{\vec R_2} \approx \vec d_{\vec R_1}
\end{equation}
for $\vert \vec R_2 - \vec R_1 \vert \sim \mathcal O(\vert \tilde{\vec a}_1 \vert)$.
The Hamiltonian matrix element between two atomic Wannier orbitals for the commensurate lattice is captured by $\bra{\vec R_1 ,\beta_1}\hat H^\text{DFT}(\vec d_{\vec R_2}) \ket{\vec R_2, \beta_2}$, so we make the approximation 
\begin{equation}\label{eq:DFT approximation}
    \bra{\vec R_1, \beta_1} \hat H_0 \ket{\vec R_2 ,\beta_2} \approx \bra{\vec R_1 ,\beta_1} \hat H^\text{DFT}(\vec d_{\vec R_2}) \ket{\vec R_2, \beta_2}
\end{equation}
and 
\begin{equation}
	\bra{\vec k_1, \beta_1} \hat H_0 \ket{\vec k_2, \beta_2} = \sum_{\vec R_1 \vec R_2} e^{- i \vec k_1 \cdot \vec R_1} \bra{\vec R_1 ,\beta_1} \hat H^\text{DFT}(\vec d_{\vec R_2}) \ket{\vec R_2, \beta_2} e^{i \vec k_2 \cdot \vec R_2}.
\end{equation}
To extract $\vec R_2$ in $\hat H^\text{DFT}(\vec d_{\vec R_2})$ for the summation, we transform $\hat H^\text{DFT}(\vec d_\vec R)$ to the momentum-space by 
\begin{equation}
    \hat H^\text{DFT}(\vec d_\vec R) = \sum_{\tilde{\vec G}}e^{-i \tilde{\vec G} \cdot \vec d _ \vec R} \hat H^\text{DFT}(\tilde{\vec G})
\end{equation}
as $\hat H^\text{DFT}(\vec d_\vec R + x \tilde{\vec a}_1 + y \tilde{\vec a}_2) = \hat H^\text{DFT}(\vec d_\vec R)$ is periodic for atomic lattice vectors ($x,y$ are integers here), as shown in \figref{fig:DFT lattice structures}(b). We also denote the atomic reciprocal lattice vector $\tilde{\vec G} = \tilde{\vec G}_{wz} = w \tilde{\vec b}_1 + z \tilde{\vec b}_2$ with integers $w,z$, so the summation over $\tilde{\vec G}$ is equivalent to the summation over $w,z$. 
$\tilde{\vec b}_{1,2}$ are atomic reciprocal lattice vectors satisfying $\tilde{\vec b}_{i}\cdot \tilde{\vec a}_j = \delta_{ij}$ for $i,j=1,2$ and shown in \figref{fig:DFT lattice structures}(c).
%Rewrite the phase factor $e^{-i \tilde{\vec G} \cdot \vec d _ \vec R}$ to extract $\vec R$ explicitly by 
%\addCXL{CX: what is wz here? }
Since
\begin{equation}\label{eq:DFT phase factor}
    \tilde{\vec G}_{wz} \cdot \vec d_\vec R  = \tilde{\vec G}_{wz} \cdot (\mathcal R (\theta)\vec R - \vec R) = ( \tilde{\vec G}_{wz} - \mathcal R (\theta) \tilde{\vec G}_{wz}) \cdot ( \mathcal R (\theta) \vec R )= \vec G_{wz} \cdot (\vec R +\vec d_\vec R) \approx \vec G_{wz} \cdot \vec R
\end{equation}
with the moir\'e reciprocal lattice vectors
$\vec G_{wz}$ given by
\begin{equation}
\begin{split}
    \vec G_{wz} & = \tilde{\vec G}_{wz} -  \mathcal R(\theta) \tilde{\vec G}_{wz} \\ 
    & =(w \tilde{\vec b}_1 + z \tilde{\vec b}_2) -  \mathcal R(\theta) (w \tilde{\vec b}_1 + z \tilde{\vec b}_2) \\
    & =w (\tilde{\vec b}_1 -  \mathcal R(\theta) \tilde{\vec b}_1) + z(\tilde{\vec b}_2 -  \mathcal R(\theta)\tilde{\vec b}_2) \\
    & = w (\tilde{\vec b}_1 - \tilde{\vec b}_1') + z(\tilde{\vec b}_2 - \tilde{\vec b}_2') \\
    & = w {\vec b_1^\text{M}} + z {\vec b_2^\text{M}},
\end{split}
\end{equation}
we have 
\begin{eqnarray} \label{eq:HDFT_Fourier2}
e^{-i \tilde{\vec G} \cdot \vec d _ \vec R}\approx e^{-i \vec G \cdot \vec R}, \quad \hat H^\text{DFT}(\vec d_\vec R) \approx  \sum_{w,z}e^{-i \vec G_{wz} \cdot \vec R} \hat H^\text{DFT}(\tilde{\vec G}_{wz}). 
\end{eqnarray}
Here $\tilde{\vec a}_{1,2}' = \mathcal R(\theta) \tilde{\vec a}_{1,2} \  (\tilde{\vec b}_{1,2}' = \mathcal R(\theta) \tilde{\vec b}_{1,2})$ are primitive (reciprocal) lattice vectors for the twisted Sb$_2$ layer as shown in \figref{fig:DFT lattice structures}(c) and $\vec b_{1,2}^\text M= \tilde{\vec b}_{1,2} - \tilde{\vec b}_{1,2}'$ are the moir\'e reciprocal lattice vectors.
The approximation in \eqnref{eq:DFT phase factor} is valid as $\vert \vec d_\vec R \vert \sim \mathcal O(\vert \tilde{\vec a}_1 \vert ) \ll \vert \vec a_1^{\text M} \vert$ and $\vec G \cdot \vec d_\vec R \ll  \vec G \cdot \vec a_{1}^\text M \sim \mathcal O(1)$.

% \begin{equation}\label{eq:DFT phase factor}
% \begin{split}
% 	\tilde{\vec G}_{wz} \cdot \vec d_\vec R & = (w \tilde{\vec b}_1 + z \tilde{\vec b}_2 ) \cdot (x \tilde{\vec a}_1'  + y \tilde{ \vec a}_2' - x \tilde{ \vec a}_1  - y \tilde{ \vec a}_2) \\
% 	& = (w \tilde{\vec b}_1 + z \tilde{\vec b}_2 - w \tilde{\vec b}_1' - z \tilde{\vec b}_2' ) \cdot (x \tilde{\vec a}_1'  + y \tilde{\vec a}_2') \\
% 	& = (w \vec b_1^\text M + z \vec b_{2}^\text M) \cdot (x \tilde{\vec a}_1 + y \tilde{\vec a}_2 + \vec d_\vec R) \\
% 	& \approx \vec G_{wz} \cdot \vec R
% \end{split}    
% \end{equation}
% where $\tilde{\vec a}_{1,2}' = \mathcal R(\theta) \tilde{\vec a}_{1,2} \  (\tilde{\vec b}_{1,2}' = \mathcal R(\theta) \tilde{\vec b}_{1,2})$ are primitive (reciprocal) lattice vectors for the twisted Sb$_2$ layer as shown in \figref{fig:DFT lattice structures}(c) and $\vec b_{1,2}^\text M= \tilde{\vec b}_{1,2} - \tilde{\vec b}_{1,2}'$ and $\vec G_{wz}$ are the moir\'e reciprocal lattice vectors.
% The reason for the approximation in the fourth line of \eqnref{eq:DFT phase factor} is $\vert \vec d_\vec R \vert \sim \mathcal O(\vert \tilde{\vec a}_1 \vert ) \ll \vert \vec a_1^{\text M} \vert$ and $\vec G \cdot \vec d_\vec R \ll  \vec G \cdot \vec a_{1}^\text M \sim \mathcal O(1)$.
Substituting \eqnref{eq:DFT approximation} and (\ref{eq:HDFT_Fourier2}) into \eqnref{eq:DFT Wannier basis} leads to
\begin{equation}\label{eq:DFT ham relation with Moire ham}
    \begin{split}
		\bra{\vec k_1 ,\beta_1} \hat H_0 \ket{\vec k_2, \beta_2} & \approx \sum_{\vec R_1 \vec R_2, w, z} e^{- i \vec k_1 \cdot \vec R_1} e^{-i \vec G_{wz} \cdot \vec R_2} \bra{\vec R_1, \beta_1}  \hat H^\text{DFT}(\tilde{\vec G}_{wz}) \ket{\vec R_2, \beta_2} e^{i \vec k_2 \cdot \vec R_2} \\
		& = \sum_{w,z} \bra{\vec k_1 ,\beta_1} \hat H^\text{DFT}(\tilde{\vec G} _{wz}) \ket{\vec k_2 -\vec G_{wz}, \beta_2} \\
		& = \sum_{w,z}\delta(\vec k _2 - \vec k_1 -\vec G_{wz}) \bra{\vec k_1, \beta_1} \hat H^\text{DFT}(\tilde{\vec G}_{wz}) \ket{\vec k_1 ,\beta_2}.
	\end{split}
\end{equation}
The last line comes from the conservation of crystal momenta of $H^\text{DFT}(\vec d)$, 
\begin{equation}\label{eq:DFT fourier transform}
    \bra{\vec k_1, \beta_1}\hat H^\text{DFT}(\tilde{\vec G})\ket{\vec k_2, \beta_2} 
    = \int \dd^2 \vec d e^{i \tilde{\vec G} \cdot \vec d} \bra{\vec k_1, \beta_1} \hat H^\text{DFT}(\vec d) \ket{\vec k_2, \beta_2} 
    = \delta(\vec k_1 - \vec k_2 ) \bra{\vec k_1, \beta_1}\hat H^\text{DFT}(\tilde{\vec G})\ket{\vec k_1, \beta_2}.
\end{equation}
\eqnref{eq:DFT ham relation with Moire ham} connects $H^\text{DFT}$ and $H_0$ in atomic Bloch states in general, which is applied to the moir\'e potential in our model next. % \addCXL{CX: why this sentence just stops here? }  

We next show the relation between the potential $\tilde{\Delta}(\vec d_\vec R)$ from DFT and the moir\'e potential $\Delta(\vec r)$ in our model by \eqnref{eq:DFT ham relation with Moire ham}.
The Fourier transform of $\bra {\vec k,\alpha}\hat H^\text{DFT}(\vec d_\vec R)\ket{\vec k,\alpha}$ in \eqnref{eq:DFT ham} by \eqnref{eq:DFT fourier transform} is
\begin{equation}
    \bra {\vec k,\beta_1}\hat H^\text{DFT}(\tilde{\vec G})\ket{\vec k,\beta_2} = 
    \begin{pmatrix}
        h_D^t(\vec k) & m s_0 \\
        m s_0 & h_D^b(\vec k)
    \end{pmatrix} \delta_{\tilde{\vec G}=0} + 
    \begin{pmatrix}
        \tilde \Delta(\tilde{\vec G})s_0 & 0 \\
        0 & \alpha \tilde \Delta(\tilde{\vec G}) s_0
    \end{pmatrix}.
\end{equation}
In atomic Bloch basis, the moir\'e Hamiltonian from Eq.1 of the main text without external electrical field is
\begin{equation}
    \bra {\vec k_1, \beta_1}\hat H_0\ket{\vec k_2,\beta_2} =  
    \begin{pmatrix}
        h_D^t(\vec k_1) & m s_0 \\
        m s_0 & h_D^b(\vec k_1)
    \end{pmatrix} \delta(\vec k_2 - \vec k_1) + 
    \begin{pmatrix}
        \Delta(\vec G)s_0 & 0 \\
        0 & \alpha \Delta(\vec G) s_0
    \end{pmatrix}\delta(\vec k_2 - \vec k_1 - \vec G).
\end{equation}
By comparison of two Hamiltonian following \eqnref{eq:DFT ham relation with Moire ham}, one obtains
\begin{equation}\label{eq:DFT relation between Delta}
    \Delta (\vec G_{wz}) = \tilde \Delta(\tilde{\vec G}_{wz})
\end{equation}
for $w,z$ as integers, $\vec G_{wz} = w \vec b_1^\text M + z \vec b_2^\text M$, and $\tilde{\vec G}_{wz} = w \tilde{\vec b}_1 + z \tilde{\vec b}_2$.
In real space, this leads to
\begin{equation}
    \Delta (\vec R) = \sum_{w,z} e^{- i \vec G_{w,z} \cdot \vec R} \Delta (\vec G_{wz}) \approx \sum_{w,z} e^{- i \tilde{\vec G}_{w,z} \cdot (\mathcal R (\theta) \vec R - \vec R)} \tilde \Delta(\tilde{\vec G}_{wz}) = \tilde \Delta(\mathcal R (\theta) \vec R - \vec R) = \tilde \Delta(\vec d_\vec R)
\end{equation}
%\addCXL{Will it be better to use $\vec R$ instead of $\vec r$ here? }
following \eqnref{eq:DFT phase factor} and \eqnref{eq:DFT relation between Delta}, reproducing Eq. (4) in the main text.
It then can be interpolated to the whole real space by replacing the atomic lattice vectors $\vec R$ by the continuous variable $\vec r$, 
\begin{equation}
    \Delta (\vec r) \approx \tilde \Delta(\mathcal R (\theta) \vec r - \vec r),
\end{equation}
because the atomic length scale is much smaller than the moir\'e length scale for small twist angles so that it is a good approximation to take the continuous limit for the atomic length scale.

\end{document}